\documentclass[]{aastex61}
\pdfoutput=1
\shorttitle{Characterization of Near-Earth Asteroids using KMTNet-SAAO}
\shortauthors{Erasmus et al.}

\begin{document}

\title{Characterization of Near-Earth Asteroids using KMTNet-SAAO}

\correspondingauthor{Nicolas Erasmus}
\email{nerasmus@saao.ac.za}

\author{N. Erasmus}
\affil{South African Astronomical Observatory, Cape Town, 7925, South Africa.}

\author{M. Mommert}
\affil{Department of Physics and Astronomy, Northern Arizona University, Flagstaff, AZ 86001, USA.}

\author{D. E. Trilling}
\affil{Department of Physics and Astronomy, Northern Arizona University, Flagstaff, AZ 86001, USA.}
\affil{South African Astronomical Observatory, Cape Town, 7925, South Africa.}

\author{A. A. Sickafoose}
\affil{South African Astronomical Observatory, Cape Town, 7925, South Africa.}
\affil{Department of Earth, Atmospheric, and Planetary Sciences, Massachusetts Institute of Technology, Cambridge, MA 02139-4307, USA.}

\author{C. van Gend}
\affil{South African Astronomical Observatory, Cape Town, 7925, South Africa.}

\author{J. L. Hora}
\affil{Harvard-Smithsonian Center for Astrophysics, Cambridge, MA 02138-1516, USA.}

\begin{abstract}

 We present here \textit{VRI} spectrophotometry of 39 near-Earth asteroids (NEAs) observed with the Sutherland, South Africa, node of the Korea Microlensing Telescope Network (KMTNet). Of the 39 NEAs, 19 were targeted, but because of KMTNet's large 2~deg~$\times$~2~deg field of view, 20 serendipitous NEAs were also captured in the observing fields. Targeted observations were performed within 44 days (median: 16 days, min: 4 days) of each NEA's discovery date. Our broadband spectrophotometry is reliable enough to distinguish among four asteroid taxonomies and we were able to confidently categorize 31 of the 39 observed targets as either a S-, C-, X- or D-type asteroid by means of a Machine Learning (ML) algorithm approach. Our data suggest that the ratio between ``stony" S-type NEAs and ``not-stony" (C+X+D)-type NEAs, with \textit{H} magnitudes between 15 and 25, is roughly 1:1.  Additionally, we report $\sim$1-hour light curve data for each NEA and of the 39 targets we were able to resolve the complete rotation period and amplitude for six targets and report lower limits for the remaining targets.

\end{abstract}


\section{Introduction} \label{sec:intro}

\begin{figure}
	\centering
	\includegraphics[width=85mm]{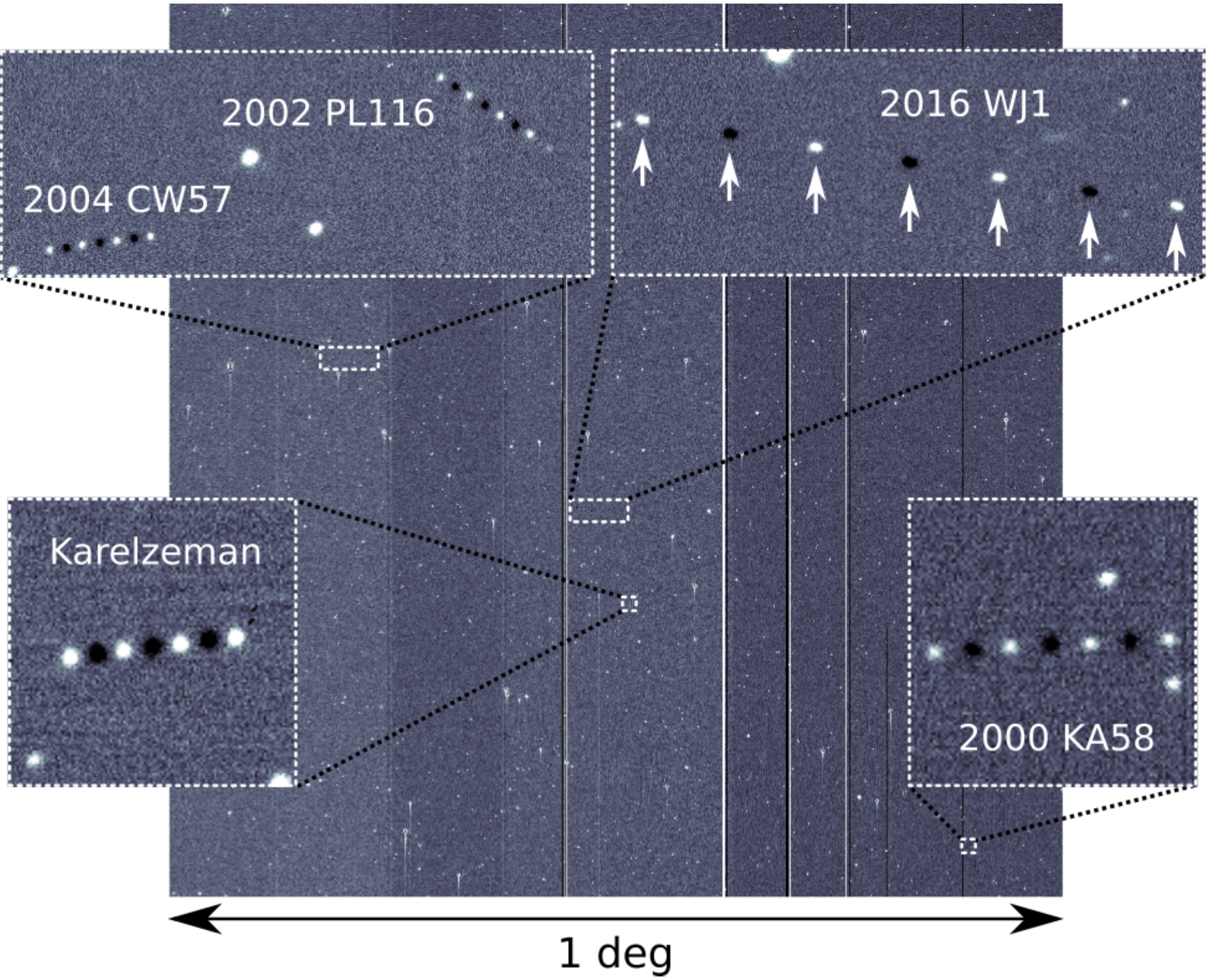}
	\caption{One of KMTNet's four CCD chips, illustrating the large 1 deg $\times$ 1 deg field of view that results in multiple serendipitous asteroids being detected. The zoomed-in areas show four serendipitously observed main belt asteroids detected in this one quadrant alone with NEA 2016 WJ1, in the center of the CCD and indicated with white arrows, the originally targeted asteroid. To visualize the asteroids in the figure seven consecutive V-filter images are stacked. The images are taken roughly three minutes apart. Even-numbered images are subtracted (asteroid appears dark) and odd-numbered images are added (asteroid appears bright)}
	\label{fig1}
\end{figure}

Near Earth Asteroids (NEAs) are a group of asteroids that have orbits that bring them close to the Earth's orbit. The majority are either Apollos or Atens which have earth-crossing orbits and these pose a direct threat to Earth as they potentially have impacting orbits, the most significant recent example being the Chelyabinsk airburst in 2013 speculated to be a fragment of the NEA 1999 NC43 \citep{Borovicka2013,Brown2013}. While more recent work has shown that this link between Chelyabinsk and 1999 NC43 is tenuous at best \citep{Reddy2015}, impacts are still inevitable in the future and mitigation strategies is a subject undergoing intense study \citep{Sanchez2009}. On the other hand, the close proximity of NEAs to Earth has its benefits. Being nearby on an astronomical scale makes them convenient small Solar System  objects to characterize with Earth-bound measurements (see Figure \ref{fig1} for an example image of our KMTNet observations). Coming close to Earth also makes them attractive objects to visit and investigate with spacecraft. This is evident with the numerous ongoing and proposed space exploration missions from the scientific community, like JAXA's Hayabusa \citep{Yano2006} and Hayabusa2 \citep{Tsuda2013} missions, and NASA's OSIRIS-REx \citep{Lauretta2017} and DART \citep{Cheng2016} missions. There is also an increasing interest in space mining endeavors from the technology and industrial community \citep{Elvis2012}. Determining the physical properties and composition of NEAs is critical to all of the above.

Other than planetary defense, space exploration, and more recently space mining, NEAs are of interest from a more fundamental scientific research perspective as well. The source of the NEA population has been a postulated topic for decades. The current most widely accepted model is that the NEA population is in steady state and is continuously being replenished by asteroids from the main asteroid belt through resonant motion with Jupiter and Saturn \citep{Bottke2000,Granvik2017}. However, these distribution models assume a uniform and size-independent compositional distribution and therefore composition survey studies like this and others \citep[e.g., ][]{Mommert2016} can help with further refinements of future NEA distribution models.

\begin{figure}
	\centering
	\includegraphics[width=85mm]{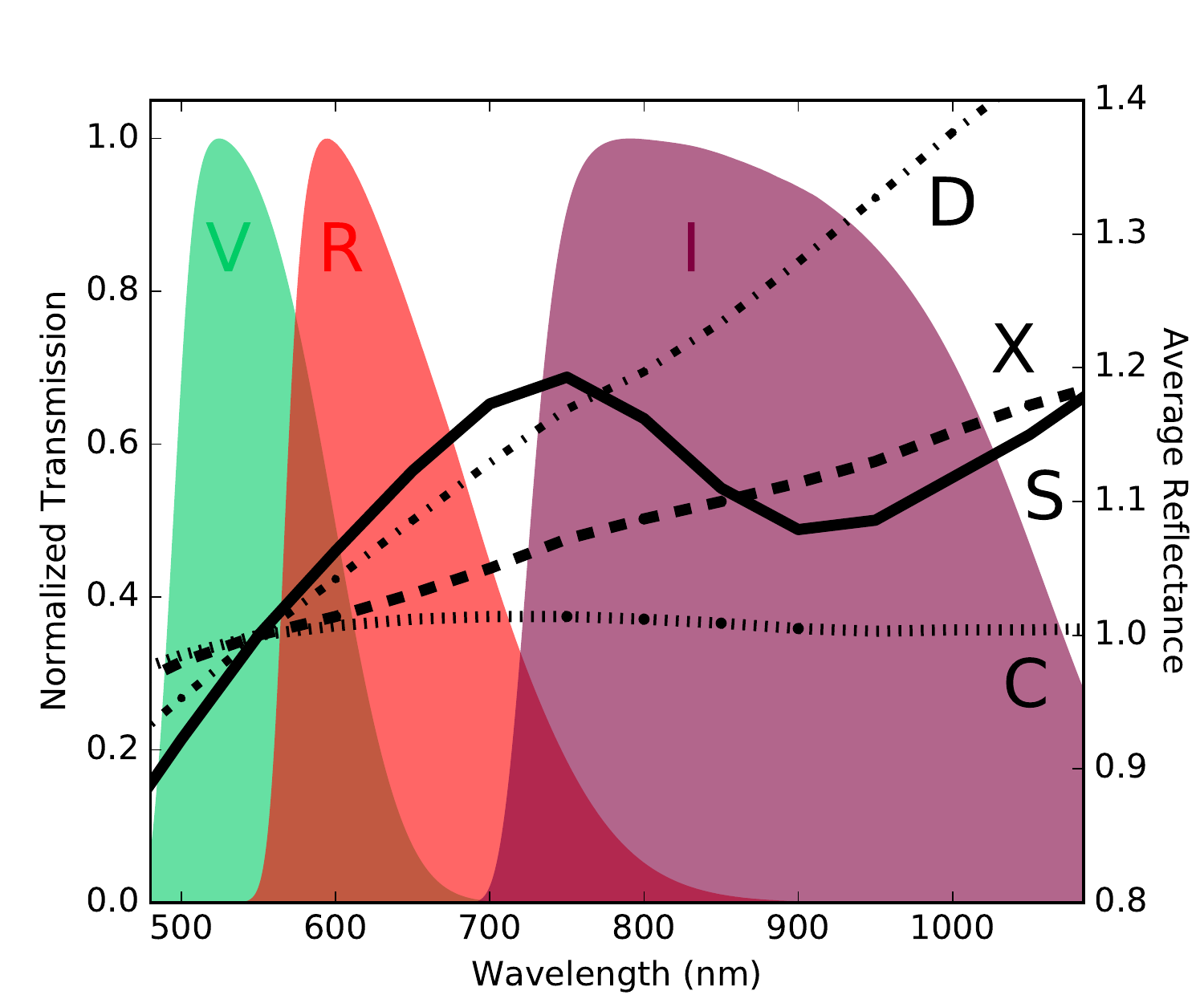}
	\caption{The typical normalized transmission of \textit{V}, \textit{R} and \textit{I} Johnson-Cousins filters are plotted together with the averaged visible wavelength reflectance spectra, normalized at 550~nm, of four taxonomic types \citep{DeMeo2009}. This shows that \textit{VRI} colors are adequate to distinguish among S-, C-, X- and D-type asteroid taxonomies. }
	\label{fig2}
\end{figure}

One of the most common methods to determine asteroid composition remotely is with spectroscopy. The reflected light spectrum of an asteroid can be compared to the reflectance spectra of known collected meteorite types and in this way the composition of asteroids can be inferred. The most widely used spectral system is the Bus-DeMeo taxonomic scheme \citep{DeMeo2009} consisting of 24 spectral types and sub-types that covers the wavelength range of 0.45--2.45~$\micron$.  Although there are many taxonomic classes, NEAs fall mainly in two groups: silicon-rich (``stony") asteroids which are classified as S-type asteroids and those that are ``not-stony" which could be one of several classes. The most common being C-type (carbonaceous) asteroids, X-type (several different possible compositions) asteroids, and lastly D-type  (even more primitive material) asteroids. These three classes all have featureless flat or reddish spectral slopes in the visible region. Figure \ref{fig2} shows reflectance spectra for these four classes. 

In some cases it is adequate to sample the reflectance spectra at strategic wavelengths using broadband filters in order to constrain the taxonomic class. This technique is referred to as spectrophotometry and is the method used in this study. In spectrophotometry broadband photometric filters are used. This has the benefit of being employable to investigate fainter asteroids. This significantly increases the observable population numbers since the size distribution is heavily skewed towards smaller asteroids \citep{Mainzer2011}. 

Here we present the \textit{VRI} spectrophotometry observations and $\sim$1-hour light curve data of 39 NEAs. In Section \ref{sec:Obs&Data} the observational details, the method for extracting serendipitously discovered NEAs, and the photometry pipeline used are explained.  In Section \ref{sec:Results} the photometric results of the 39 NEAs observed are presented. Section \ref{sec:Analysis} describes how the color, classification and light curve of each NEA was determined from the photometric data. Sections \ref{Discussion} and \ref{Conclusion} conclude this study by discussing the analyzed data, comparing to other published work, and presenting our conclusions. 

\section{Observations and Data Reduction} \label{sec:Obs&Data}

Observations were made with the Sutherland, South Africa, node of the Korea Microlensing Telescope Network (KMTNet) \citep{Kim2016}. The telescope has a primary mirror of 1.6 m in diameter and is fitted with four 9k $\times$ 9k CCDs, mosaicking the 2 deg $\times$ 2 deg field of view. Each CCD covers 1 deg $\times$ 1 deg of sky with a plate-scale of 0.40 arcsec/pixel (Figure \ref{fig1}). The vertical and horizontal gaps between the CCDs are $184\arcsec$ and $373\arcsec$, respectively. Johnson-Cousins \textit{B, V, R} and \textit{I} filters are installed and only sidereal tracking is available. The minimum recommended exposure time for an efficient cyclical observation is 60 seconds due to the large readout time of $\sim$75 seconds.

Observations were completed over 23 nights falling between 25 October 2016 -- 20 February 2017 and newly discovered NEAs were specifically targeted for this study. On average 1--2 new NEAs are discovered every night \citep{Galache2015} and recorded in the Minor Planet Center's (MPC) NEA database\footnote{\url{http://minorplanetcenter.net/}}. On each observing night the database was queried for any new discoveries observable from Sutherland and the latest and most promising candidates were selected. This approach meant targeted observations were performed within 44 days (median: 16 days, min: 4 days) of each NEA's discovery date. 

Telescope pointing was offset to place the asteroid in the center of the top-right CCD in order to avoid the gaps between the CCDs. Observations were performed alternating between \textit{V, R} and \textit{I} filters in the sequence\textit{VRVI}, repeating the sequence continuously for roughly one hour per target. NEAs typically move non-sidereally between 0.5--5~arcsec/minute mostly dependent on how close they are to Earth. The target lists included only objects moving slower than than 2~arcsec/minute to prevent any significant streaking in the chosen 60-second exposure. This exposure time limited this study to asteroids with magnitudes $V\lesssim$ 21 to achieve the desired SNR in a single exposure of at least 10.

All observed fields were scanned for serendipitously observed asteroids by calculating the ephemerides for all $\sim$700,000 known asteroids in the MPC's database\footnote{\url{http://minorplanetcenter.net/data/}} at the respective mid-time of the observation. Ephemerides were calculated using PyEphem\footnote{\url{http://rhodesmill.org/pyephem/}} and any asteroid falling within the observed field's coordinate perimeter was flagged. Flagged asteroids were queried on JPL's Horizons interface\footnote{\url{https://ssd.jpl.nasa.gov/horizons.cgi}} using the \texttt{Python} module CALLHORIZONS\footnote{\url{https://github.com/mommermi/callhorizons}} and discarded if $V>$ 21. The uncertainties in coordinates given by Horizons were also recorded. In total 2650 serendipitous asteroids were captured and the majority were main belt asteroids with only 20 being NEAs with uncertainties in the coordinates of $<$10 arcseconds. 

Photometry and light curve extraction was achieved using PHOTOMETRYPIPELINE developed by \cite{Mommert2017}. The pipeline utilizes the widely used Source Extractor software\footnote{\url{http://www.astromatic.net/software/sextractor/}} for source identification and aperture photometry. SCAMP\footnote{\url{http://www.astromatic.net/software/scamp/}} is used for image registration. Both image registration and photometric calibration are based on matching field stars with star catalogs from the Sloan Digital Sky Survey, the AAVSO Photometric All-Sky Survey, and GAIA.

\begin{figure*}
	\gridline{\fig{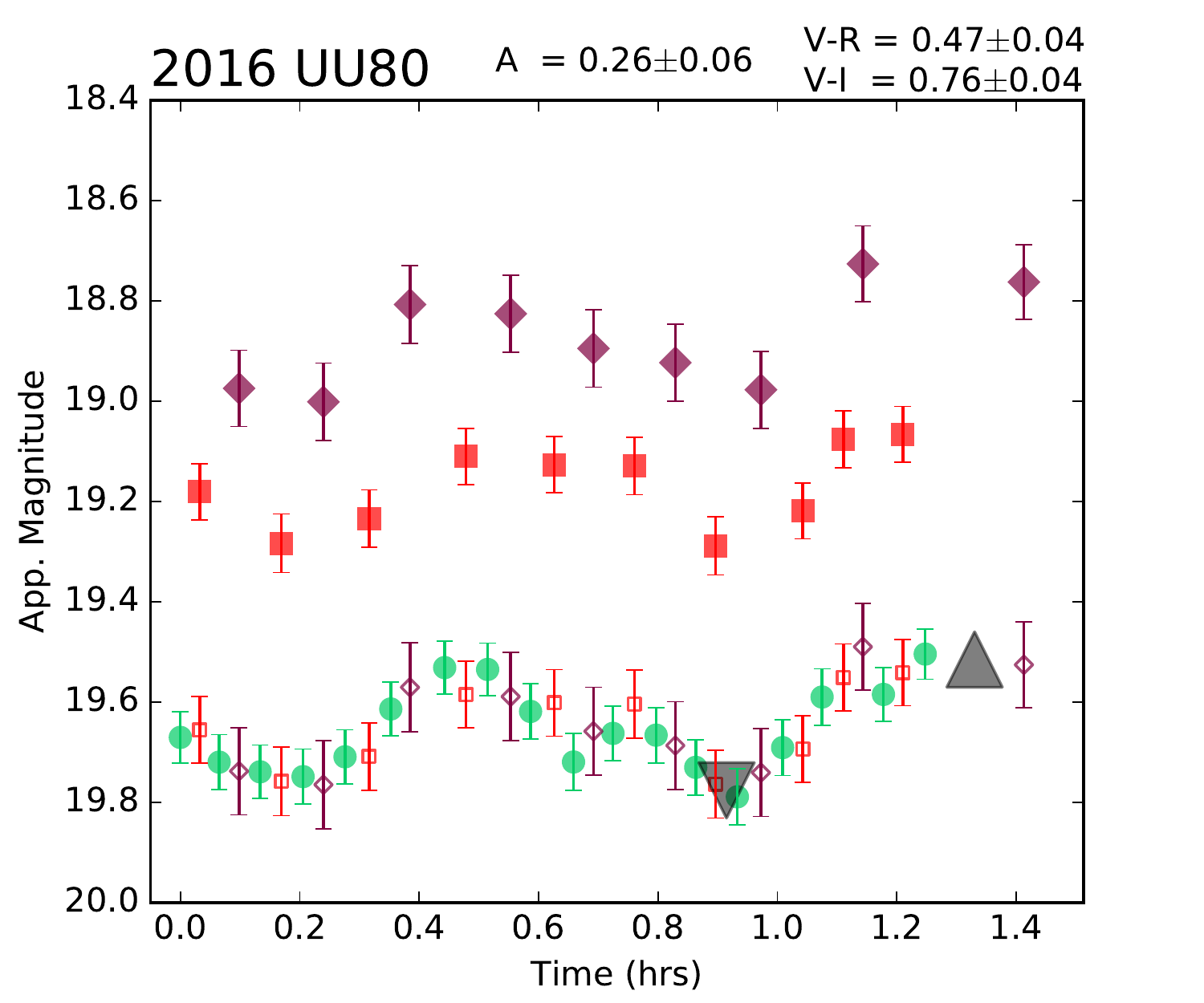}{0.5\textwidth}{(a)}
		\fig{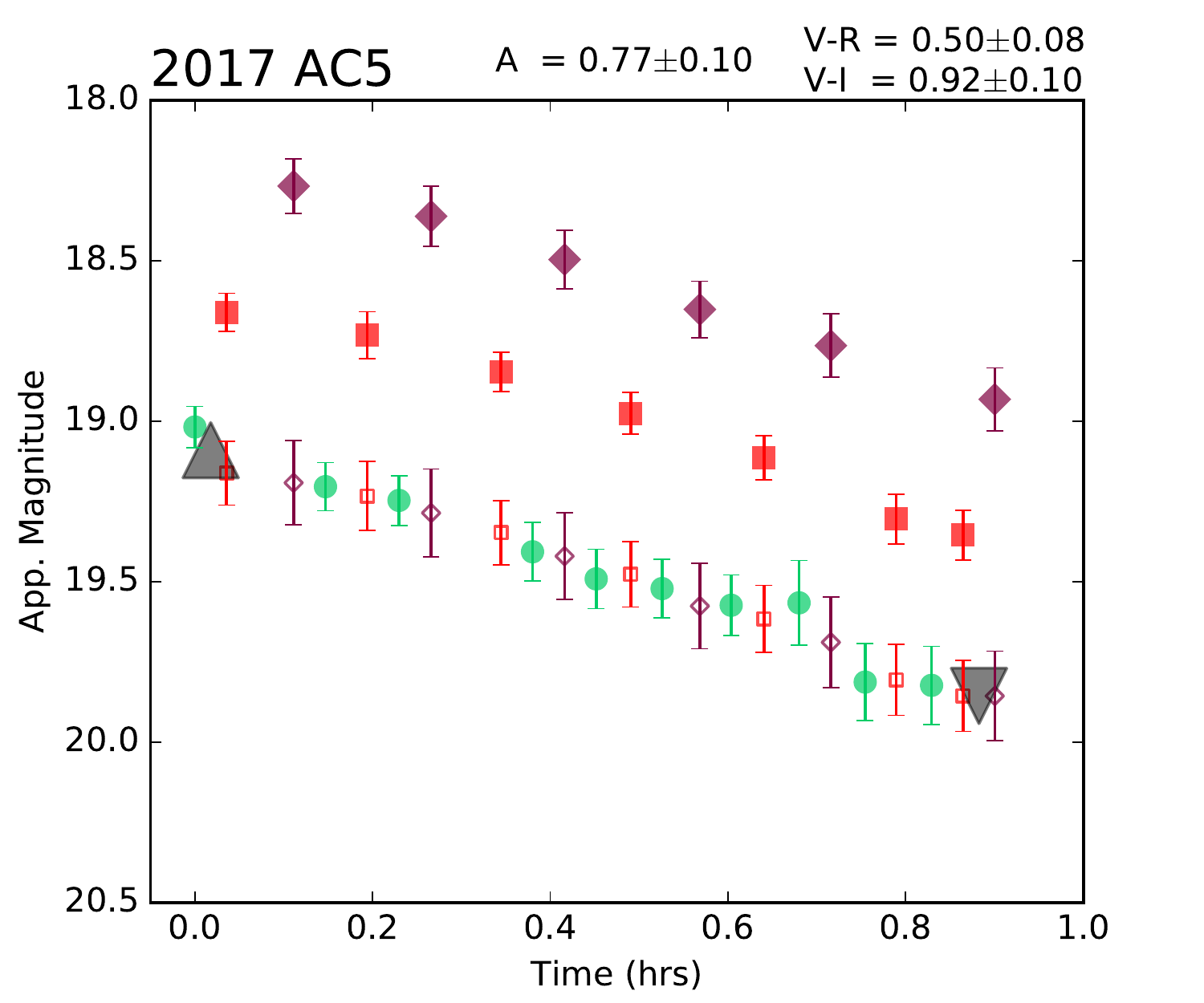}{0.5\textwidth}{(b)}
	}
	\gridline{\fig{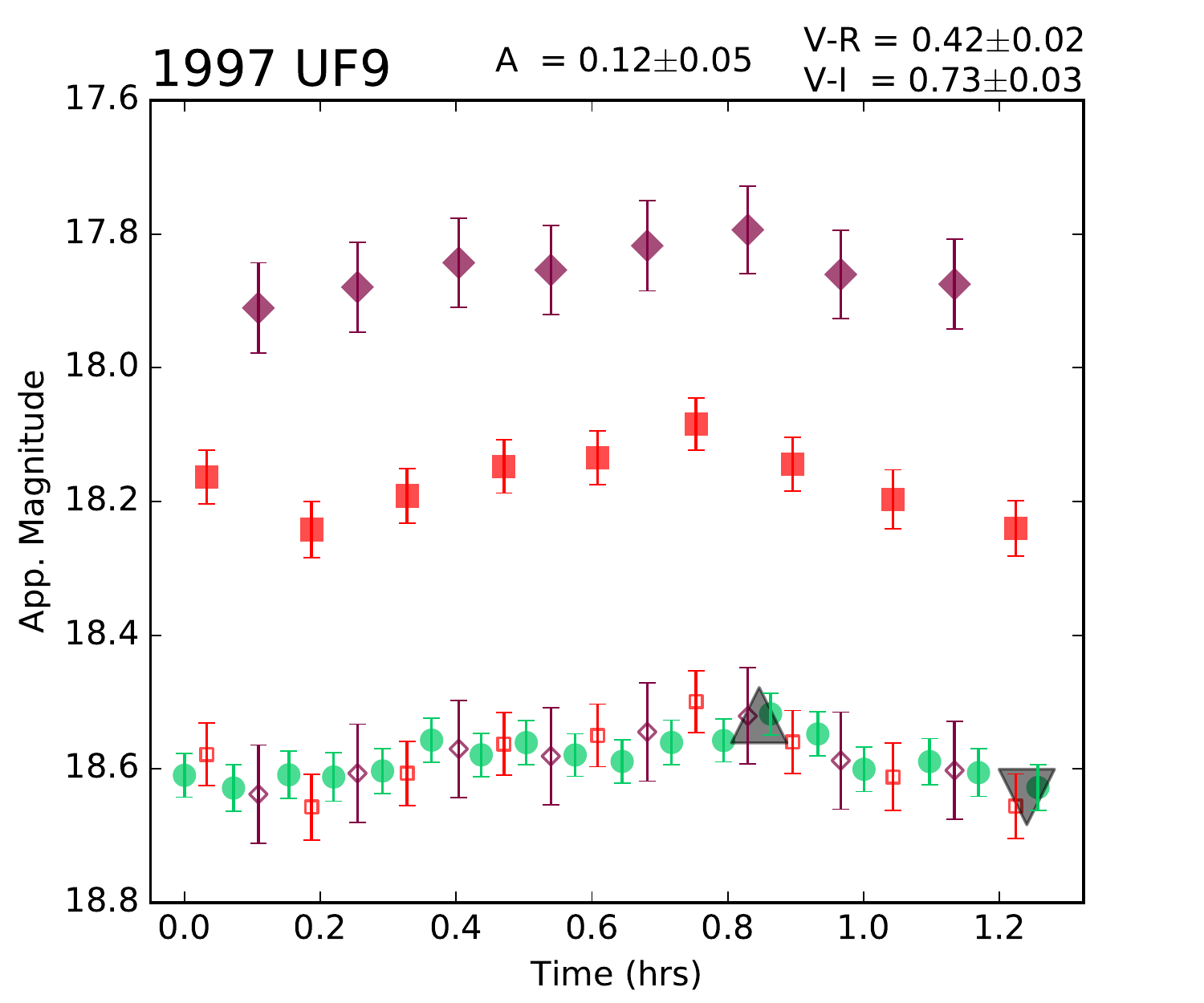}{0.5\textwidth}{(c)}
		\fig{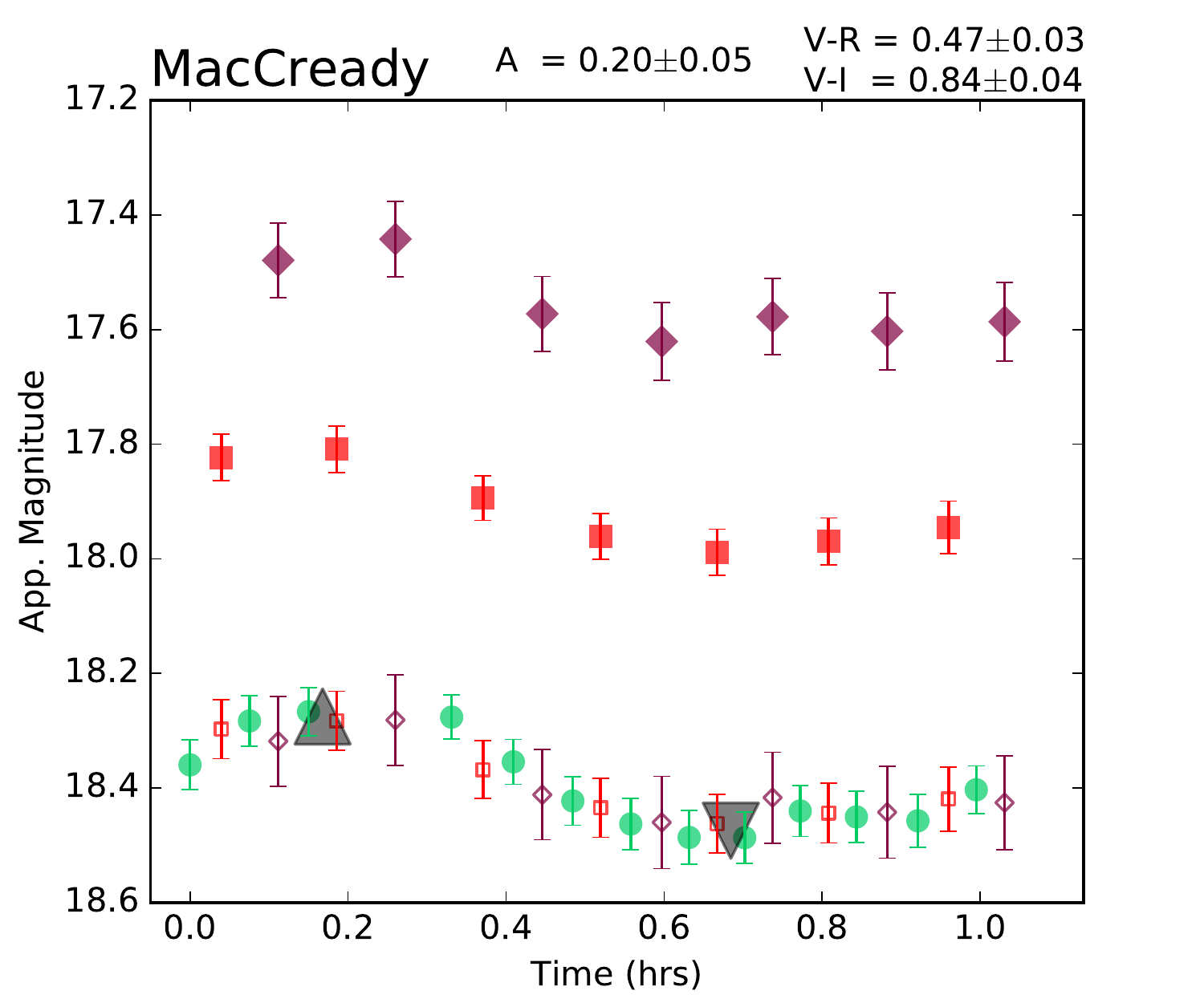}{0.5\textwidth}{(d)}
	}
	
	\caption{The spectrophotometric data for two of the 19 targeted NEAs ((a) \& (b)) and two of the 20 serendipitous NEAs ((c) \& (d)). The targets are (a) 2016 UU80, (b) 2017 AC5, (c) 1997 UF9 and (d) 1984 SS (MacCready). The data are the calibrated photometric results generated by the PHOTOMETRYPIPELINE  developed by \cite{Mommert2017}. The \textit{V} (green circles), the \textit{R} (red squares) and the \textit{I} (burgundy diamonds) filter data are shown. A final light curve (bottom set of data in each plot) is constructed by offsetting \textit{R} and \textit{I} data with a constant value (see Section \ref{subsec:LCurved}). The adjusted data points are displayed as smaller, hollow symbols interspersed with the \textit{V} data. The lower limit on the amplitude is calculated using the difference between magnitudes highlighted with the gray triangle symbols. The triangles are situated where the mean of two adjacent data points have the minimum/maximum magnitude. See text for further description. See the Appendix for data of all 39 observed NEAs.} 
	\label{fig3}
\end{figure*}

 \section{Results} \label{sec:Results}
 
Figure \ref{fig3} shows the spectrophotometric data for two of the 19 targeted NEAs and two of the 20 serendipitous NEAs (see the electronic edition of Figure \ref{fig3} for all observed NEAs). The data are the calibrated photometric results  generated by the PHOTOMETRYPIPELINE from the \textit{V}-, \textit{R}- and \textit{I}-filter images. Adjusted \textit{R} and \textit{I} data are also shown (inter-spaced between \textit{V} data) which results in a final light curve (bottom set of data in each plot). See Section \ref{subsec:Colors} \& \ref{subsec:LCurved} for details regarding the process for adjusting the \textit{R} and \textit{I} data points. Photometric uncertainties account for instrumental uncertainties as well as statistical uncertainties when compared to the photometric catalog. 

For a single exposure a SNR of 30--40 was achieved for targets with \textit{V}$\approx$18 and around 10 for \textit{V}$\approx$21. The SNR in the \textit{R} band was slightly better and in the \textit{I} band slightly worse. 
 
For each image the PHOTOMETRYPIPELINE also outputs cropped thumbnail images centered around  the predicted position of the target. The aperture size and aperture position is also indicated. These thumbnail images were manually inspected for all data points. Data points were manually deleted where in the corresponding image it was clear that the wrong source was selected, or that the target intersected a CCD defect or a star.  
   
\section{Analysis}\label{sec:Analysis}
\begin{figure*}
 	\gridline{\fig{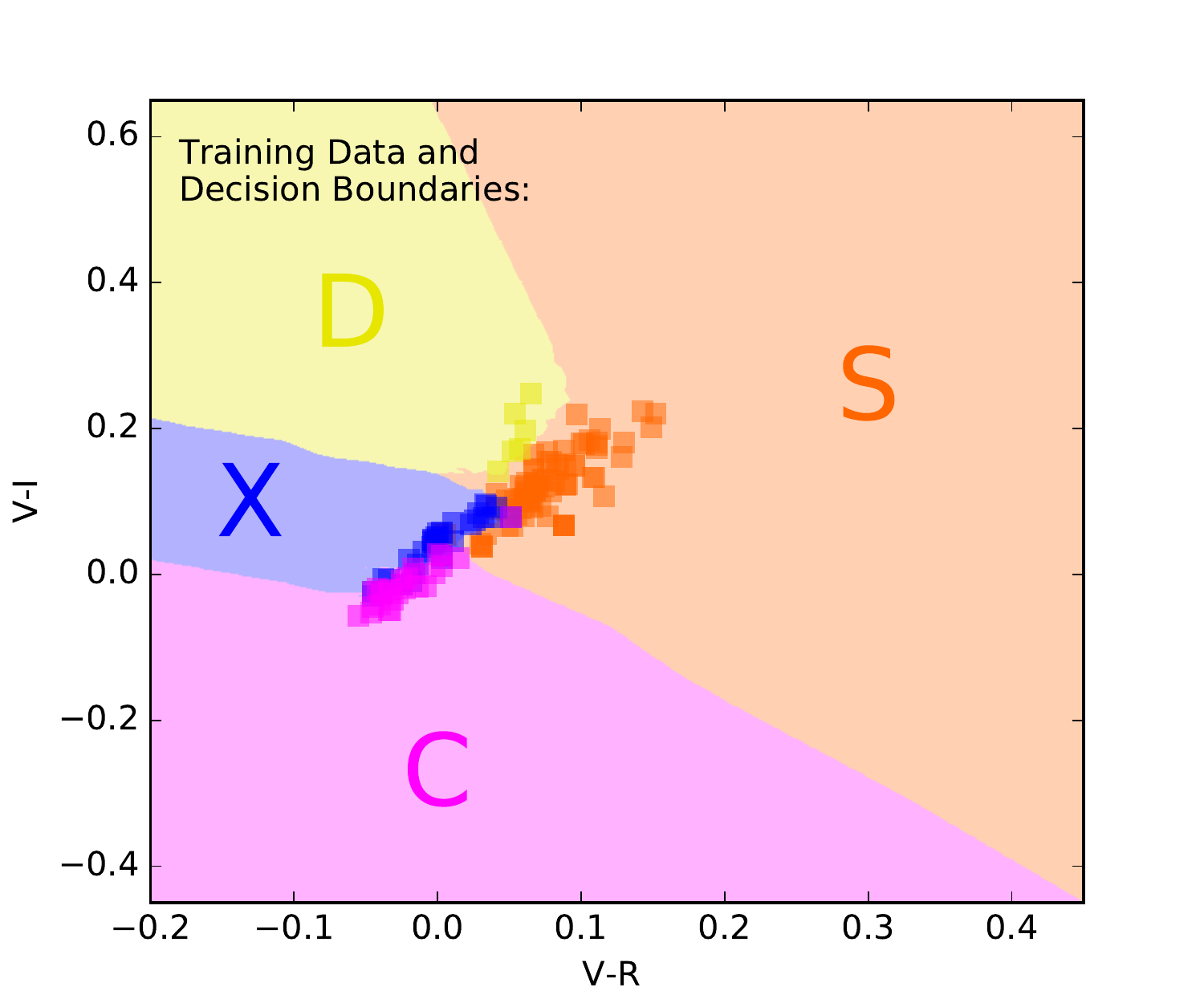}{0.5\textwidth}{(a)}
 		\fig{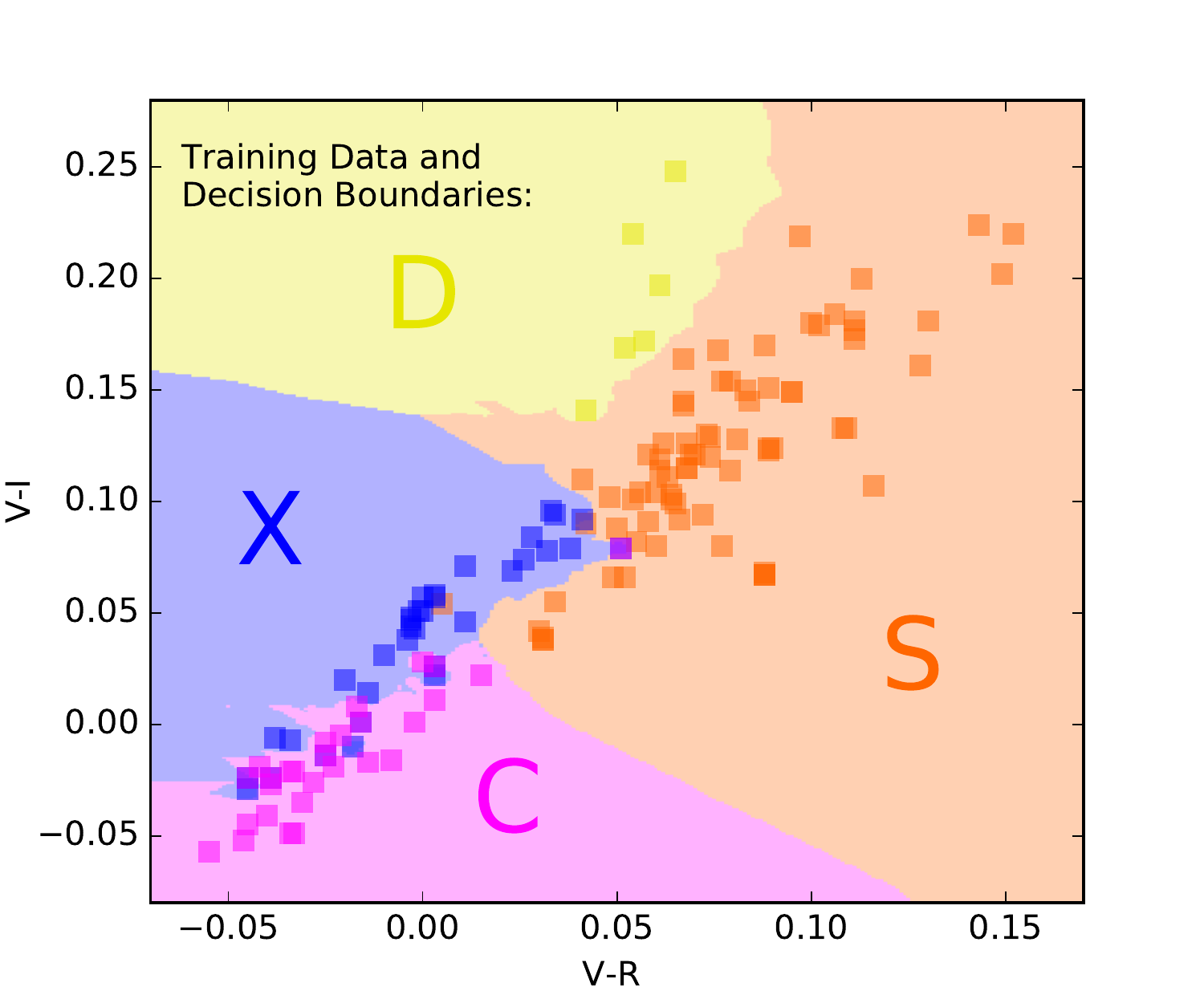}{0.5\textwidth}{(b)}
 	}
 	\gridline{\fig{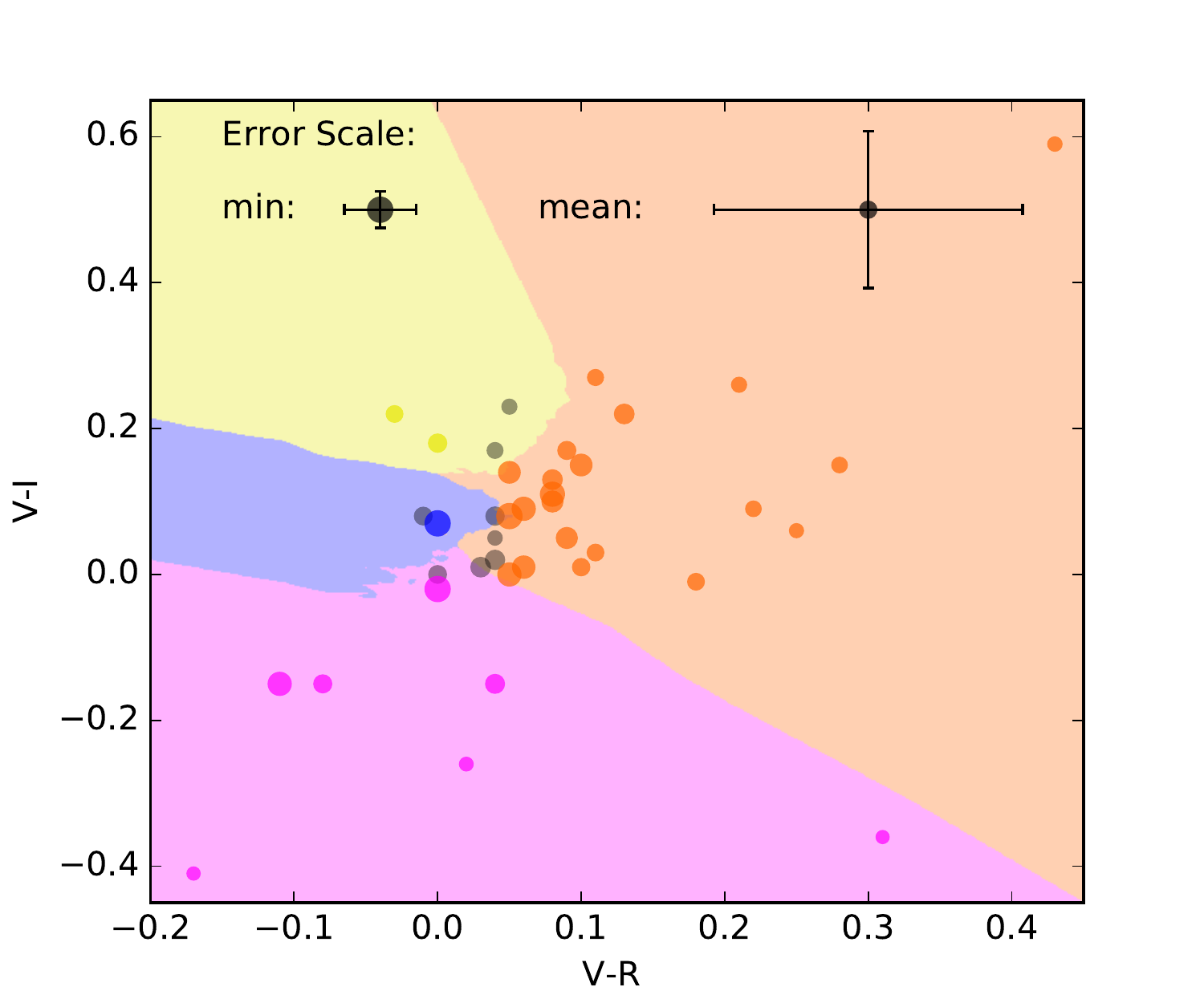}{0.5\textwidth}{(c)}
 		\fig{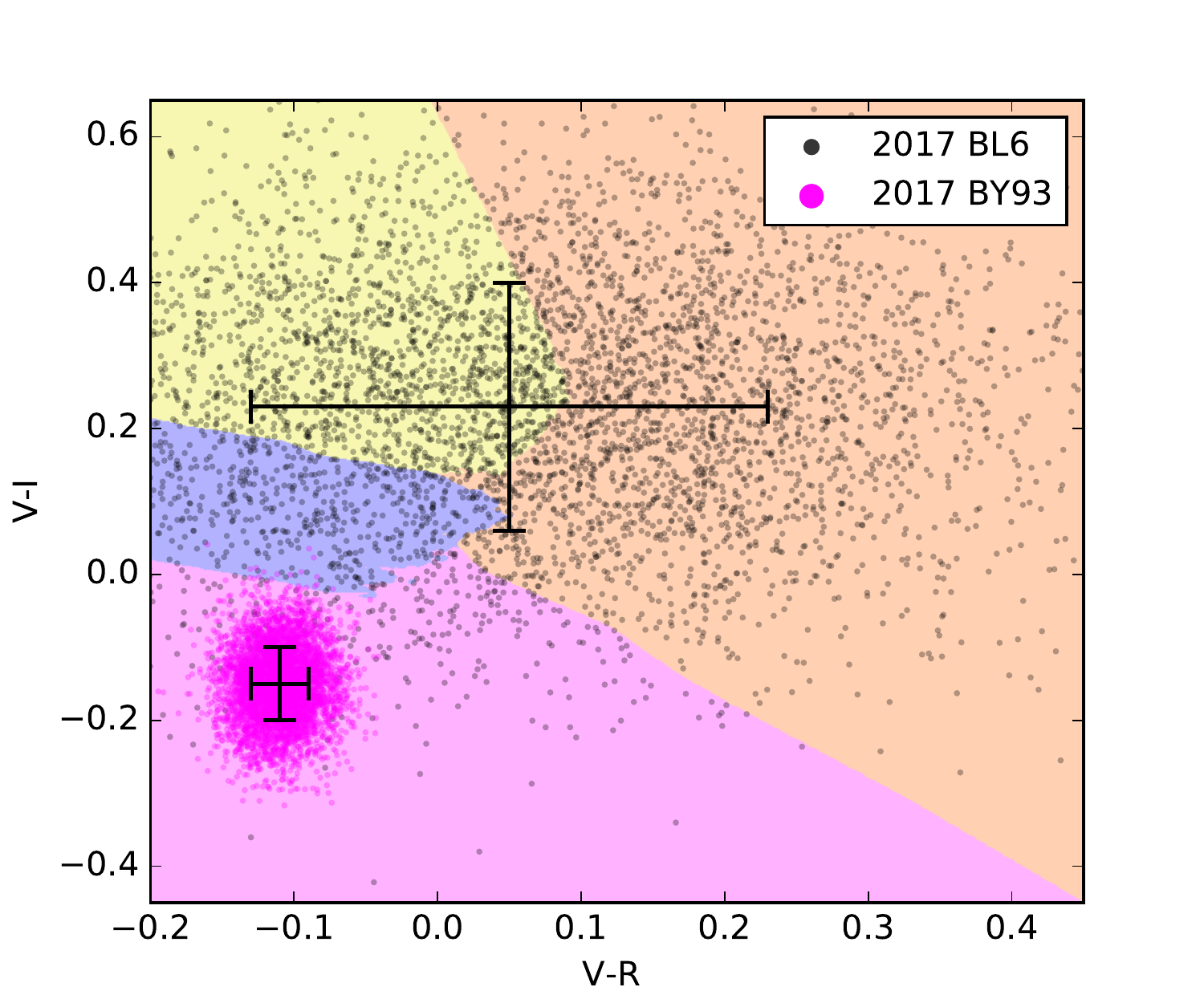}{0.5\textwidth}{(d)}
 	}
 	\caption{(a) Shown are the training data synthesized from MIT-UH-IRTF spectra (see Section \ref{subsec:Classification}) and resultant decision boundaries generated by the ML algorithm. (b) Zoomed-in version of (a). (c) Plotted are the calculated colors of all observed NEAs (see Section \ref{subsec:Colors} for detail) with the decision boundaries superimposed. NEAs in orange, blue, pink and yellow were classified by the ML algorithm (see Section \ref{subsec:Classification}) as S-, X- and C- and D-type asteroids, respectively. NEAs in gray remain unclassified since the highest taxonomic probability was less than 50\% . Sizes of the data points are inversely proportional to the uncertainty (scale is shown). (d) 5000 of the $10^6$ pseudo-measured colors are shown for two example targets. These pseudo-measured colors are individually classified using the ML algorithm and the frequency of each classification type tallied to determine the final classification probability. Note that 2017 BL6 is colored gray since the largest fraction of random pseudo-measured colors lie within the S-type zone but this fraction is 0.47 and below 50\%. Therefore this NEA's classification remains undetermined. The probabilities for each type (that is, fraction of all $10^6$ trials) are given in Table 1.
 	}
 	\label{fig4}
\end{figure*}

\subsection{Colors} \label{subsec:Colors}
The \textit{V}$-$\textit{R} and \textit{V}$-$\textit{I} colors of a NEA are calculated by simply subtracting the calibrated \textit{I} and \textit{R} magnitudes from the \textit{V} magnitude. However, the magnitudes can vary in time because asteroids are rotating and not perfectly spherical bodies. They may also have non-uniformity in surface color. Therefore, the subtraction calculation is only strictly correct if the different filter magnitudes are measured at the exact same time. For this study, sequential filter images were taken (see Section \ref{sec:Obs&Data})  with the time difference between the mid-time approximately three minutes on average. On this time-scale the respective filter magnitude could change significantly since some  NEAs have rotation periods on the order of minutes \citep{Thirouin2016}. To account for this, a linear interpolation was performed between adjacent \textit{V}  data points as described by \cite{Mommert2016}. The interpolation was used to obtain a corrected  \textit{V} magnitude  at times of inter-spaced non-\textit{V} observations. Respective colors were derived by subtracting the non-\textit{V} magnitudes from these interpolated \textit{V} magnitudes. The final color is the weighted average (by error in respective data points)  of all retrievable subtraction calculations in an observation. The final colors were also corrected for solar colors by subtracting the respective  \textit{V}$-$\textit{R} and \textit{V}$-$\textit{I} solar colors \citep{Binney1998}. The solar corrected colors of all observed NEAs are shown in Table \ref{table:results} and plotted in Figure \ref{fig4} (c).

\subsection{Taxonomic Classification}\label{subsec:Classification}
To classify each observed NEA as one of S-, X-, C- or D-type asteroid, a Machine Learning (ML) algorithm approach was employed as described by \cite{Mommert2016}. The ML algorithm was trained using asteroid colors that were synthesized from measured asteroid spectra taken from the MIT-UH-IRTF Joint Campaign for NEO Spectral Reconnaissance\footnote{\url{http://smass.mit.edu/minus.html}}. The database contains spectra of over 600 NEAs obtained with \textit{SpeX} \citep{Rayner2003} on NASA's Infrared Telescope Facility. While most spectra range from 0.8--2.5~$\micron$, visible wavelength spectral data are also included from the MIT SMASS observing program\footnote{\url{http://smass.mit.edu/smass.html}}. The spectra were classified using the Bus-DeMeo taxonomy spectrum classification online routine\footnote{\url{http://smass.mit.edu/busdemeoclass.html}} and in the end 74, 34, 28 and 6 reliable \textit{V}$-$\textit{R} and \textit{V}$-$\textit{I}  colors of S-, X-, C- and D-type asteroids, respectively, could be synthesized by using the above described approach.  

The synthesized colors are plotted and shown in Figure \ref{fig4} (a) and a zoomed-in version shown in Figure \ref{fig4} (b). The resultant classification boundaries generated by the ML algorithm are  also indicated in these figures. The training of the ML algorithm was executed using the \texttt{scikit-learn} module \citep{Pedregosa2012} for \texttt{Python}. The k-nearest neighbor (k=5) method was implemented, as in \cite{Mommert2016}. 

In order to account for uncertainties in the measurement of each of the observed NEA's color, a Monte-Carlo approach was used where $10^6$ pseudo-measured colors were randomly generated with a Gaussian distribution within the corresponding 1$\sigma$ uncertainties of each NEA's real measured color. To illustrate this process two examples are shown in Figure \ref{fig4} (d). Each of the $10^6$ instances of each NEA was classified using the trained ML algorithm and the overall frequency of each taxonomic type was tallied. The classification probability for each taxonomic type was then calculated.  The most likely (i.e. highest probability) taxonomic type was assigned to the NEA only if this probability was larger than 50\% (see Table \ref{table:results} for probabilities where bold text indicates the most likely classification).  The assigned classifications are indicated in Figure \ref{fig4} (c) by means of the color of the data point (orange = S-type, blue = X-type, pink = C-type, yellow = D-type asteroid and gray = undetermined).

\begin{deluxetable*}{|l|c|c|c|c|c|c|r|r|c|c|cccc|c|}
	\floattable
	\rotate
	\tabletypesize{\tiny}		
	\tablecaption{Observations and Results \label{table:results}}
	\tablecolumns{16}
	\tablenum{1}
	\tablewidth{0pt}
	\tablehead{
		\colhead{Object} &\colhead{Obs. Midtime} & \colhead{Disc. Date} & \colhead{$\Delta$D\tablenotemark{\textit{a}}} &\colhead{Obs. Duration} & \colhead{H\tablenotemark{\textit{b}}} & \colhead{V}& \colhead{V-R}& \colhead{V-I} & \colhead{Amplitude\tablenotemark{\textit{c}}} &\colhead{Rot. Period\tablenotemark{\textit{c}}}& \colhead{S-type}& \colhead{X-type}&\colhead{C-type}&\colhead{D-type}&\colhead{PHA} \\
		\colhead{} &\colhead{(UT)} & \colhead{} & \colhead{(days)} &\colhead{(h:mm)} & \colhead{(mag)} & \colhead{(mag)}&  \colhead{(mag)} & \colhead{(mag)}& \colhead{(mag)}  &\colhead{(min)}& \multicolumn{4}{c}{(probability)}&\colhead{}
	}
	\startdata
	\multicolumn{16}{c}{Targeted Observations}\\
	\hline
	2016 XT1 & 07/12/2016 22:49:53 & 04/12/2016 & 4 & 1:30 & 19.7 & 20.12 & 0.09$\pm$0.05 & 0.05$\pm$0.05 & $\geq$0.21 & $\geq$90 & \textbf{0.84} & 0.08 & 0.07 & 0.01 & N \\
	2017 CP32 & 20/02/2017 20:13:56 & 15/02/2017 & 6 & 1:42 & 23.9 & 20.14 & -0.01$\pm$0.09 & 0.08$\pm$0.10 & $\geq$0.60 & $\geq$102 & 0.26 & 0.38 & 0.18 & 0.19 & N \\
	2017 AE3 & 11/01/2017 20:29:12 & 02/01/2017 & 10 & 0:59 & 21.8 & 19.55 & 0.28$\pm$0.15 & 0.15$\pm$0.16 & $\geq$1.13 & $\geq$59 & \textbf{0.91} & 0.01 & 0.03 & 0.04 & Y \\
	2017 CJ1 & 12/02/2017 19:32:38 & 02/02/2017 & 11 & 1:08 & 19.2 & 20.81 & -0.17$\pm$0.26 & -0.41$\pm$0.31 & $\geq$0.65 & $\geq$68 & 0.05 & 0.04 & \textbf{0.90} & 0.02 & N \\
	2016 WJ1 & 29/11/2016 22:06:04 & 19/11/2016 & 11 & 1:47 & 21.3 & 18.08 & 0.00$\pm$0.02 & 0.07$\pm$0.03 & 0.20$\pm$0.04 & 20.0$\pm$0.9 & 0.09 & \textbf{0.85} & 0.05 & 0.01 & Y \\
	2016 WO1 & 01/12/2016 20:37:53 & 19/11/2016 & 13 & 1:02 & 19.9 & 20.10 & 0.18$\pm$0.09 & -0.01$\pm$0.15 & $\geq$0.40 & $\geq$62 & \textbf{0.75} & 0.01 & 0.22 & 0.02 & N \\
	2016 WP & 01/12/2016 19:30:24 & 18/11/2016 & 14 & 1:08 & 21.9 & 20.82 & 0.02$\pm$0.19 & -0.26$\pm$0.30 & $\geq$0.89 & $\geq$68 & 0.17 & 0.06 & \textbf{0.73} & 0.04 & N \\
	2016 YM1 & 05/01/2017 19:49:43 & 22/12/2016 & 15 & 1:18 & 22.1 & 20.41 & 0.11$\pm$0.14 & 0.27$\pm$0.14 & 1.07$\pm$0.19 & 28.2$\pm$2.1 & \textbf{0.63} & 0.05 & 0.01 & 0.30 & N \\
	2017 CS & 16/02/2017 22:08:12 & 02/02/2017 & 15 & 1:43 & 19.4 & 20.54 & 0.04$\pm$0.17 & 0.05$\pm$0.24 & $\geq$0.79 & $\geq$103 & 0.37 & 0.13 & 0.33 & 0.17 & Y \\
	2017 CO5 & 18/02/2017 00:31:53 & 02/02/2017 & 16 & 0:59 & 20.1 & 20.52 & 0.22$\pm$0.13 & 0.09$\pm$0.18 & $\geq$0.31 & $\geq$59 & \textbf{0.83} & 0.02 & 0.10 & 0.05 & N \\
	2017 BL6 & 11/02/2017 19:34:03 & 26/01/2017 & 17 & 1:07 & 21.1 & 20.36 & 0.05$\pm$0.18 & 0.23$\pm$0.17 & $\geq$0.83 & $\geq$67 & 0.47 & 0.13 & 0.06 & 0.34 & N \\
	2017 BG85 & 17/02/2017 21:40:52 & 28/01/2017 & 21 & 1:00 & 18.9 & 20.04 & 0.04$\pm$0.08 & -0.15$\pm$0.07 & $\geq$0.43 & $\geq$60 & 0.08 & 0.01 & \textbf{0.91} & 0.00 & N \\
	2017 BY93 & 17/02/2017 23:06:37 & 26/01/2017 & 23 & 1:45 & 23.2 & 18.31 & -0.11$\pm$0.02 & -0.15$\pm$0.05 & $\geq$0.14 & $\geq$105 & 0.00 & 0.00 & \textbf{1.00} & 0.00 & N \\
	2016 VU3 & 29/11/2016 20:00:35 & 06/11/2016 & 24 & 2:04 & 20.6 & 19.97 & -0.08$\pm$0.08 & -0.15$\pm$0.10 & $\geq$0.68 & $\geq$124 & 0.01 & 0.07 & \textbf{0.92} & 0.00 & N \\
	2016 YM & 11/01/2017 21:29:00 & 18/12/2016 & 25 & 0:55 & 22.1 & 19.92 & -0.03$\pm$0.12 & 0.22$\pm$0.12 & $\geq$0.14 & $\geq$55 & 0.22 & 0.21 & 0.03 & \textbf{0.53} & N \\
	2016 VP4 & 06/12/2016 00:11:51 & 09/11/2016 & 27 & 1:02 & 24.2 & 20.65 & 0.25$\pm$0.24 & 0.06$\pm$0.21 & $\geq$0.92 & $\geq$62 & \textbf{0.72} & 0.05 & 0.15 & 0.07 & N \\
	2016 UJ101 & 08/12/2016 00:11:11 & 31/10/2016 & 38 & 0:55 & 20.4 & 19.76 & 0.13$\pm$0.06 & 0.22$\pm$0.07 & $\geq$0.18 & $\geq$55 & \textbf{0.82} & 0.01 & 0.00 & 0.17 & N \\
	2016 UU80 & 07/12/2016 21:13:55 & 29/10/2016 & 40 & 1:24 & 21.1 & 19.66 & 0.06$\pm$0.04 & 0.01$\pm$0.04 & 0.25$\pm$0.06 & 88.3$\pm$16.2 & \textbf{0.65} & 0.08 & 0.28 & 0.00 & N \\
	2017 AC5 & 15/02/2017 23:09:38 & 03/01/2017 & 44 & 0:54 & 19.9 & 19.47 & 0.09$\pm$0.08 & 0.17$\pm$0.10 & $\geq$0.77 & $\geq$54 & \textbf{0.64} & 0.08 & 0.02 & 0.26 & N \\
	\hline
	\multicolumn{16}{c}{Serendipitously Observed Targets}\\
	\hline
	2016 XX23 & 15/02/2017 19:02:34 & 15/12/2016 & 63 & 1:07 & 17.6 & 20.95 & 0.31$\pm$0.25 & -0.36$\pm$0.34 & $\geq$0.96 & $\geq$67 & 0.43 & 0.01 & \textbf{0.55} & 0.01 & N \\
	2016 VC & 28/10/2016 23:21:38 & 16/03/2015 & 593 & 2:24 & 17.5 & 19.96 & 0.03$\pm$0.07 & 0.01$\pm$0.06 & $\geq$0.40 & $\geq$144 & 0.39 & 0.25 & 0.35 & 0.01 & N \\
	2011 LD20 & 11/01/2017 19:25:05 & 08/06/2011 & 2045 & 1:04 & 16.4 & 19.54 & 0.04$\pm$0.08 & 0.08$\pm$0.09 & $\geq$0.20 & $\geq$64 & 0.46 & 0.25 & 0.14 & 0.15 & N \\
	2009 RW6 & 01/12/2016 20:37:53 & 10/09/2009 & 2640 & 1:02 & 17.9 & 21.08 & 0.43$\pm$0.17 & 0.59$\pm$0.25 & $\geq$0.61 & $\geq$62 & \textbf{0.99} & 0.00 & 0.00 & 0.01 & N \\
	2007 PM4 & 30/10/2016 00:04:04 & 16/07/2007 & 3394 & 2:00 & 17.4 & 20.42 & 0.04$\pm$0.06 & 0.02$\pm$0.08 & $\geq$0.45 & $\geq$120 & 0.42 & 0.21 & 0.34 & 0.04 & N \\
	2006 SX217 & 13/01/2017 22:03:31 & 30/09/2006 & 3759 & 1:54 & 18.9 & 20.75 & 0.00$\pm$0.08 & 0.00$\pm$0.11 & $\geq$0.41 & $\geq$114 & 0.22 & 0.26 & 0.45 & 0.07 & N \\
	2003 UE7 & 29/11/2016 22:06:04 & 17/10/2003 & 4793 & 1:47 & 17.6 & 19.49 & 0.08$\pm$0.03 & 0.11$\pm$0.03 & $\geq$0.10 & $\geq$107 & \textbf{0.91} & 0.06 & 0.00 & 0.03 & N \\
	2003 OX5 & 28/10/2016 21:06:08 & 22/07/2003 & 4848 & 2:01 & 20.0 & 20.69 & 0.00$\pm$0.06 & 0.18$\pm$0.11 & $\geq$0.30 & $\geq$121 & 0.18 & 0.23 & 0.05 & \textbf{0.55} & N \\
	2002 WP & 29/10/2016 01:04:28 & 18/11/2002 & 5094 & 0:55 & 18.4 & 18.00 & 0.04$\pm$0.21 & 0.17$\pm$0.08 & $\geq$0.50 & $\geq$55 & 0.47 & 0.26 & 0.02 & 0.26 & N \\
	2002 NQ7 & 06/12/2016 00:11:51 & 09/07/2002 & 5264 & 1:02 & 17.5 & 20.12 & 0.21$\pm$0.15 & 0.26$\pm$0.20 & $\geq$0.44 & $\geq$62 & \textbf{0.83} & 0.02 & 0.03 & 0.11 & N \\
	2002 KL6 & 01/12/2016 19:30:24 & 27/05/2002 & 5303 & 1:08 & 17.5 & 19.40 & 0.05$\pm$0.03 & 0.00$\pm$0.04 & $\geq$0.48 & $\geq$68 & \textbf{0.53} & 0.05 & 0.41 & 0.00 & N \\
	2001 TN41 & 30/10/2016 00:04:04 & 14/10/2001 & 5495 & 2:00 & 16.5 & 18.80 & 0.10$\pm$0.04 & 0.15$\pm$0.05 & $\geq$0.10 & $\geq$120 & \textbf{0.86} & 0.02 & 0.00 & 0.13 & N \\
	2000 VJ61 & 12/02/2017 19:32:38 & 02/11/2000 & 5947 & 1:08 & 16.0 & 16.51 & 0.08$\pm$0.04 & 0.10$\pm$0.06 & 0.22$\pm$0.06 & 77.3$\pm$14.7 & \textbf{0.81} & 0.07 & 0.02 & 0.09 & N \\
	2000 GG162 & 30/10/2016 23:12:24 & 07/04/2000 & 6051 & 1:00 & 15.1 & 17.90 & 0.05$\pm$0.04 & 0.14$\pm$0.05 & $\geq$0.04 & $\geq$60 & \textbf{0.53} & 0.15 & 0.00 & 0.32 & N \\
	1999 RO42 & 18/02/2017 00:31:53 & 14/09/1999 & 6367 & 0:59 & 15.9 & 20.16 & 0.10$\pm$0.09 & 0.01$\pm$0.12 & $\geq$0.27 & $\geq$59 & \textbf{0.58} & 0.08 & 0.28 & 0.05 & N \\
	1997 UF9 & 30/10/2016 21:59:18 & 29/10/1997 & 6942 & 1:20 & 16.2 & 18.59 & 0.00$\pm$0.02 & -0.02$\pm$0.03 & 0.12$\pm$0.05 & 144.0$\pm$25.9 & 0.03 & 0.11 & \textbf{0.86} & 0.00 & N \\
	1996 VK37 & 30/10/2016 23:12:24 & 11/11/1996 & 7294 & 1:00 & 17.2 & 20.11 & 0.08$\pm$0.06 & 0.13$\pm$0.07 & $\geq$0.22 & $\geq$60 & \textbf{0.69} & 0.10 & 0.02 & 0.19 & N \\
	1996 VX & 01/12/2016 19:30:24 & 03/11/1996 & 7334 & 1:08 & 16.3 & 20.49 & 0.11$\pm$0.10 & 0.03$\pm$0.14 & $\geq$0.44 & $\geq$68 & \textbf{0.61} & 0.08 & 0.25 & 0.07 & N \\
	1984 SS & 11/01/2017 19:25:05 & 28/09/1984 & 11794 & 1:04 & 14.0 & 18.40 & 0.06$\pm$0.03 & 0.09$\pm$0.04 & 0.20$\pm$0.05 & 123.7$\pm$22.9 & \textbf{0.78} & 0.16 & 0.01 & 0.05 & N \\
	1982 RA1 & 13/01/2017 22:03:31 & 13/09/1982 & 12542 & 1:54 & 16.5 & 18.74 & 0.05$\pm$0.02 & 0.08$\pm$0.03 & $\geq$0.24 & $\geq$114 & \textbf{0.75} & 0.23 & 0.01 & 0.01 & N \\
	\enddata
	\tablenotetext{\textit{a}}{Days between when object was discovered and when it was observed for this study}
	\tablenotetext{\textit{b}}{Obtained from \url{https://ssd.jpl.nasa.gov/horizons.cgi}}	
	\tablenotetext{\textit{c}}{Lower limit shown in cases where observational duration was insufficient to observe entire light-curve period}
\end{deluxetable*}

\subsection{Light Curves}\label{subsec:LCurved}

\begin{figure}
	\gridline{\fig{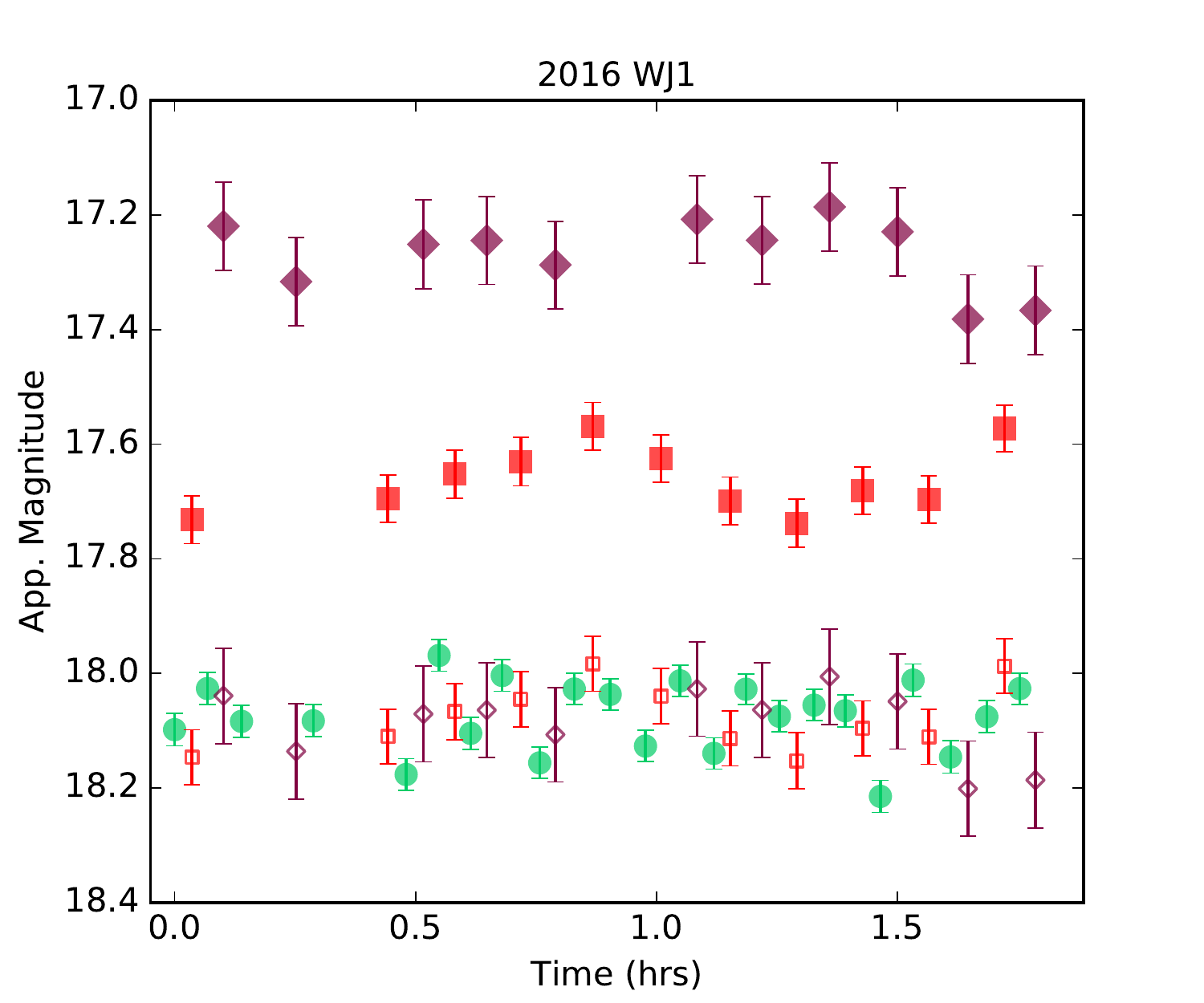}{0.5\textwidth}{(a)}}
	\gridline{\fig{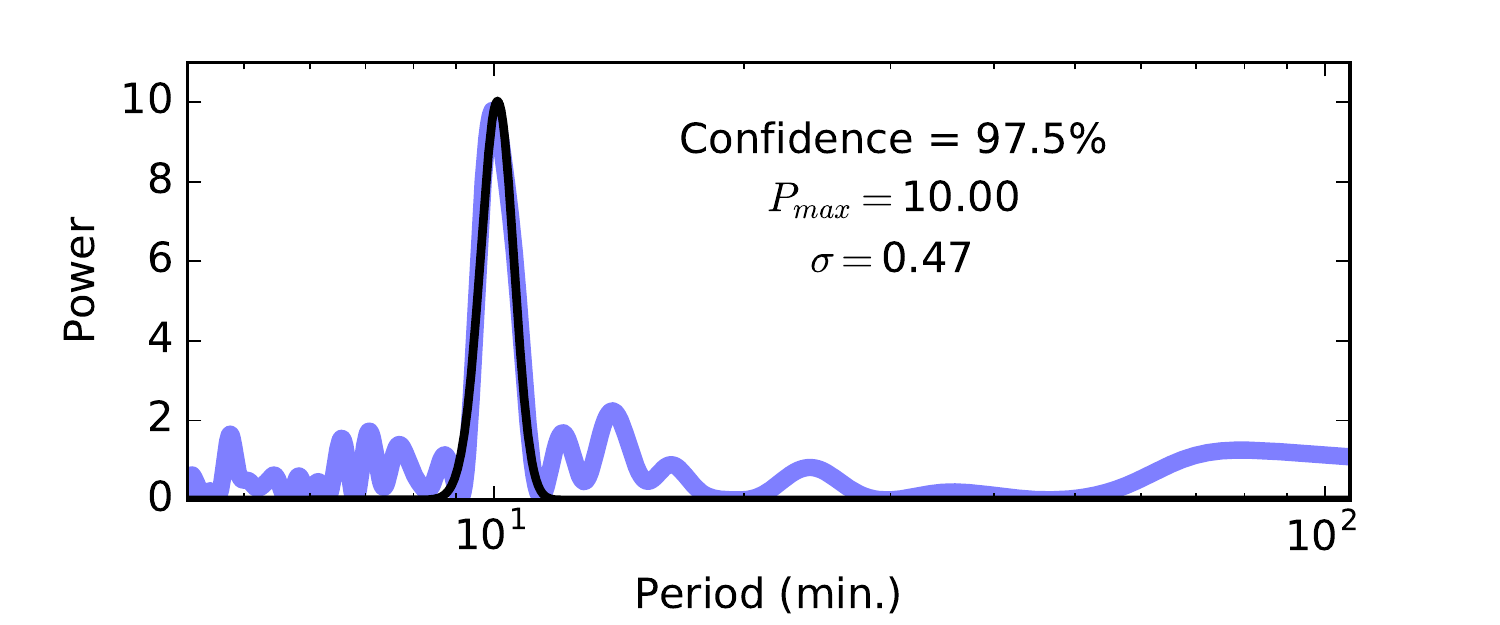}{0.5\textwidth}{(b)}}
	\gridline{\fig{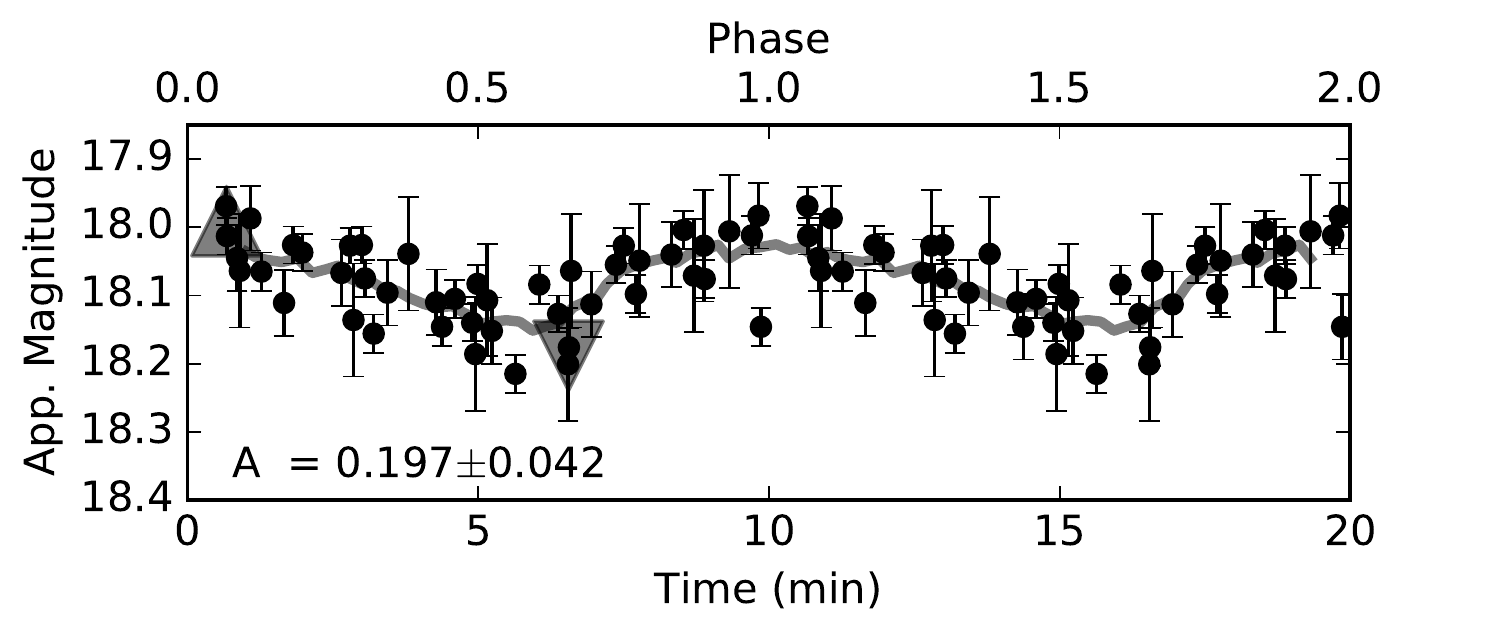}{0.5\textwidth}{(c)}}
	\caption{(a) The calibrated spectrophotometric data from the PHOTOMETRYPIPELINE of observed NEA 2016 WJ1. (b) Periodogram of 2016 WJ1's light curve data, bottom set of data in (a), showing a distinct peak at the position P$_{max}=$10.00~min with a RMS width $\sigma= $0.47~min. The confidence in the suggested light-curve period is 97.5$\%$. (c) Folded representation of the original light curve data using the 10.00 minute folding interval. Note that the folded data span two of the best-fit light-curve periods. The amplitude (A) is displayed with the gray triangles, situated where the mean of two adjacent data points have the minimum/maximum value, indicating the values used for determining A. See the Appendix for data of all 6 NEAs that produced periodograms with a distinct peak.}
	\label{fig5}
\end{figure}

Light curves were produced by combining the \textit{V} magnitude data points with adjusted \textit{I} and \textit{R} magnitude data points. Adjustments were implemented by normalizing the original \textit{I} and \textit{R} magnitude data points to \textit V magnitudes by adding the derived colors (see Section \ref{subsec:Colors} for detail on how colors were derived). Examples of resultant light curves are shown as the bottom curves in Figure \ref{fig3} (a)-(d) with the original \textit{V} data and adjusted \textit{R} and \textit{I} data inter-spaced. 

Only a fraction of the observed NEAs (e.g. 2016 UU80 in Figure \ref{fig3} (a)) had rotational periods short enough to produce a light curve with a resolvable light-curve period within the $\sim$1-hour observation. Most observations only produced light curves covering a fraction of the rotational period (e.g. 2017 AC5 in Figure \ref{fig3} (b)), and therefore in most cases only lower limits on both the light curve amplitude (A) and rotation period (P) could be determined. However, six of the 39 targets had resolvable light-curve periods within the observation time. 

To determine which targets had resolvable light-curve periods the light curve data for each NEA were analyzed using the  NASA Exoplanet Science Institute's periodogram online tool\footnote{\url{https://exoplanetarchive.ipac.caltech.edu/cgi-bin/Pgram/nph-pgram}}. Designed to generate perdiograms from the Kepler and K2 light curve data archive, the tool is also able to generate periodograms of user-uploaded data. The tool makes use of the Lomb-Scargle method, a least-squares spectral analysis, to estimate a frequency spectrum and thereby generates a periodogram. Figure \ref{fig5} (b) shows the periodogram generated from the light curve data of 2016 WJ1 with a distinct peak at 10 minutes suggesting a possible periodicity at this interval. The uncertainty of this periodicity is determined by fitting a Gaussian function to the periodogram peak and using the RMS width ($\sigma$) as the uncertainty (see superimposed black curve and $\sigma$ of the fitted function). The confidence in the periodogram peak is also indicated and calculated using the false alarm probability formulation from \cite{Zechmeister2009}. Figure \ref{fig5} (c) shows the folded representation of the original light curve data using the 10.00 minute folding interval resulting in a light curve with a convincing completely resolvable oscillation. Using this folded data, the amplitude is calculated using the difference between magnitudes highlighted with the gray triangle symbols. The triangles are situated where the mean of two adjacent data points have the minimum/maximum magnitude. (See Appendix for all six targets that produced periodograms with a distinct peak.) The light-curve periods obtained from the periodogram peak position and the amplitudes determined from the folded data are recorded in Table \ref{table:results} for these six targets. Note that in Table \ref{table:results}  the rotation period is indicated which is double the light-curve period. For the remaining targets the lower limit of the rotation period (i.e. the observational duration) and the lower limit of the amplitude are used to complete Table \ref{table:results}. In the latter case, the amplitude lower limited is calculated in the same way as before but instead the original unfolded data are used (see gray triangles in Figure \ref{fig3}).

\section{Discussion}
\label{Discussion}
\subsection{Compositional Distribution}

\begin{figure}
	\centering
	\includegraphics[width=85mm]{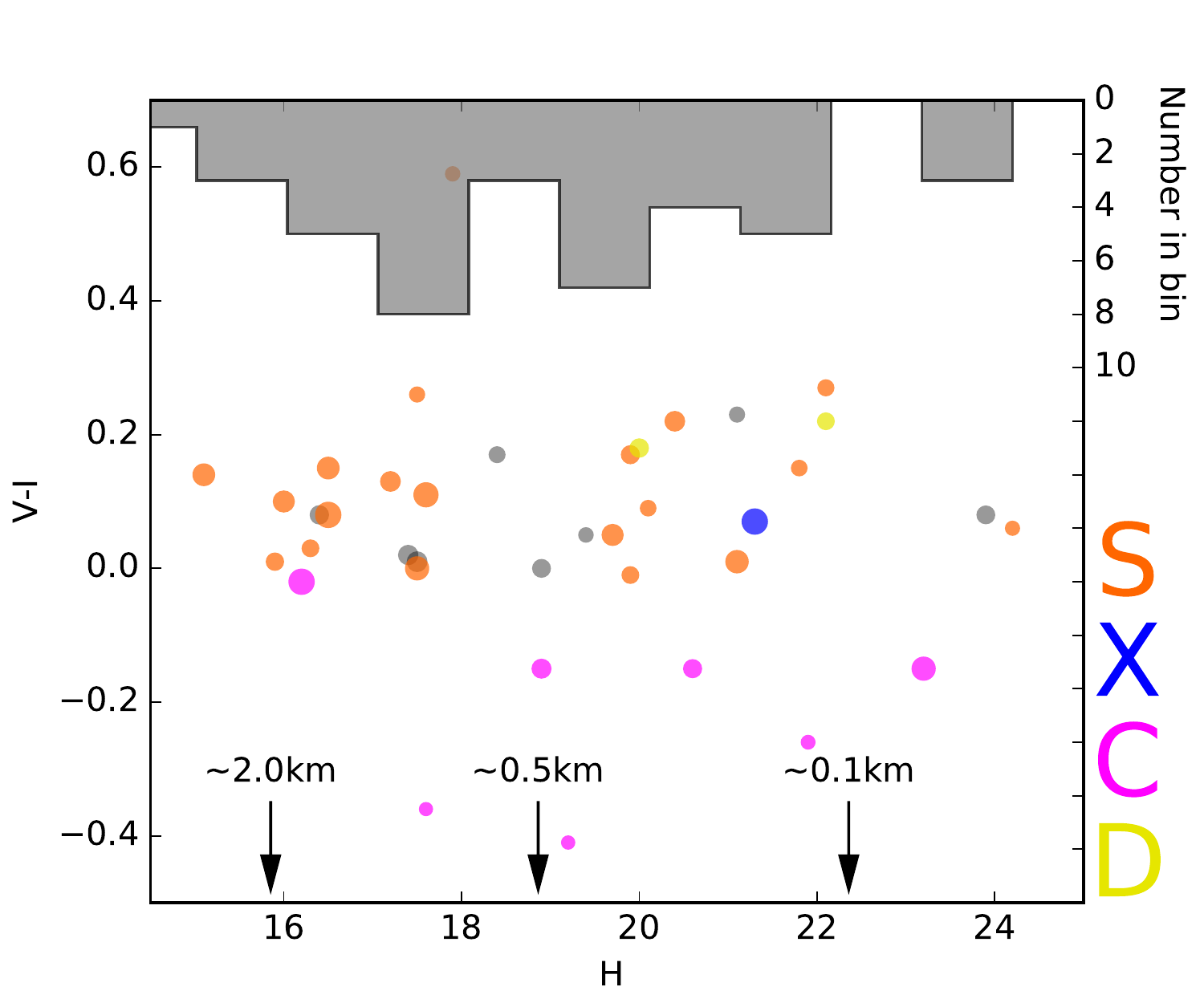}
	\caption{Plotted are the \textit{V-I} colors of all observed NEAs as a function of \textit{H} magnitude, which is roughly proportional to size. NEAs in orange, blue, pink and yellow were classified by the ML algorithm as S-, X-, C- and D-type, respectively. NEAs in gray remain unclassified since the highest taxonomic probability was less than 50\%. Sizes of the data point is inversely proportional to the uncertainty (see Figure \ref{fig4} (c) for scale). Typical asteroid diameter at the corresponding \textit{H} magnitude is indicated with arrows with an albedo of 0.2 assumed.}
	\label{fig6}
\end{figure}

One of the primary aims of this study is to derive the compositional distribution of NEAs as a function of size. Figure \ref{fig6} shows the classified target data points from Figure \ref{fig4} (c) plotted as a function of absolute magnitude (\textit{H}). The size estimates at various \textit{H} magnitudes (based on an albedo of 0.2) are also indicated. This relatively sparse plot suggests that there is no significant difference in ratio between ``stony" S-type and ``not-stony" (C+X+D)-type asteroids across the observed asteroid size range. The plot does suggest that the ratio is slightly in favor of S-type for larger asteroids and that X- and D-type could perhaps favor smaller asteroids. 

\begin{figure}
	\centering
	\includegraphics[width=85mm]{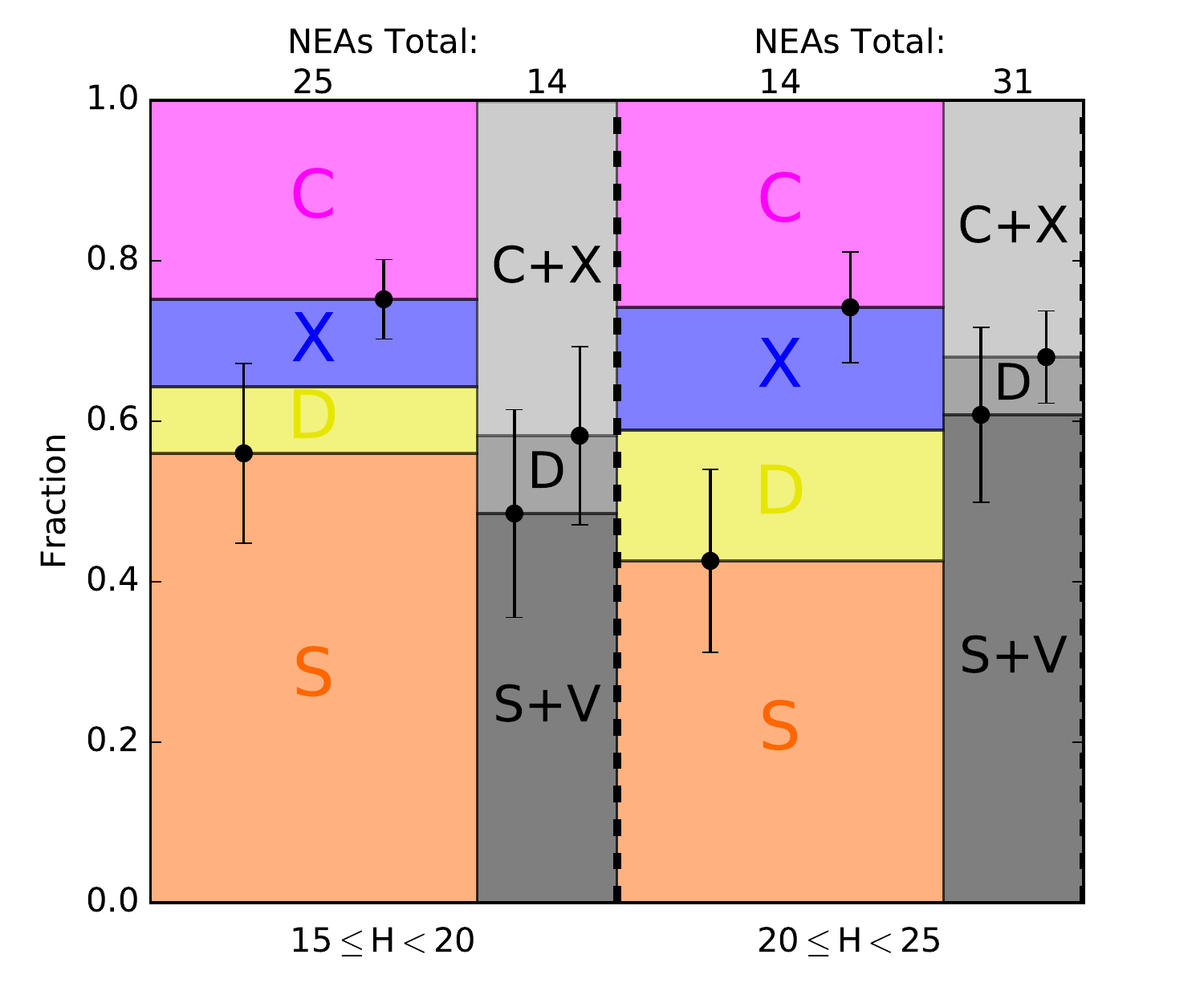}
	\caption{Compositional fractional distribution of all 39 observed targets as a function of two H magnitude ranges.  We show similar data from \cite{Mommert2016} in shades of gray. Within the error bars the distributions agree. We find that the fraction of NEAs that are S-types is around 50\% and, within our error bars, see no evidence for any dependence on size.
	}
	\label{fig7}
\end{figure}

However, Figure \ref{fig6} is based on discrete classification, i.e., only the highest classification probability in Table \ref{table:results} is considered. A more statistically correct distribution is obtained by calculating the fractions based on the summation of the probabilities in Table \ref{table:results} for each respective class type. The result of this summation is shown in Figure \ref{fig7}. The fraction of S-type asteroids in our observed population is $\sim$56$\%$ for large (15$\leq$H$<$20) NEAs and $\sim$43$\%$ for smaller (20$\leq$H$<$25) NEAs. Within the error bars, there is no difference between these two values, and in both size bins our result is in agreement with \cite{Mommert2016}. Therefore our data suggest a roughly 50/50 split in numbers between "stony" S-type and "not-stony" (C+X+D)-type asteroids across the observed asteroid size range. 

The fraction of meteorites that have compositions similar to S-type asteroids is more than 80$\%$ \citep{Harvey1989}, well above our observed $\sim$50$\%$ for S-type asteroids. Most meteorites are generated from NEAs that are around 10 meters in size (i.e. \textit{H}$\approx$27), which is slightly smaller than the smallest size bin shown in Figure \ref{fig7}.  However, \cite{Mommert2016} show a third bin for 25$\leq$H$<$30 showing a S-type fraction of around 40$\%$ and \cite{Hinkle2014} also find that the S-type fraction is low, and decreasing, at small NEA sizes. It has been proposed \citep{Reddy2016} that the discrepancy between meteorites and optically observed distributions could be a result of atmospheric filtering with the stony-like material in S-type asteroids more likely to survive atmospheric entry, and reach the ground, than that of the more primitive and fragile material present in D- and C-type asteroids. Our results strengthen this argument, as they are not consistent with an S-type fraction of 80$\%$. 

Both \cite{Mommert2016} and our results do not account for bias inherent to our target sample. Selecting from optically discovered targets means our sample targets mostly have medium to high surface albedos. This favors S-type over C-type asteroids \citep{Thomas2011}. This effect would lead to an overestimation of the fraction of S-type asteroids, which strengthens our argument even further.

\subsection{Light Curve Data}

Out of the 39 observed NEAs, lower limits on the light curve amplitude and rotation period are determined for 33 targets.  The rotation period lower limits of these 33 targets that are shown in Table \ref{table:results}  are all $\gtrsim$1~hour and are generally consistent with the overall distribution of previously reported NEA light curve properties: many asteroids with rotational periods larger than 1 hour have been reported for asteroids in our size range (15$\leq$H$\leq$25) \citep{Thirouin2016}.

For the six targets that had resolvable light curves, the three smallest targets --- 2016 YM1 ($P_{rot}=28.2$~min, \textit{H}=22.1), 2016 WJ1 ($P_{rot}=20.0$~min, \textit{H}=21.3) and 2016 UU80 ($P_{rot}=88.3$~min, \textit{H}=21.1) --- have periods and sizes consistent with previously reported observations and fall below the 100~Pa lower limit cohesion curve for lunar regolith \citep{Thirouin2016}.  Two of the three larger targets have rotation periods larger than the theoretical breakup rotation period of $\sim$2 hours and are also consistent with previously reported observations \citep{Thirouin2016}. The remaining large target, 2000 VJ61 ($P_{rot}=77.3$~min, \textit{H}=16.0), has a rotation period shorter than any previously observed targets of this size, although it still falls below the 1000~Pa higher limit cohesion curve for lunar regolith.  However, the calculated light-curve period is only half the observational duration, which suggests some aliasing and that we have not detected the entire period, and the confidence in the periodogram peak is also only 72.9$\%$. Further observations of this target for longer times may prove interesting because confirming this unexpectedly short rotation period would be extraordinary for such large asteroid.

Our data also shows that 2016 YM1, which is rotating relatively fast, also has a large amplitude (A$\approx$1~mag) in the light curve. An amplitude of 1.0 would suggest an a/b ratio (longest axis to middle axis) of 2.5 which would mean 2016 YM1 is not only fast rotating but an elongated object as well.

\section{Conclusion}
\label{Conclusion}
The \textit{VRI} spectrophotometric results of 39 NEAs observed with the Sutherland, South Africa, node of KMTNet have been presented. Our broadband spectrophotometry in combination with a ML algorithm allow us to classify 31 of the 39 observed target as either a S-, C-, X- or D-type asteroid. Our data suggest that the ratio between ``stony" S-type NEAs and ``not-stony" (C+X+D)-type NEAs, with \textit{H} magnitudes between 15 and 25, is roughly 1:1, which is an agreement with previously published survey studies. However, this ratio is not consistent with a S-type fraction of 80$\%$ shown in collected meteorites and therefore our results supports the idea of atmospheric filtering that leads to a preference for S-type meteorite survival through the atmosphere.

Our $\sim$1-hour light curve data of each observed NEA allowed us to place lower limit constraints on the rotational period of 33 NEAs and resolve the rotational periods of six NEAs. The lower limit rotational periods, together with the size of the relevant NEAs, are consistent with previously published observations of NEAs with similar sizes. For three small NEAs with resolvable periods our data are also consistent with previously published observations of NEAs with similar sizes. For the three large NEAs with resolvable periods our data are consistent for two of the targets. For the remaining large target our results are inconsistent. However, the confidence in the periodogram peak is relatively low for this target. Therefore there are justified doubts on the legitimacy of this large NEAs' calculated rotation period.
\newline

This study was facilitated by observations made at the South African Astronomical Observatory (SAAO) and this work is partially supported by the South African National Research Foundation (NRF). This work is supported in part by the National Aeronautics and Space Administration (NASA) under grant number NNX15AE90G issued through the SSO Near Earth Object Observations Program and in part by a grant from NASA's Office of the Chief Technologist. We thank the referee, Vishnu Reddy, for a useful review.

\bibliography{bibliography}

\begin{thebibliography}{}
\expandafter\ifx\csname natexlab\endcsname\relax\def\natexlab#1{#1}\fi

\bibitem[{Binney \& Merrifield(1998)}]{Binney1998}
Binney, J., \& Merrifield, M. 1998, {Galactic Astronomy}

\bibitem[{Borovi{\v{c}}ka {et~al.}(2013)Borovi{\v{c}}ka, Spurn{\'{y}}, Brown,
  Wiegert, Kalenda, Clark, \& Shrben{\'{y}}}]{Borovicka2013}
Borovi{\v{c}}ka, J., Spurn{\'{y}}, P., Brown, P., {et~al.} 2013, Nature, 503,
  235

\bibitem[{Bottke {et~al.}(2000)Bottke, Jedicke, Morbidelli, Petit, \&
  Gladman}]{Bottke2000}
Bottke, W.~F., Jedicke, R., Morbidelli, A., Petit, J.-M., \& Gladman, B. 2000,
  Science, 288, 2190

\bibitem[{Brown {et~al.}(2013)Brown, Assink, Astiz, Blaauw, Boslough,
  Borovi{\v{c}}ka, Brachet, Brown, Campbell-Brown, Ceranna, Cooke,
  de~Groot-Hedlin, Drob, Edwards, Evers, Garces, Gill, Hedlin, Kingery, Laske,
  {Le Pichon}, Mialle, Moser, Saffer, Silber, Smets, Spalding, Spurn{\'{y}},
  Tagliaferri, Uren, Weryk, Whitaker, \& Krzeminski}]{Brown2013}
Brown, P.~G., Assink, J.~D., Astiz, L., {et~al.} 2013, Nature, 503, 238

\bibitem[{Cheng {et~al.}(2016)Cheng, Michel, Jutzi, Rivkin, Stickle, Barnouin,
  Ernst, Atchison, Pravec, \& Richardson}]{Cheng2016}
Cheng, A., Michel, P., Jutzi, M., {et~al.} 2016, Planetary and Space Science,
  121, 27

\bibitem[{DeMeo {et~al.}(2009)DeMeo, Binzel, Slivan, \& Bus}]{DeMeo2009}
DeMeo, F.~E., Binzel, R.~P., Slivan, S.~M., \& Bus, S.~J. 2009, Icarus, 202,
  160

\bibitem[{Elvis(2012)}]{Elvis2012}
Elvis, M. 2012, Nature, 485, 549

\bibitem[{Galache {et~al.}(2015)Galache, Beeson, McLeod, \&
  Elvis}]{Galache2015}
Galache, J., Beeson, C., McLeod, K., \& Elvis, M. 2015, Planetary and Space
  Science, 111, 155

\bibitem[{Granvik {et~al.}(2017)Granvik, Morbidelli, Vokrouhlick{\'{y}},
  Bottke, Nesvorn{\'{y}}, \& Jedicke}]{Granvik2017}
Granvik, M., Morbidelli, A., Vokrouhlick{\'{y}}, D., {et~al.} 2017, Astronomy
  {\&} Astrophysics, 598, A52

\bibitem[{Harvey \& Cassidy(1989)}]{Harvey1989}
Harvey, R.~P., \& Cassidy, W.~A. 1989, Meteoritics, 24, 9

\bibitem[{Hinkle {et~al.}(2014)Hinkle, Moskovitz, \& Trilling}]{Hinkle2014}
Hinkle, M., Moskovitz, N., \& Trilling, D. 2014, in AAS/Division for Planetary
  Sciences Meeting Abstracts, Vol.~46, AAS/Division for Planetary Sciences
  Meeting Abstracts, 213.07

\bibitem[{Kim {et~al.}(2016)Kim, Lee, Park, Kim, Cha, Lee, Han, Chun, \&
  Yuk}]{Kim2016}
Kim, S.~L., Lee, C.~U., Park, B.~G., {et~al.} 2016, Journal of the Korean
  Astronomical Society, 49, 37

\bibitem[{Lauretta {et~al.}(2017)Lauretta, Beshore, Boynton, Drouet, Goddard,
  \& Flight}]{Lauretta2017}
Lauretta, D.~S., Beshore, E., Boynton, W.~V., {et~al.} 2017

\bibitem[{Mainzer {et~al.}(2011)Mainzer, Grav, Bauer, Masiero, McMillan, Cutri,
  Walker, Wright, Eisenhardt, Tholen, Spahr, Jedicke, Denneau, DeBaun, Elsbury,
  Gautier, Gomillion, Hand, Mo, Watkins, Wilkins, Bryngelson, {Del Pino
  Molina}, Desai, Camus, Hidalgo, Konstantopoulos, Larsen, Maleszewski, Malkan,
  Mauduit, Mullan, Olszewski, Pforr, Saro, Scotti, \& Wasserman}]{Mainzer2011}
Mainzer, A., Grav, T., Bauer, J., {et~al.} 2011, The Astrophysical Journal,
  743, 156

\bibitem[{Mommert(2017)}]{Mommert2017}
Mommert, M. 2017, Astronomy and Computing, doi:10.1016/j.ascom.2016.11.002

\bibitem[{Mommert {et~al.}(2016)Mommert, Trilling, Borth, Jedicke, Butler,
  Reyes-Ruiz, Pichardo, Petersen, Axelrod, \& Moskovitz}]{Mommert2016}
Mommert, M., Trilling, D.~E., Borth, D., {et~al.} 2016, The Astronomical
  Journal, 151, 98

\bibitem[{Pedregosa {et~al.}(2011)Pedregosa, Varoquaux, Gramfort, Michel,
  Thirion, Grisel, Blondel, Louppe, Prettenhofer, Weiss, Dubourg, Vanderplas,
  Passos, Cournapeau, Brucher, Perrot, \& Duchesnay}]{Pedregosa2012}
Pedregosa, F., Varoquaux, G., Gramfort, A., {et~al.} 2011, Journal of Machine
  Learning Research, 12, 2825

\bibitem[{Rayner {et~al.}(2003)Rayner, Toomey, Onaka, Denault, Stahlberger,
  Vacca, Cushing, \& Wang}]{Rayner2003}
Rayner, J., Toomey, D., Onaka, P., {et~al.} 2003, Publications of the
  Astronomical Society of the Pacific, 115, 362

\bibitem[{Reddy {et~al.}(2015)Reddy, Vokrouhlick{\'{y}}, Bottke, Pravec,
  Sanchez, Gary, Klima, Cloutis, Gal{\'{a}}d, Guan, Hornoch, Izawa,
  Ku{\v{s}}nir{\'{a}}k, {Le Corre}, Mann, Moskovitz, Skiff, \&
  Vra{\v{s}}til}]{Reddy2015}
Reddy, V., Vokrouhlick{\'{y}}, D., Bottke, W.~F., {et~al.} 2015, Icarus, 252,
  129

\bibitem[{Reddy {et~al.}(2016)Reddy, Sanchez, Bottke, Thirouin,
  Rivera-Valentin, Kelley, Ryan, Cloutis, Tegler, Ryan, Taylor, Richardson,
  Moskovitz, \& {Le Corre}}]{Reddy2016}
Reddy, V., Sanchez, J.~A., Bottke, W.~F., {et~al.} 2016, The Astronomical
  Journal, 152, 162

\bibitem[{Sanchez {et~al.}(2009)Sanchez, Colombo, Vasile, \&
  Radice}]{Sanchez2009}
Sanchez, P., Colombo, C., Vasile, M., \& Radice, G. 2009, Journal of Guidance,
  Control, and Dynamics, 32, 121

\bibitem[{Thirouin {et~al.}(2016)Thirouin, Moskovitz, Binzel, Christensen,
  DeMeo, Person, Polishook, Thomas, Trilling, Willman, Hinkle, Burt, Avner, \&
  Aceituno}]{Thirouin2016}
Thirouin, A., Moskovitz, N., Binzel, R.~P., {et~al.} 2016, The Astronomical
  Journal, 152, 163

\bibitem[{Thomas {et~al.}(2011)Thomas, Trilling, Emery, Mueller, Hora, Benner,
  Bhattacharya, Bottke, Chesley, Delb{\'{o}}, Fazio, Harris, Mainzer, Mommert,
  Morbidelli, Penprase, Smith, Spahr, \& Stansberry}]{Thomas2011}
Thomas, C.~A., Trilling, D.~E., Emery, J.~P., {et~al.} 2011, The Astronomical
  Journal, 142, 85

\bibitem[{Tsuda {et~al.}(2013)Tsuda, Yoshikawa, Abe, Minamino, \&
  Nakazawa}]{Tsuda2013}
Tsuda, Y., Yoshikawa, M., Abe, M., Minamino, H., \& Nakazawa, S. 2013, Acta
  Astronautica, 91, 356

\bibitem[{Yano {et~al.}(2006)Yano, Kubota, Miyamoto, Okada, Scheeres, Takagi,
  Yoshida, Abe, Abe, Fujiwara, Hasegawa, Hashimoto, Ishiguro, Kato, Kawaguchi,
  Mukai, Saito, Sasaki, \& Yoshikawa}]{Yano2006}
Yano, H., Kubota, T., Miyamoto, H., {et~al.} 2006, 312, 1350

\bibitem[{Zechmeister \& K{\"{u}}rster(2009)}]{Zechmeister2009}
Zechmeister, M., \& K{\"{u}}rster, M. 2009, 584, 577

\end{thebibliography}



\appendix
\section{All Photometry Data}
\label{All_results}
\begin{figure}[htb]
	\gridline{\fig{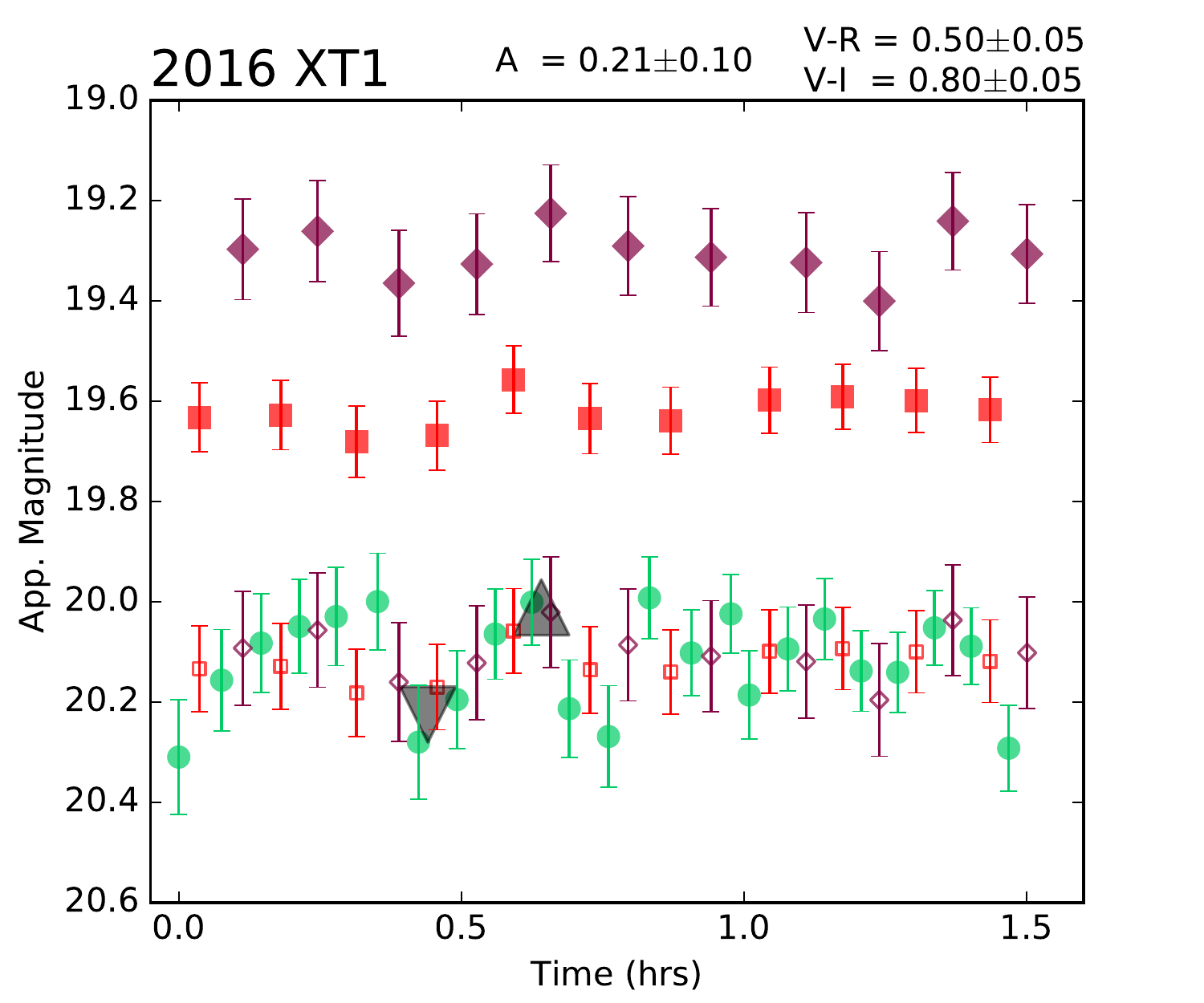}{0.3\textwidth}{(1)}
		\fig{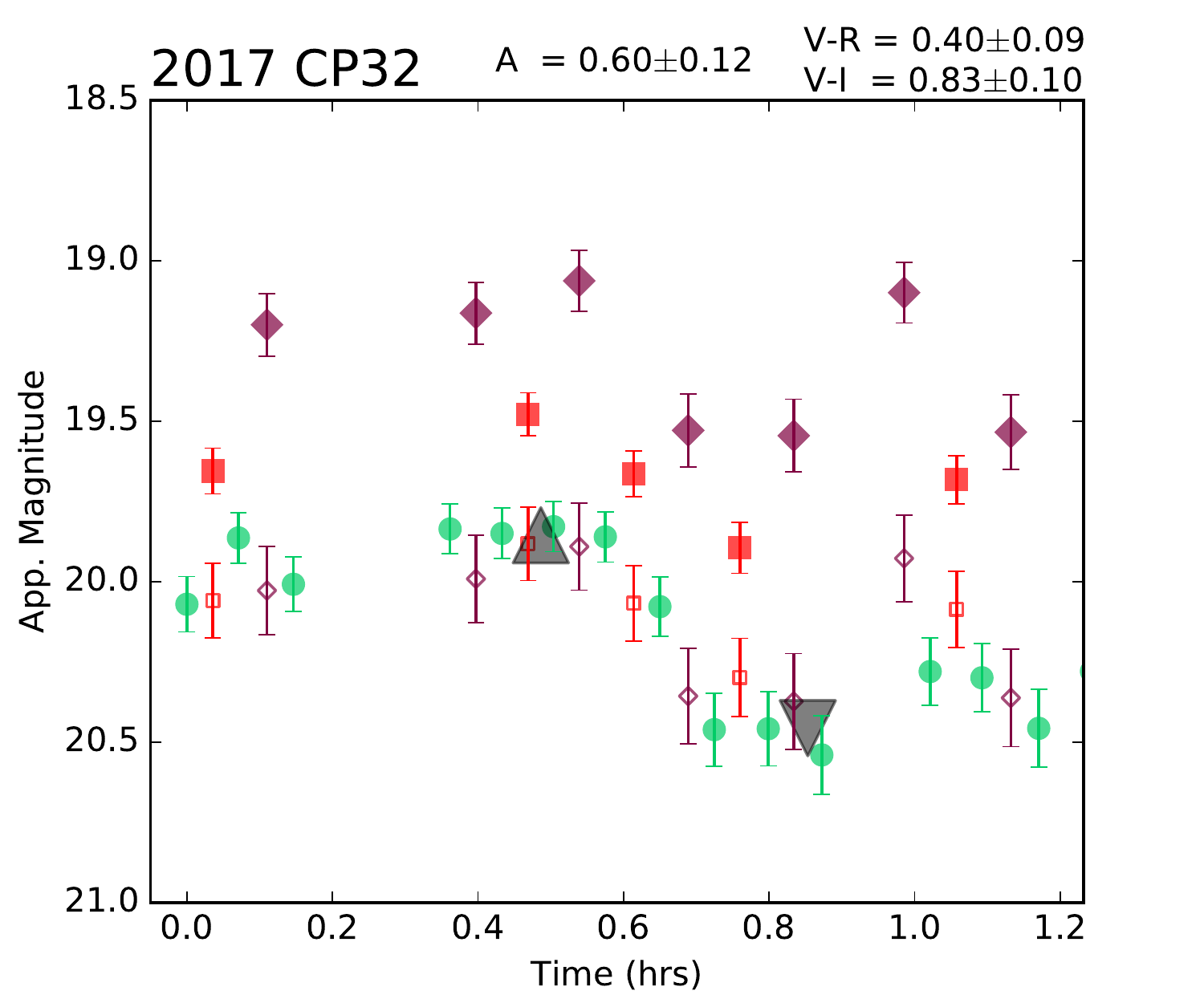}{0.3\textwidth}{(2)}
		\fig{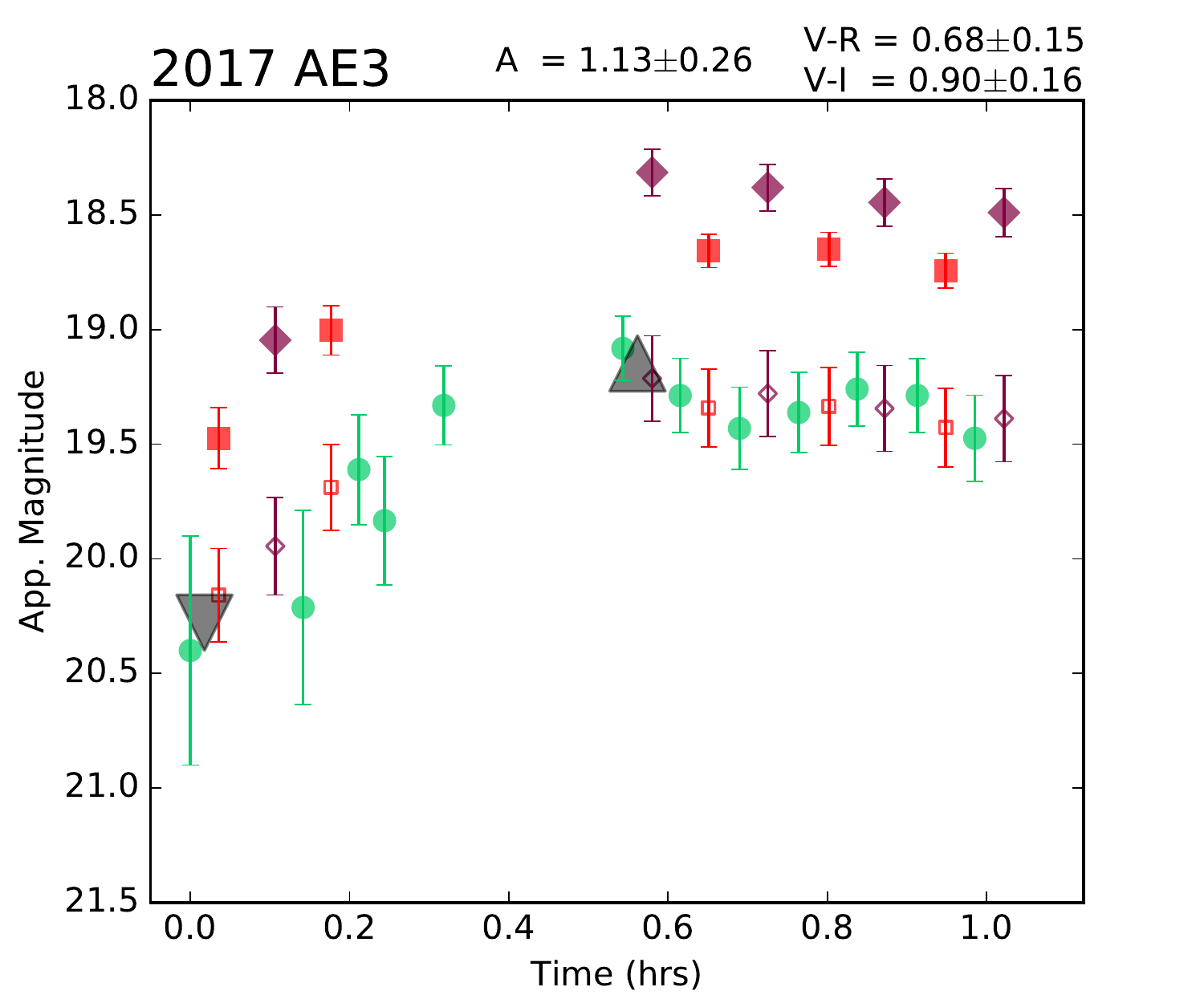}{0.3\textwidth}{(3)}
	}
	\gridline{\fig{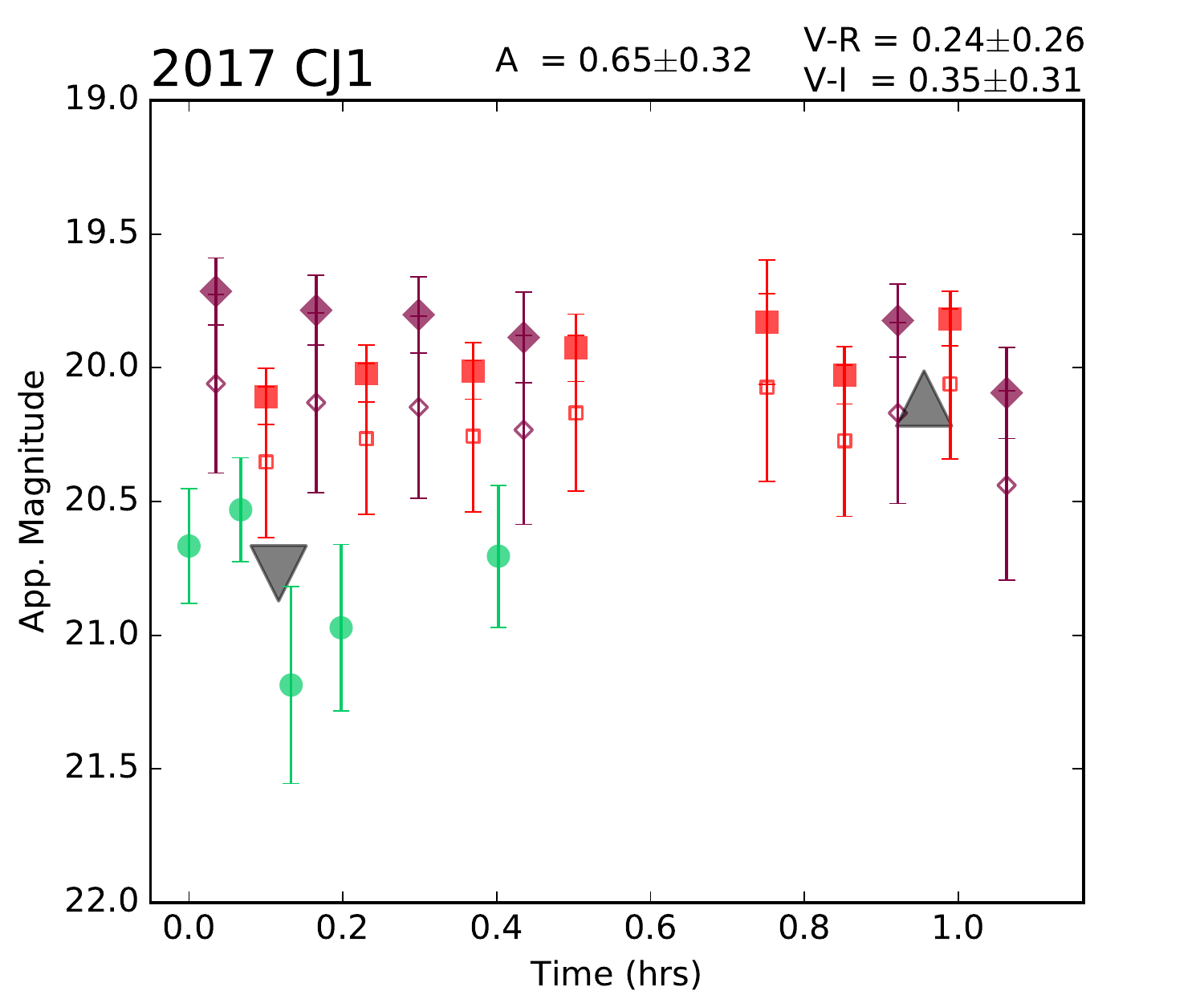}{0.3\textwidth}{(4)}
		\fig{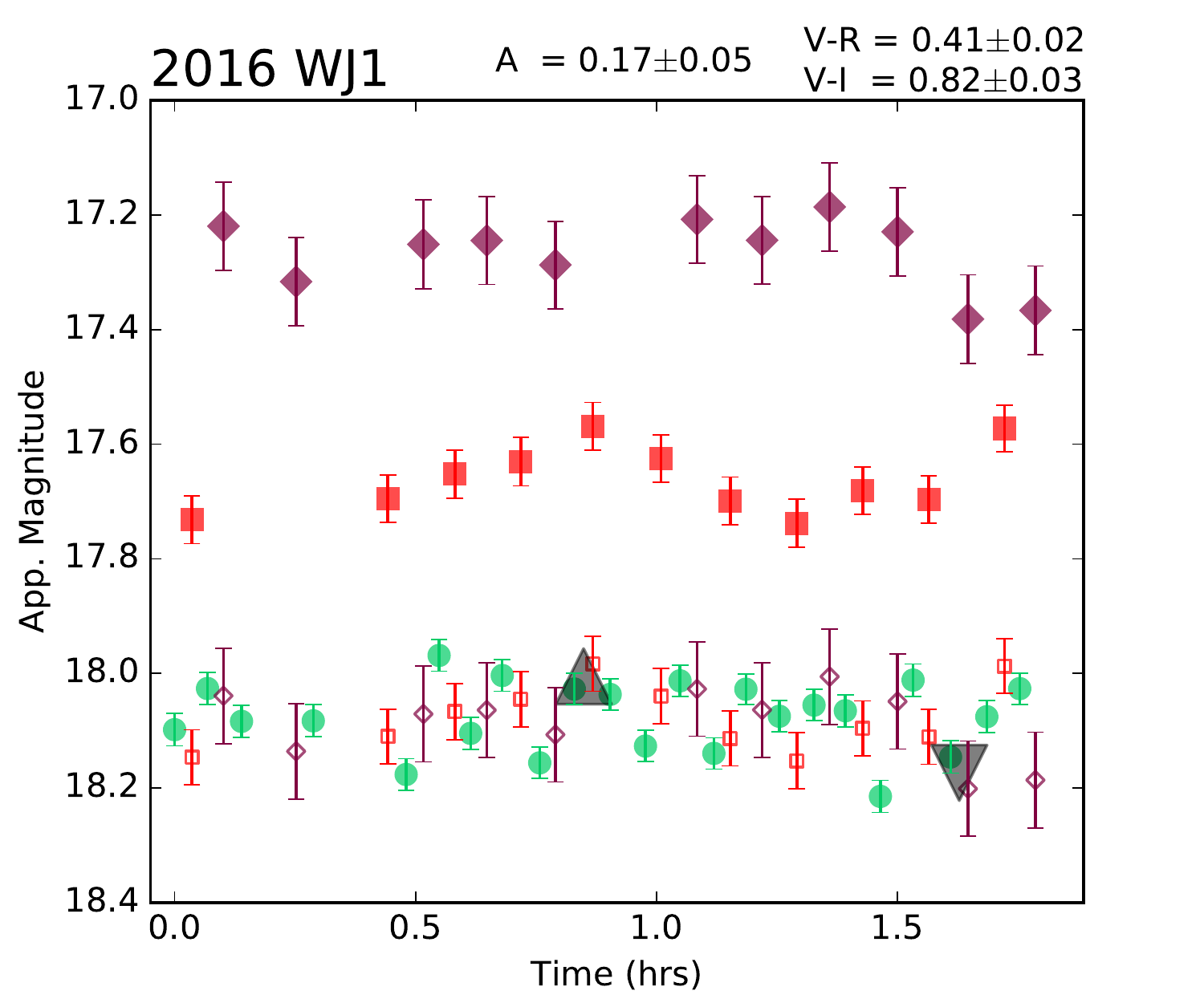}{0.3\textwidth}{(5)}
		\fig{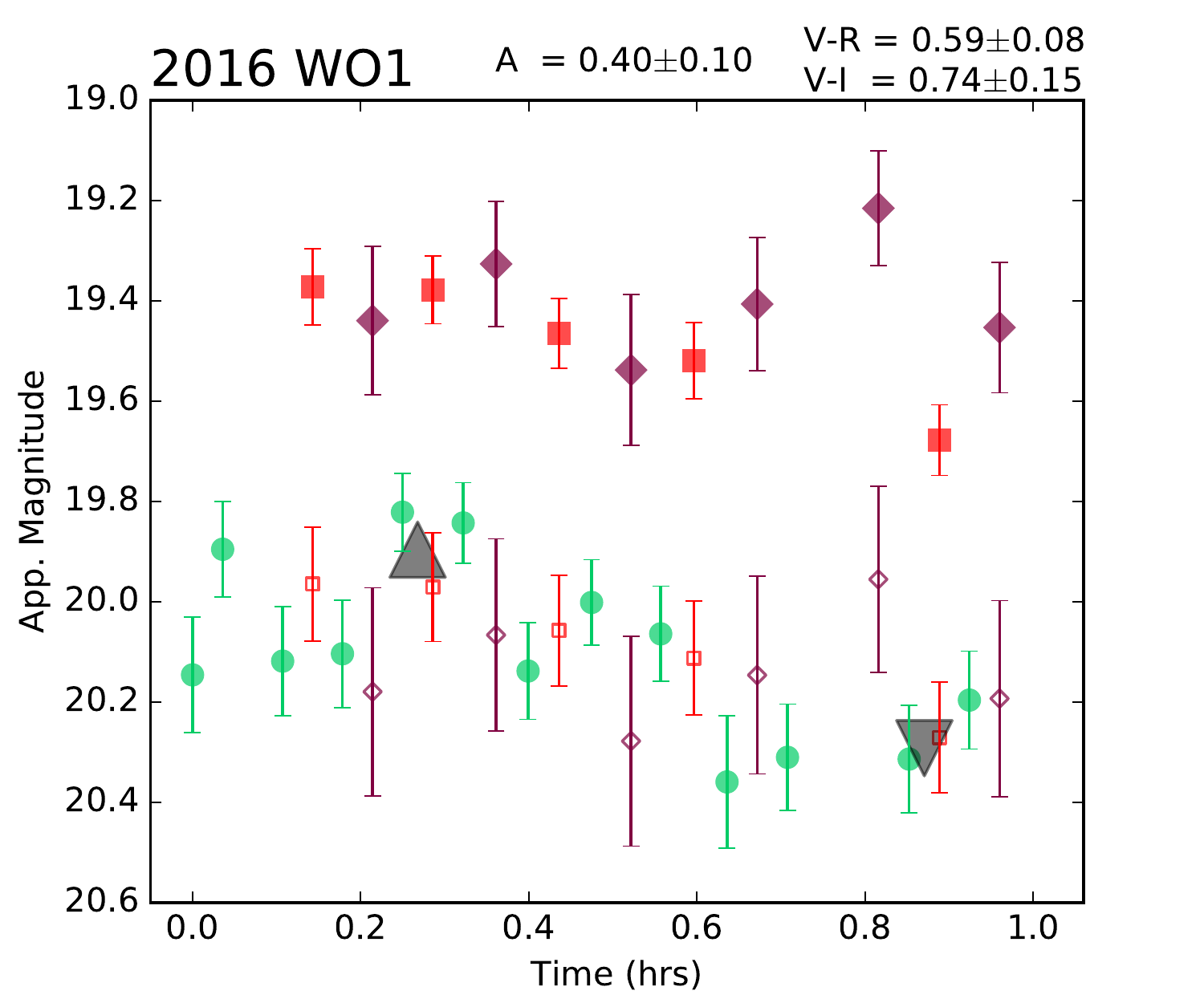}{0.3\textwidth}{(6)}
	}
	\gridline{\fig{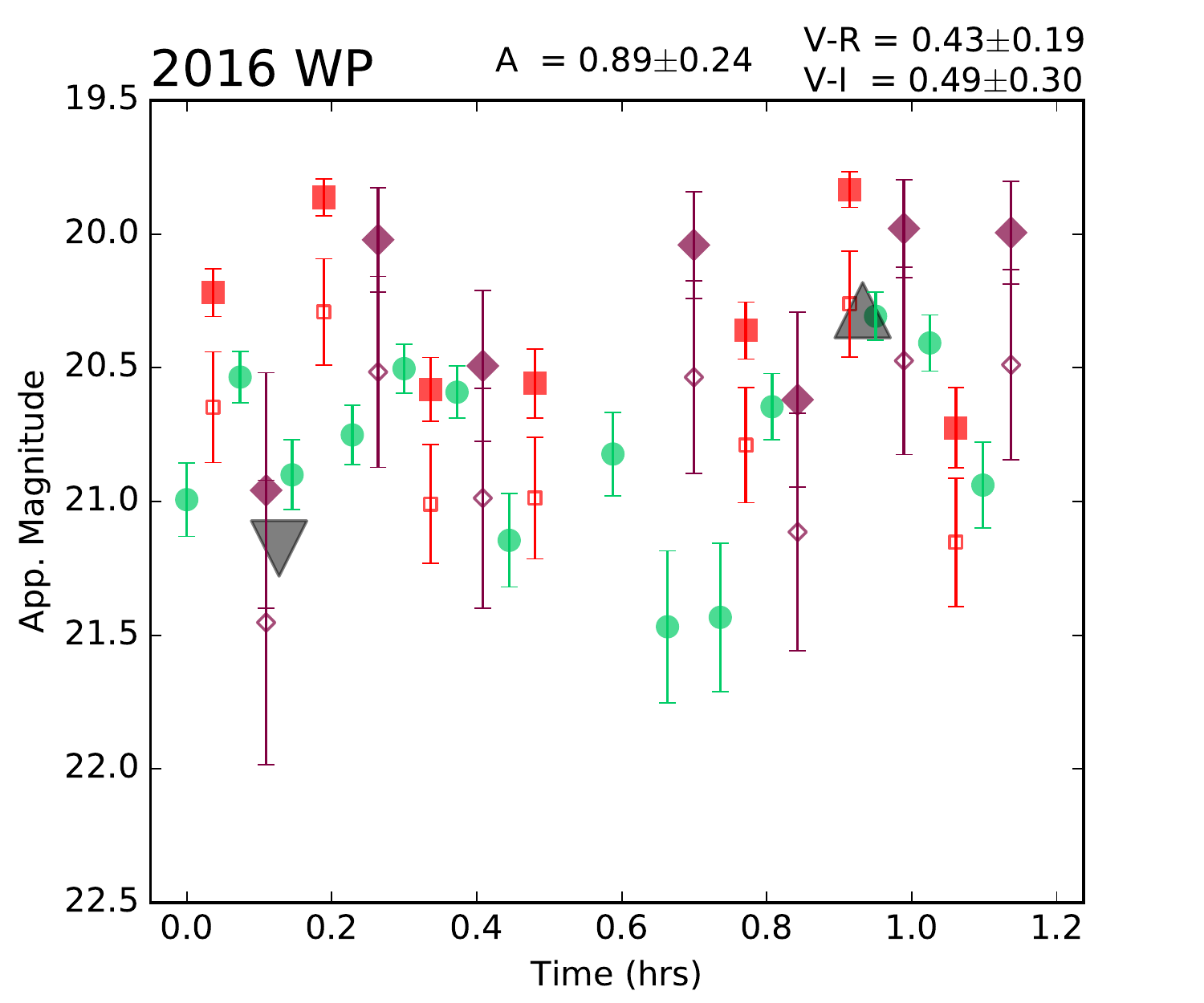}{0.3\textwidth}{(7)}
		\fig{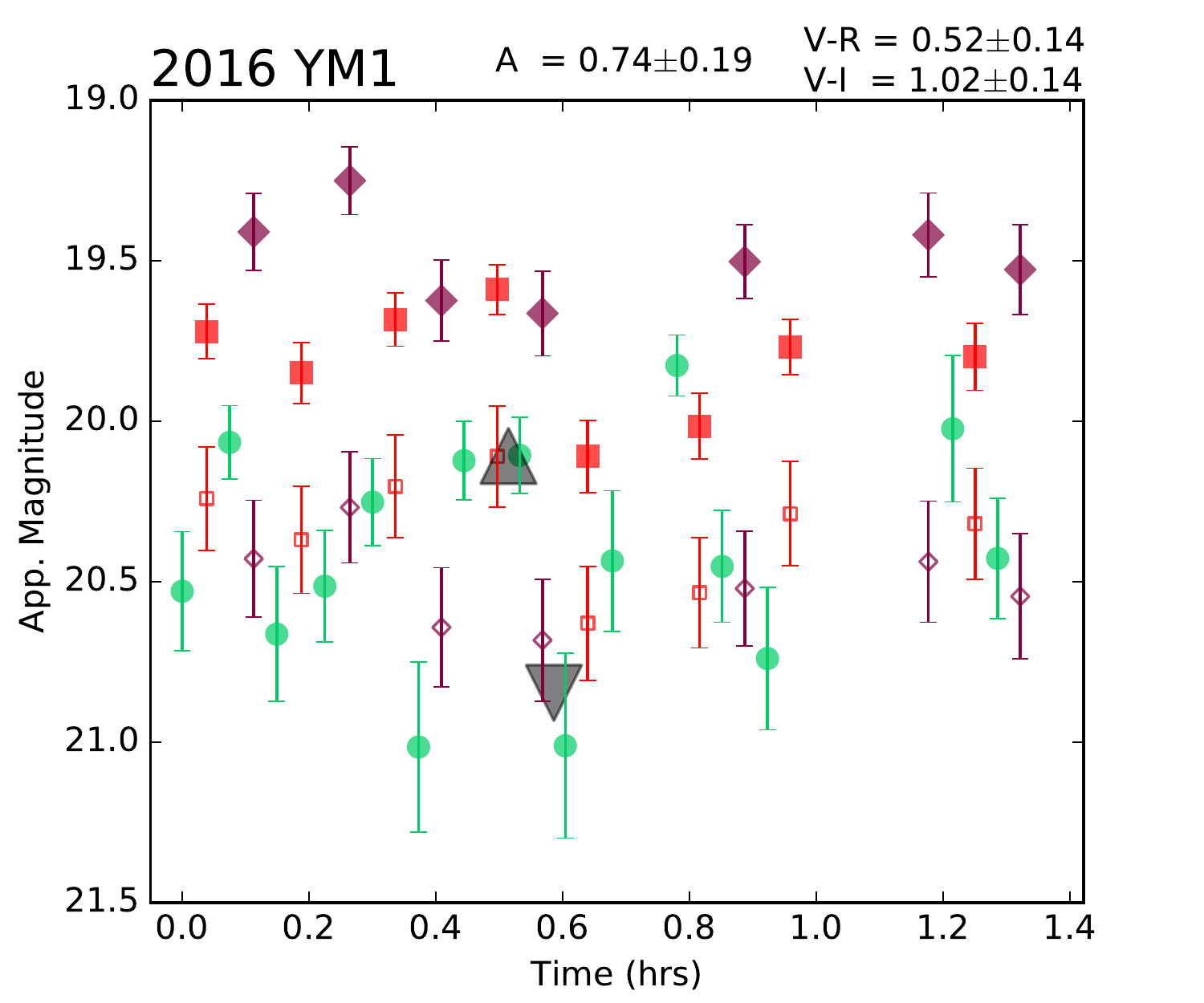}{0.3\textwidth}{(8)}
		\fig{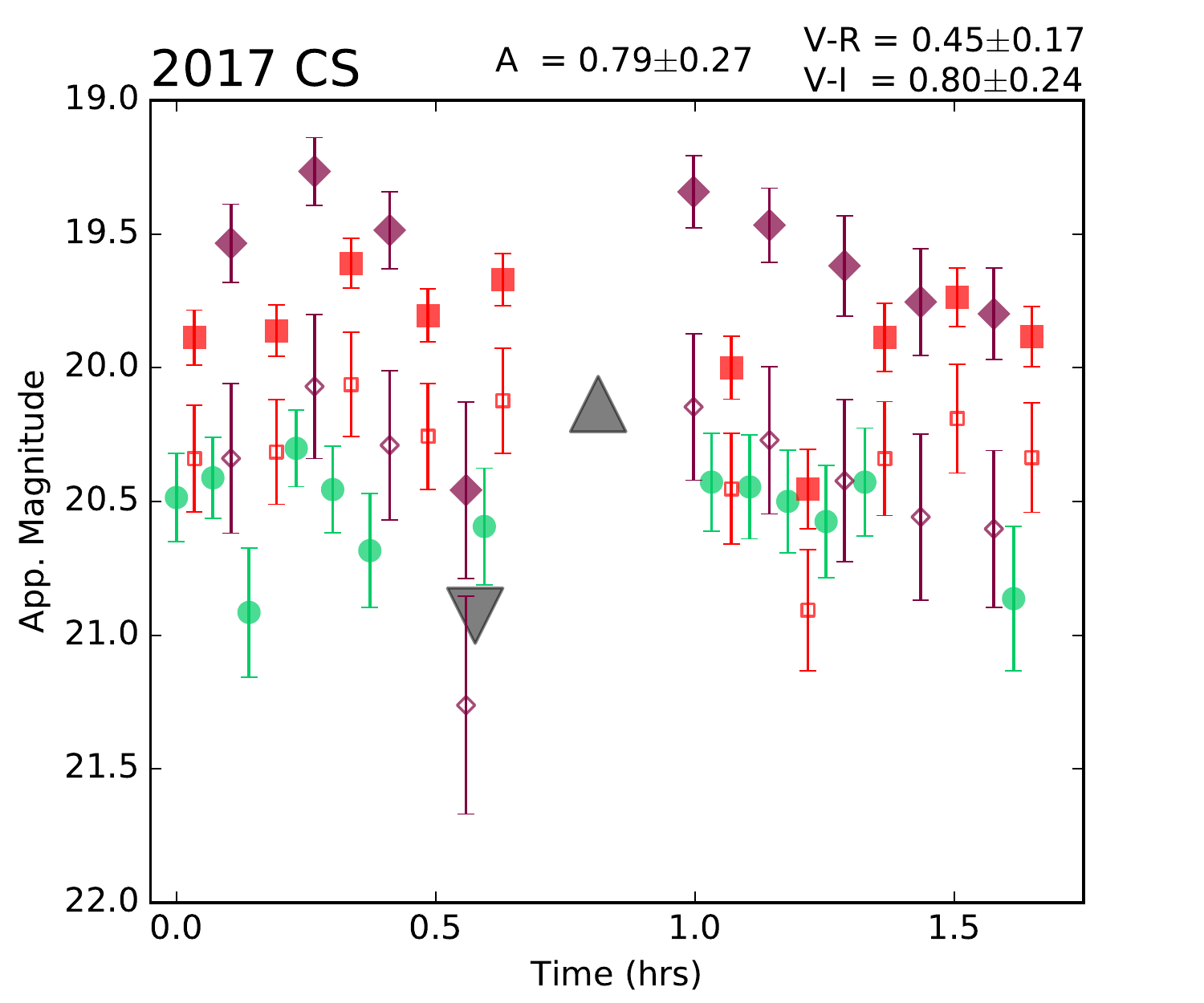}{0.3\textwidth}{(9)}
	}
\end{figure}
\begin{figure}
	\gridline{\fig{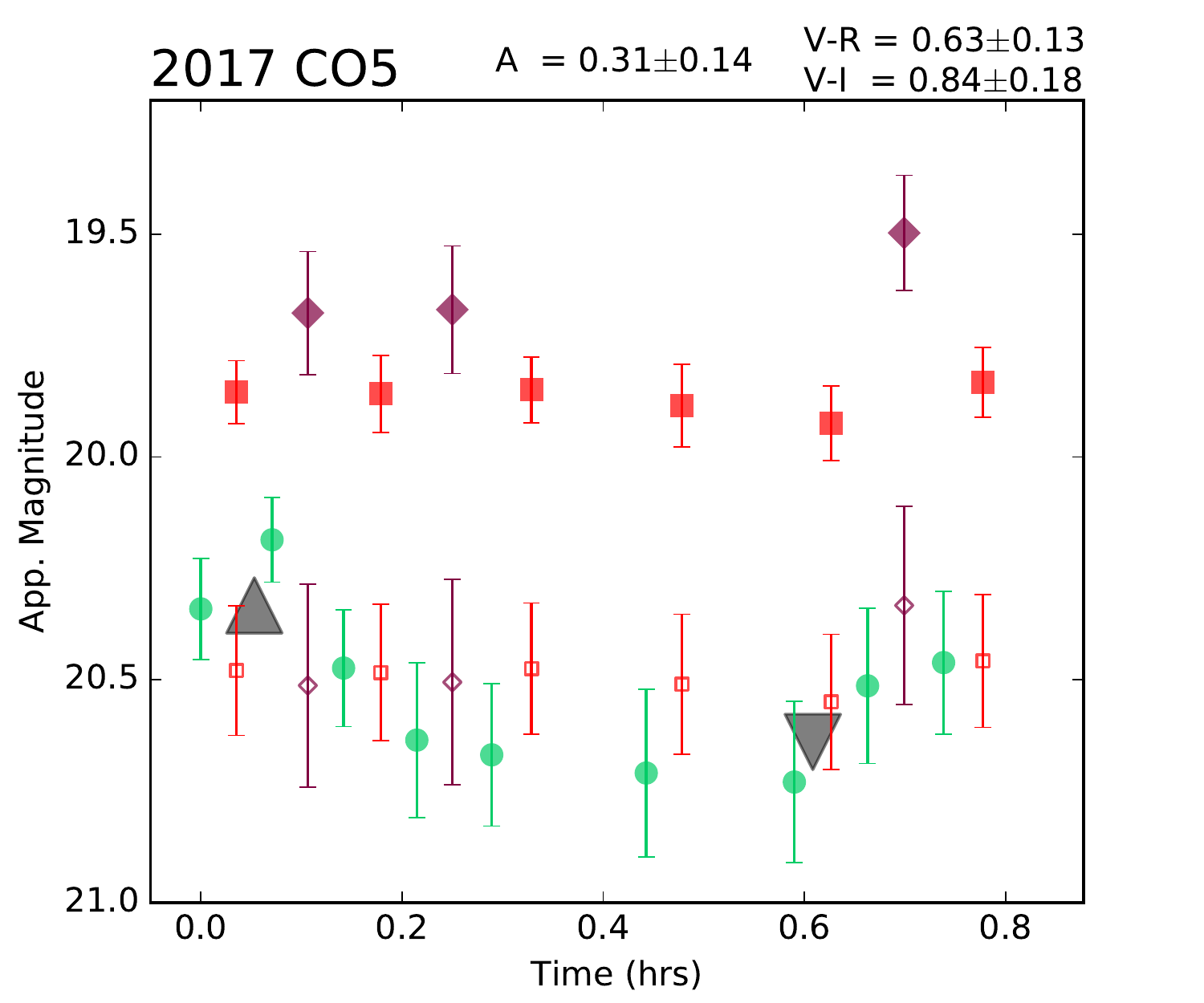}{0.3\textwidth}{(10)}
		\fig{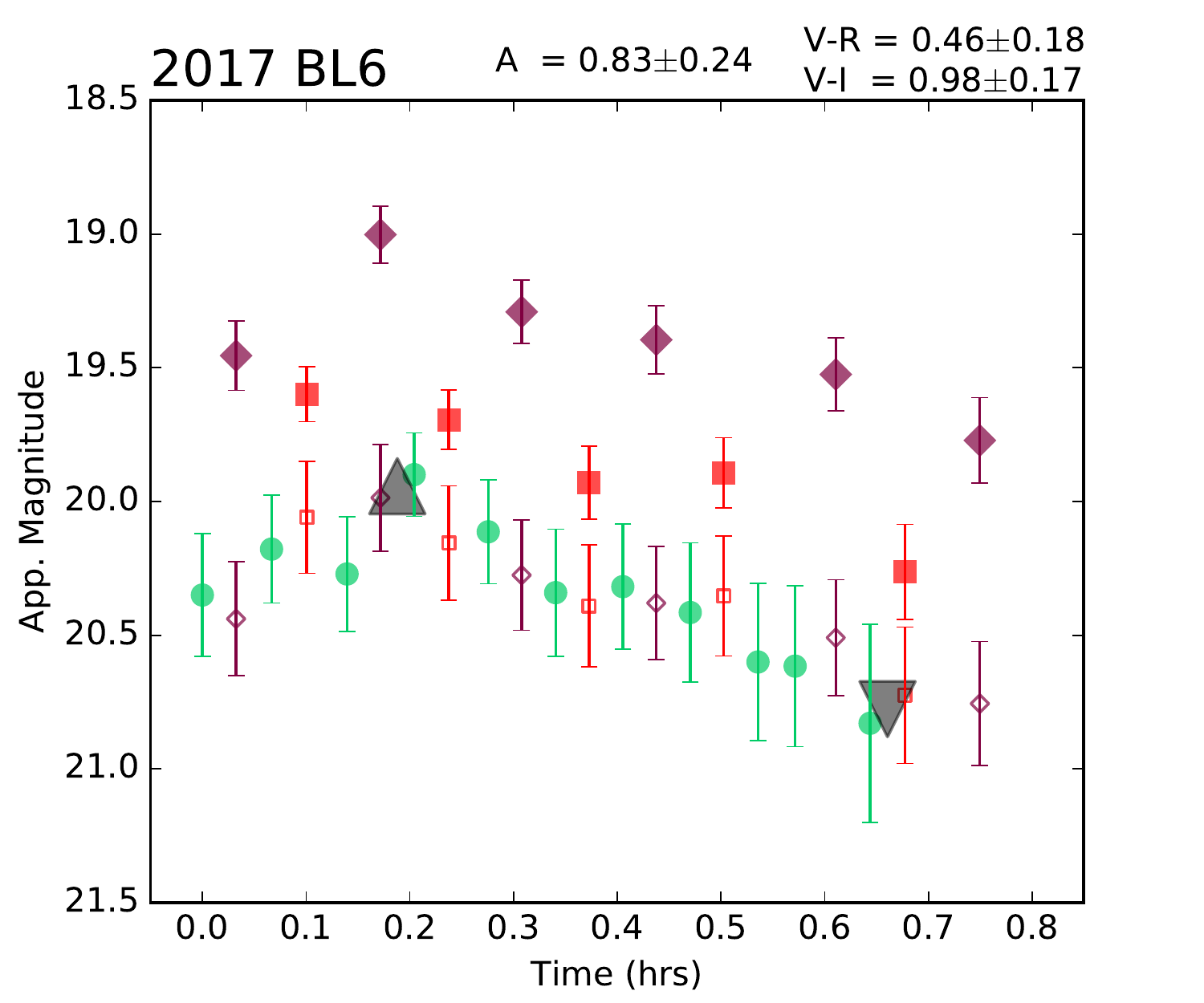}{0.3\textwidth}{(11)}
		\fig{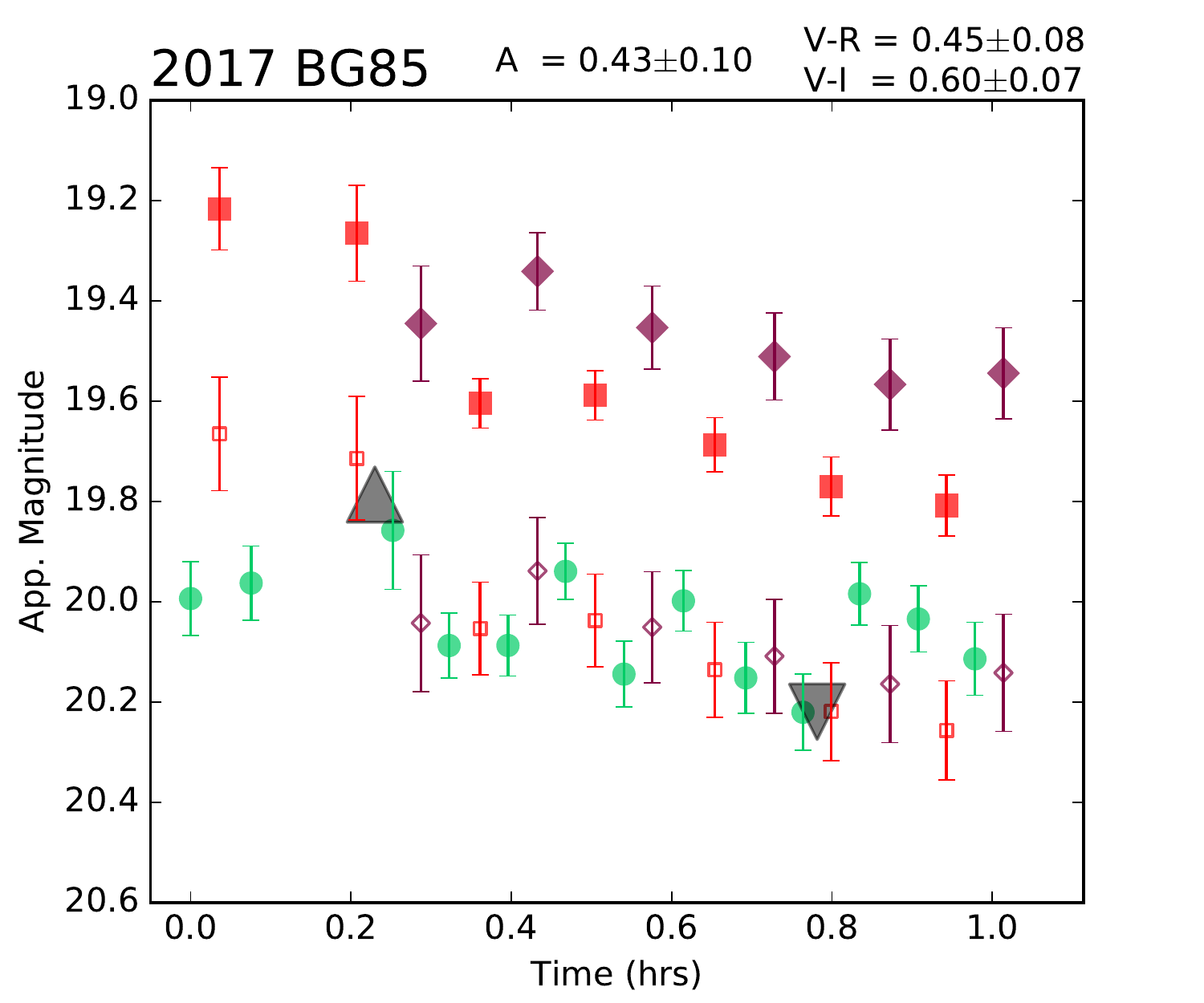}{0.3\textwidth}{(12)}
	}
	\gridline{\fig{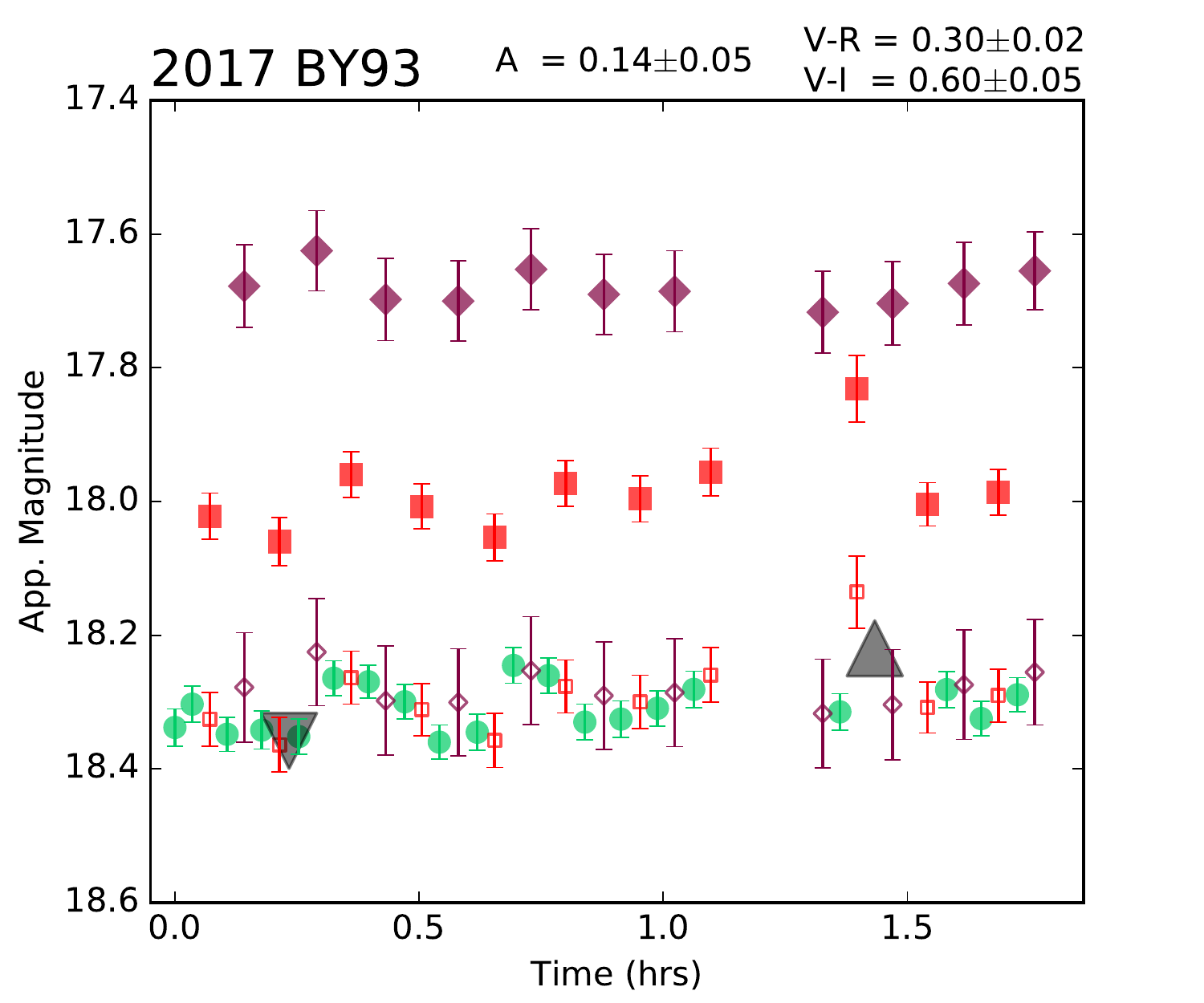}{0.3\textwidth}{(13)}
		\fig{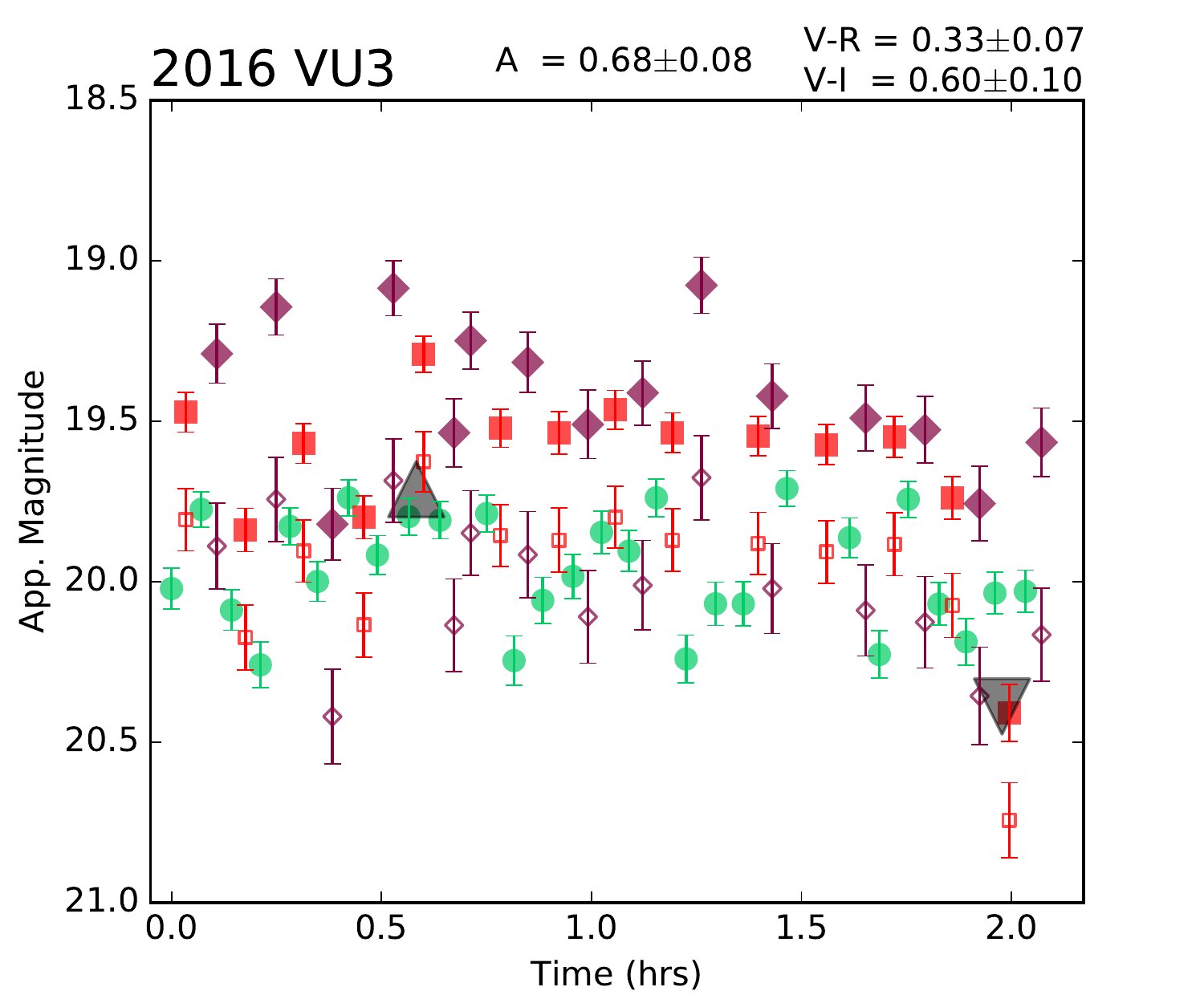}{0.3\textwidth}{(14)}
		\fig{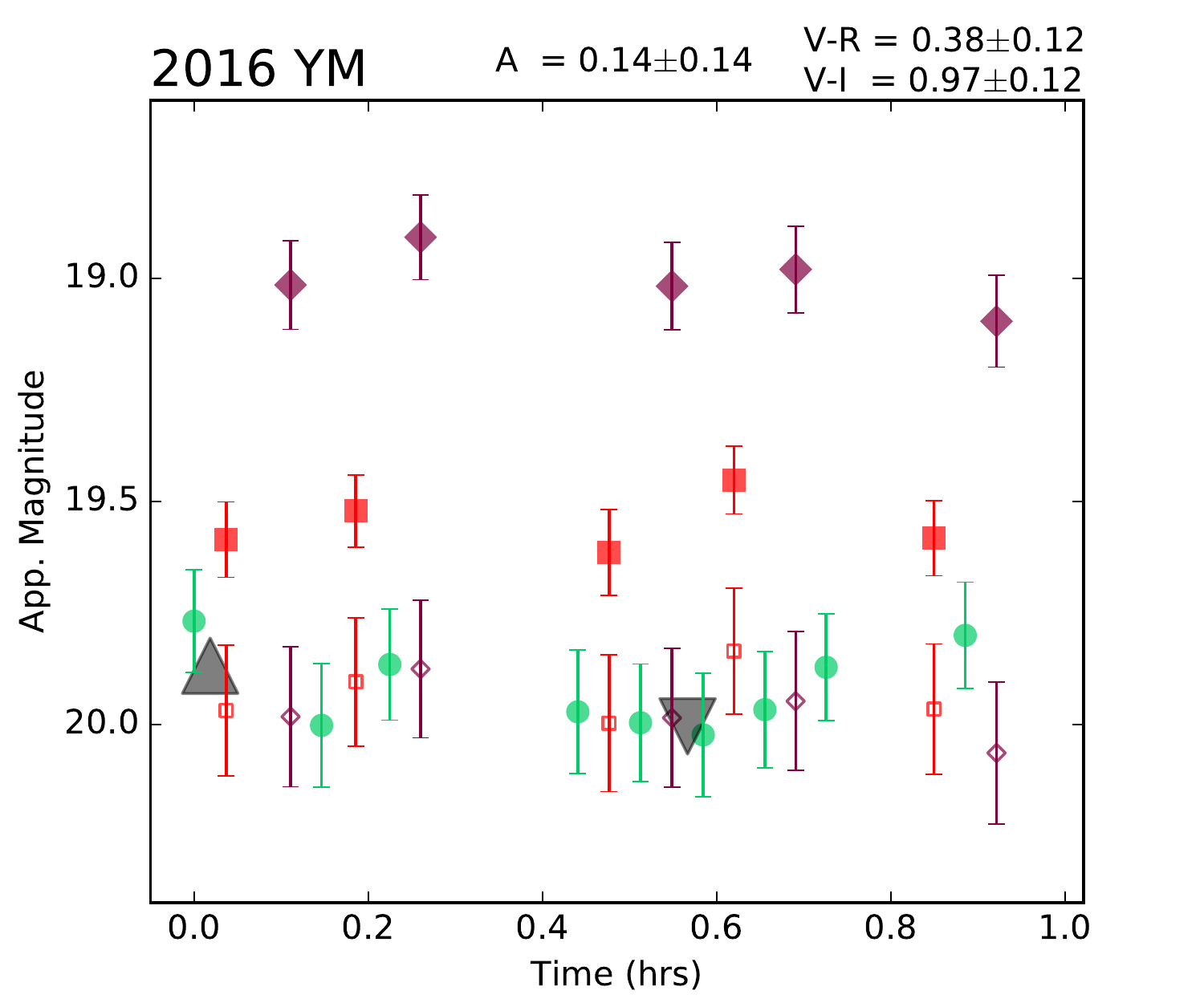}{0.3\textwidth}{(15)}
	}
	\gridline{\fig{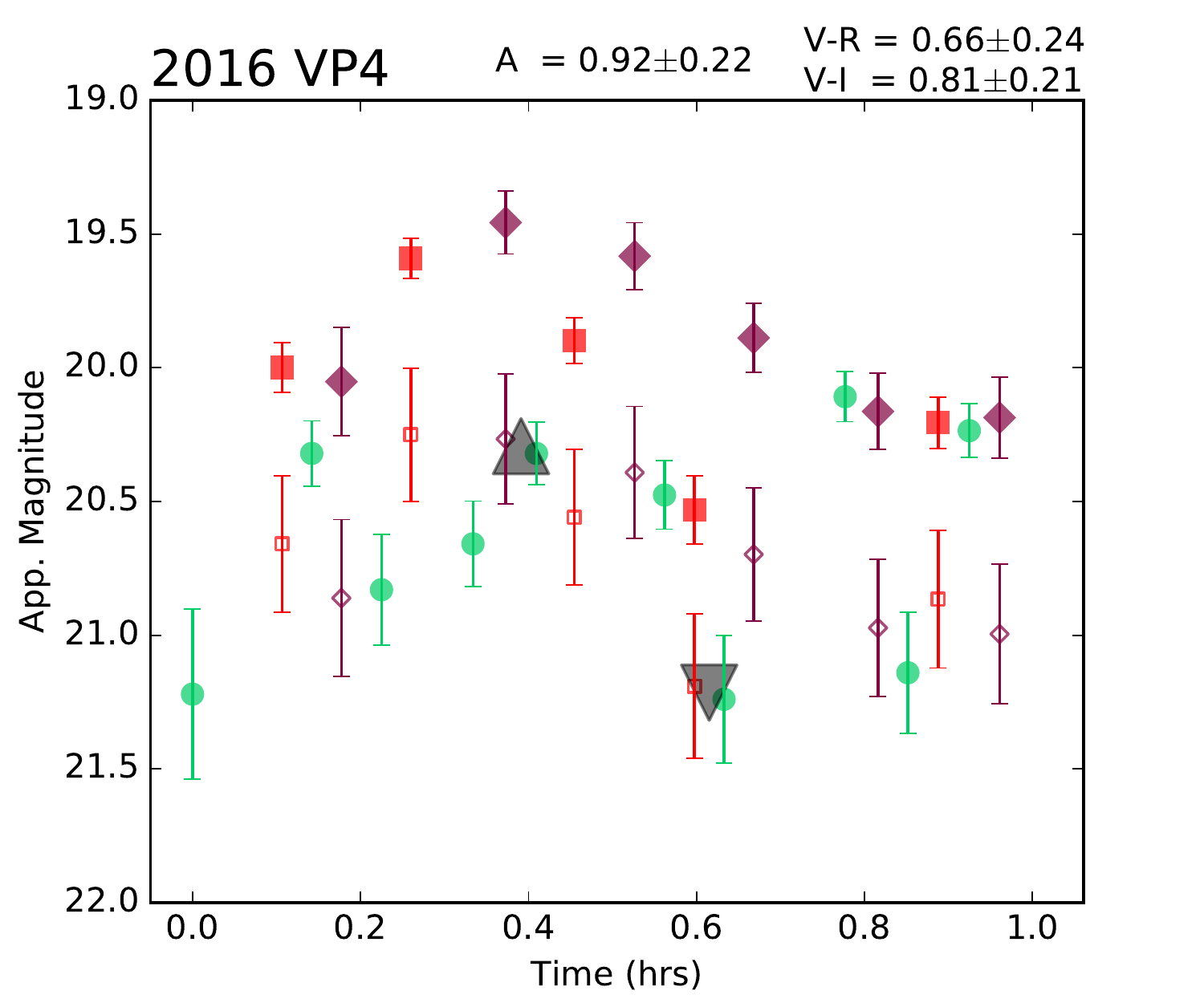}{0.3\textwidth}{(16)}
		\fig{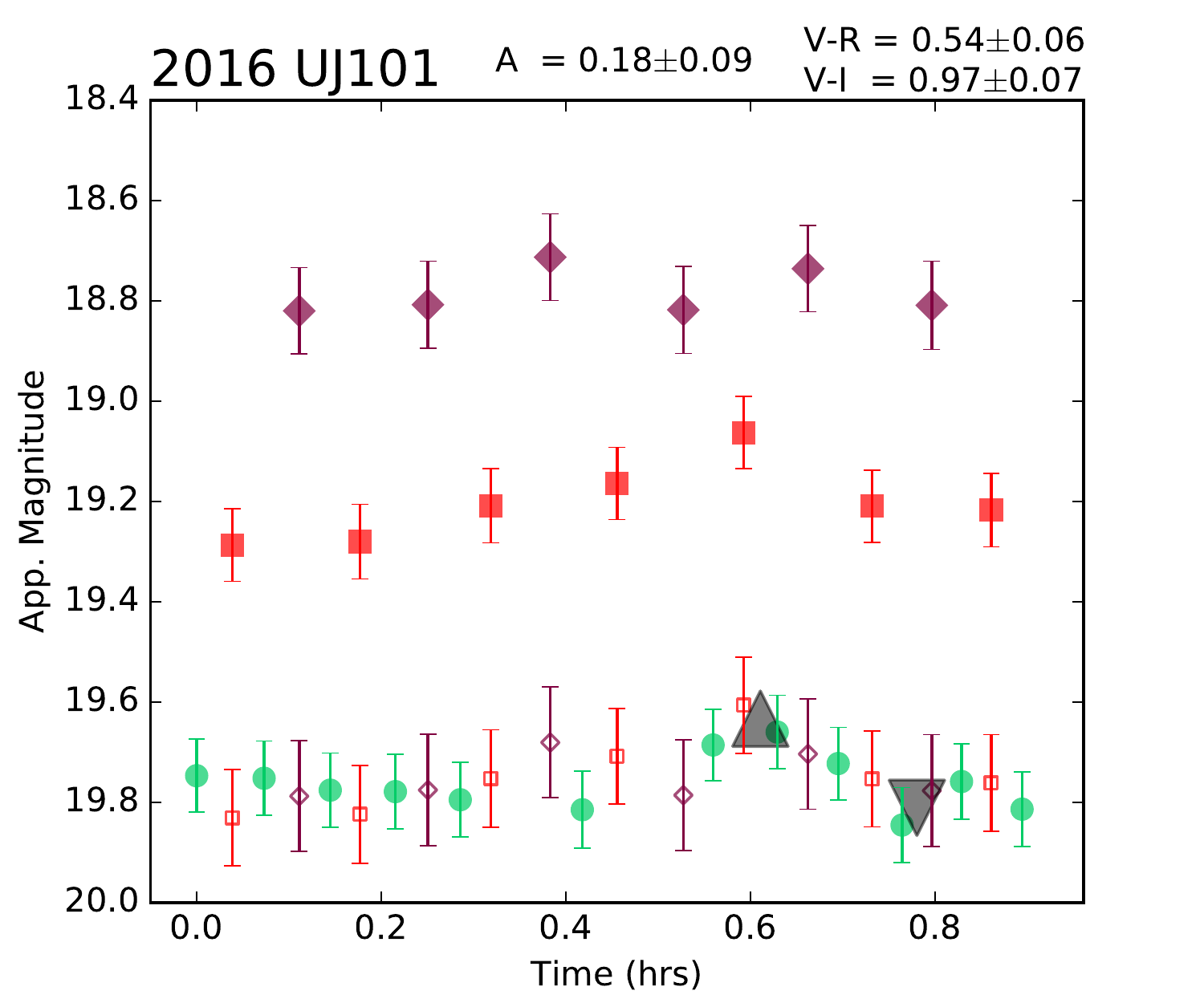}{0.3\textwidth}{(17)}
		\fig{figures/f3_18.pdf}{0.3\textwidth}{(18)}
	}
	\gridline{\fig{figures/f3_19.pdf}{0.3\textwidth}{(19)}
		\fig{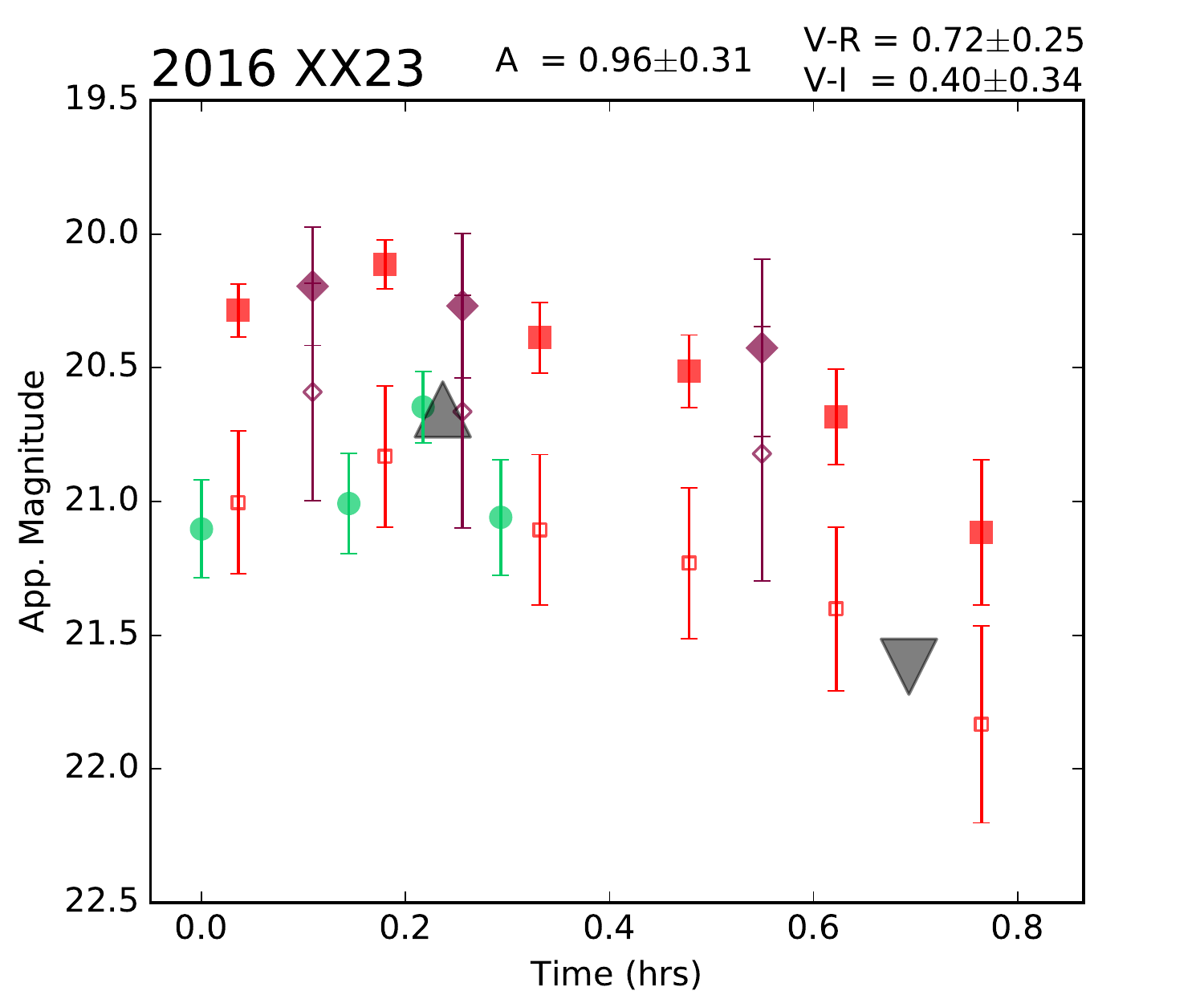}{0.3\textwidth}{(20)}
		\fig{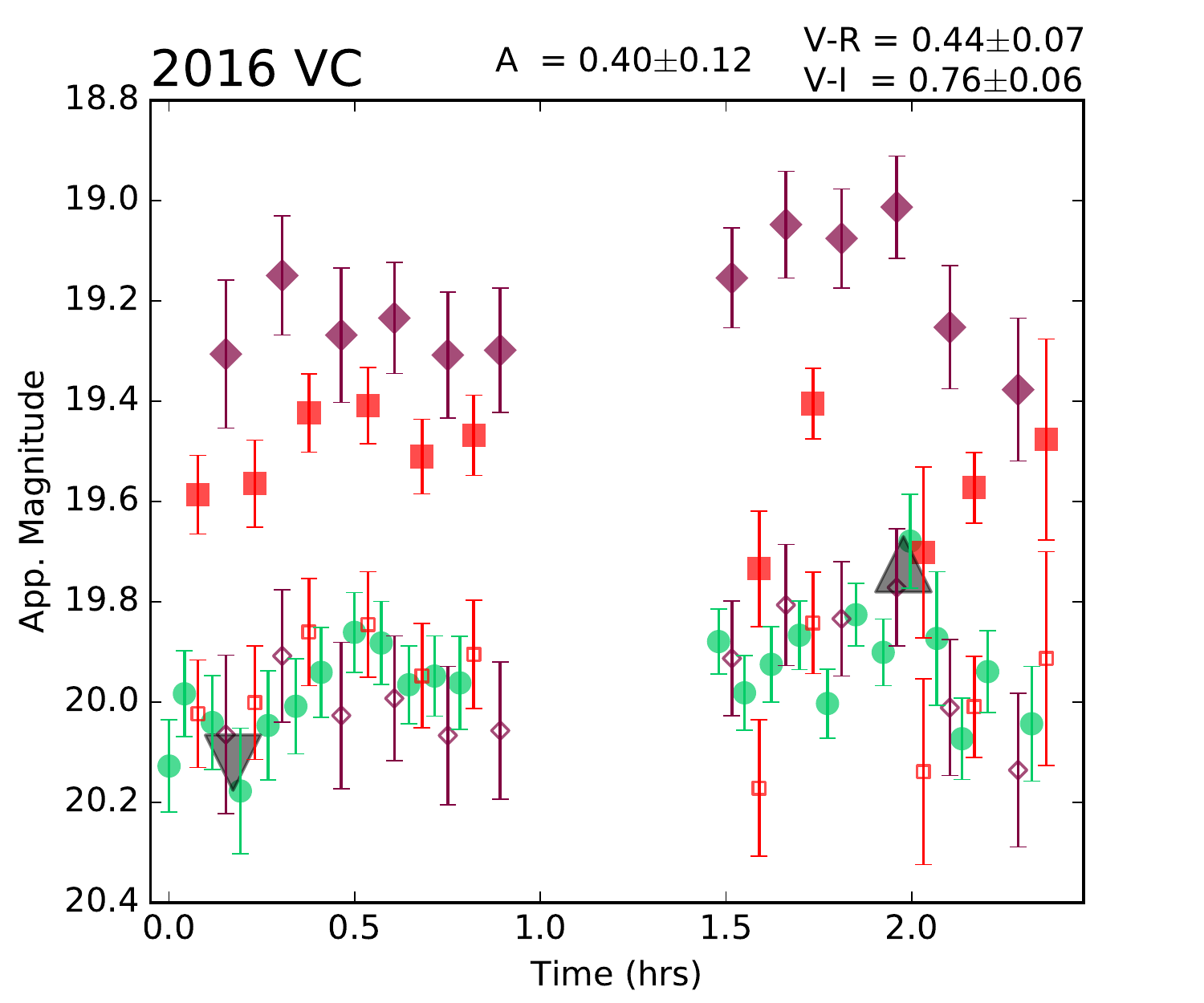}{0.3\textwidth}{(21)}
	}
\end{figure}
\begin{figure}
	\gridline{\fig{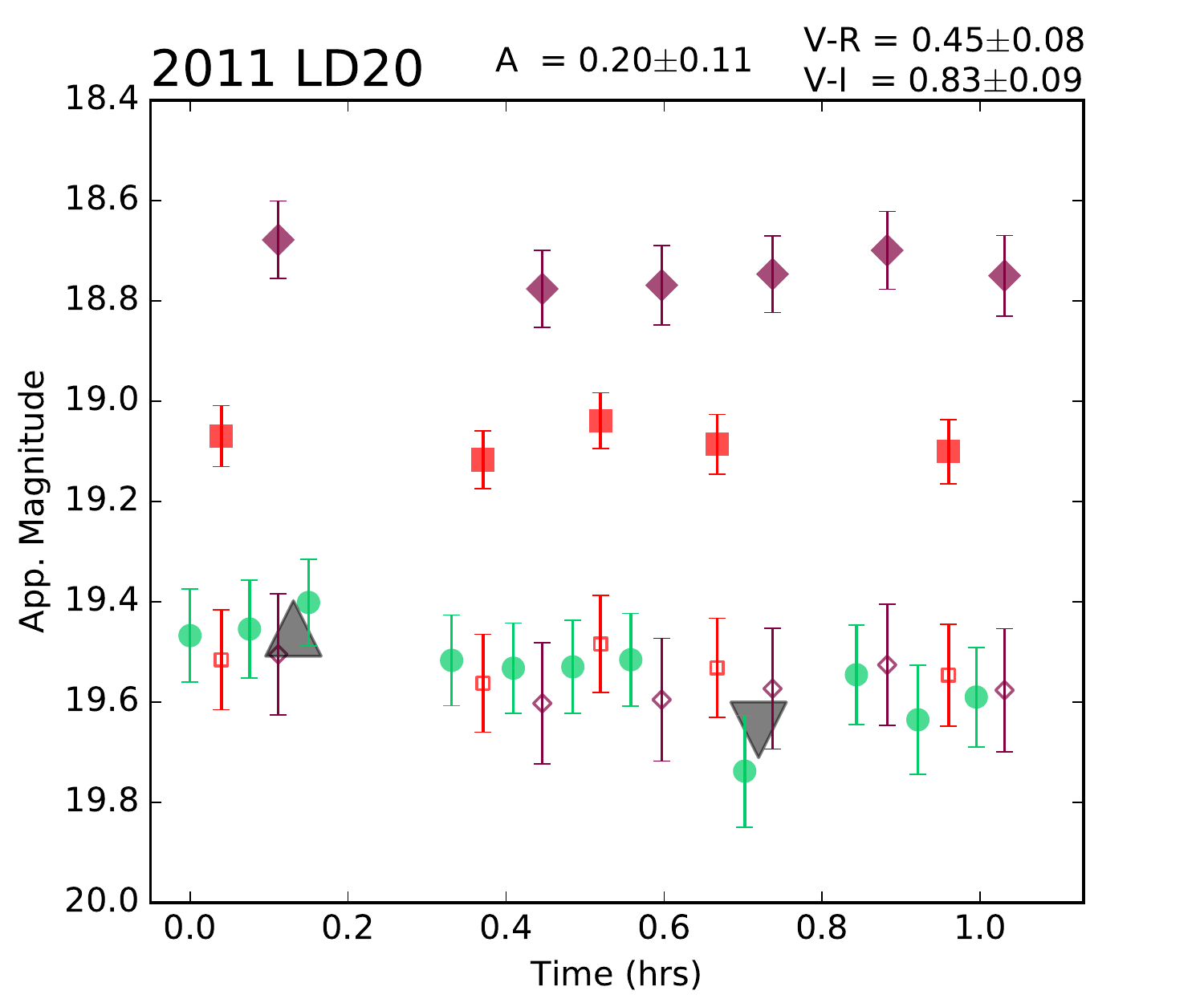}{0.3\textwidth}{(22)}
		\fig{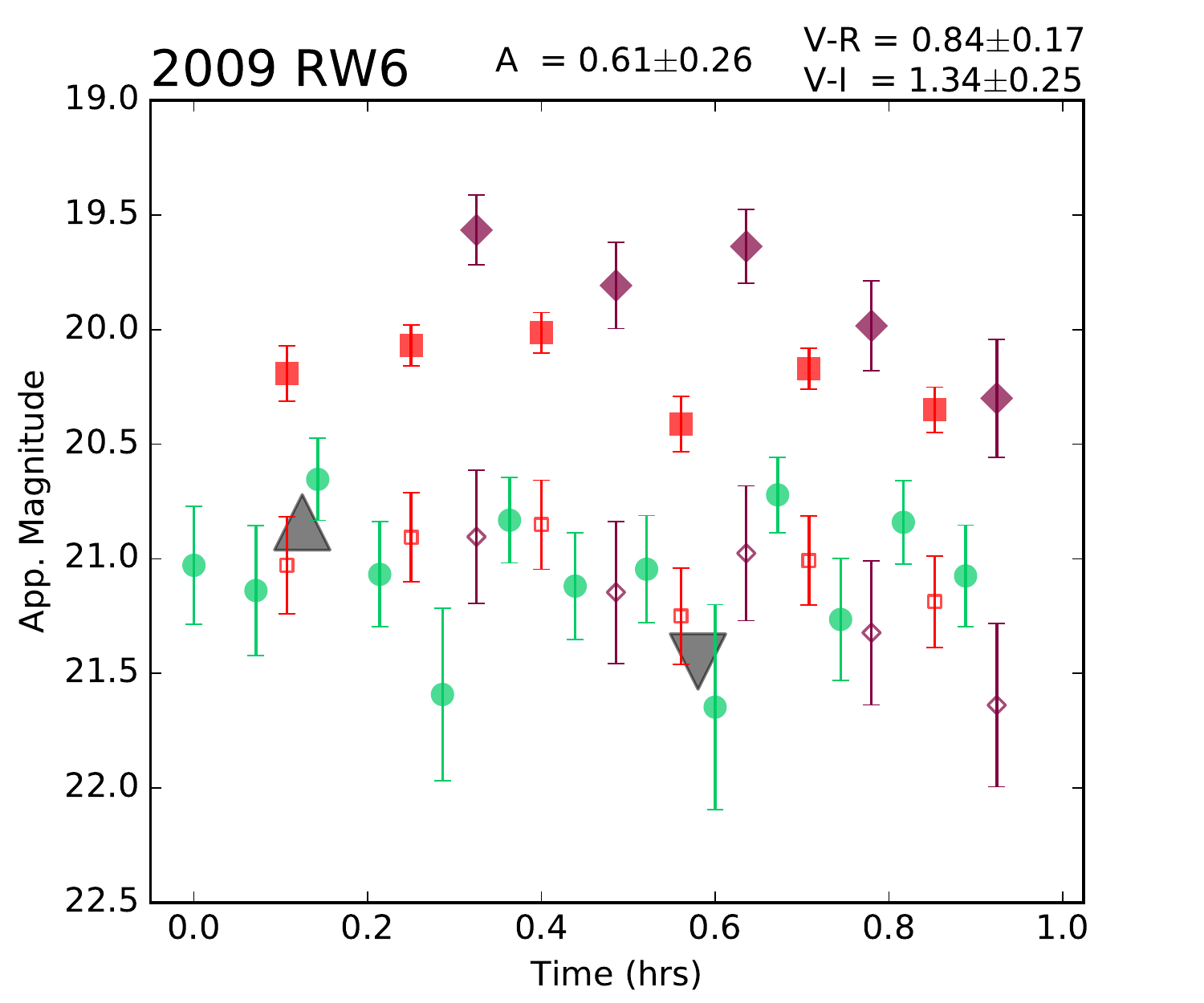}{0.3\textwidth}{(23)}
		\fig{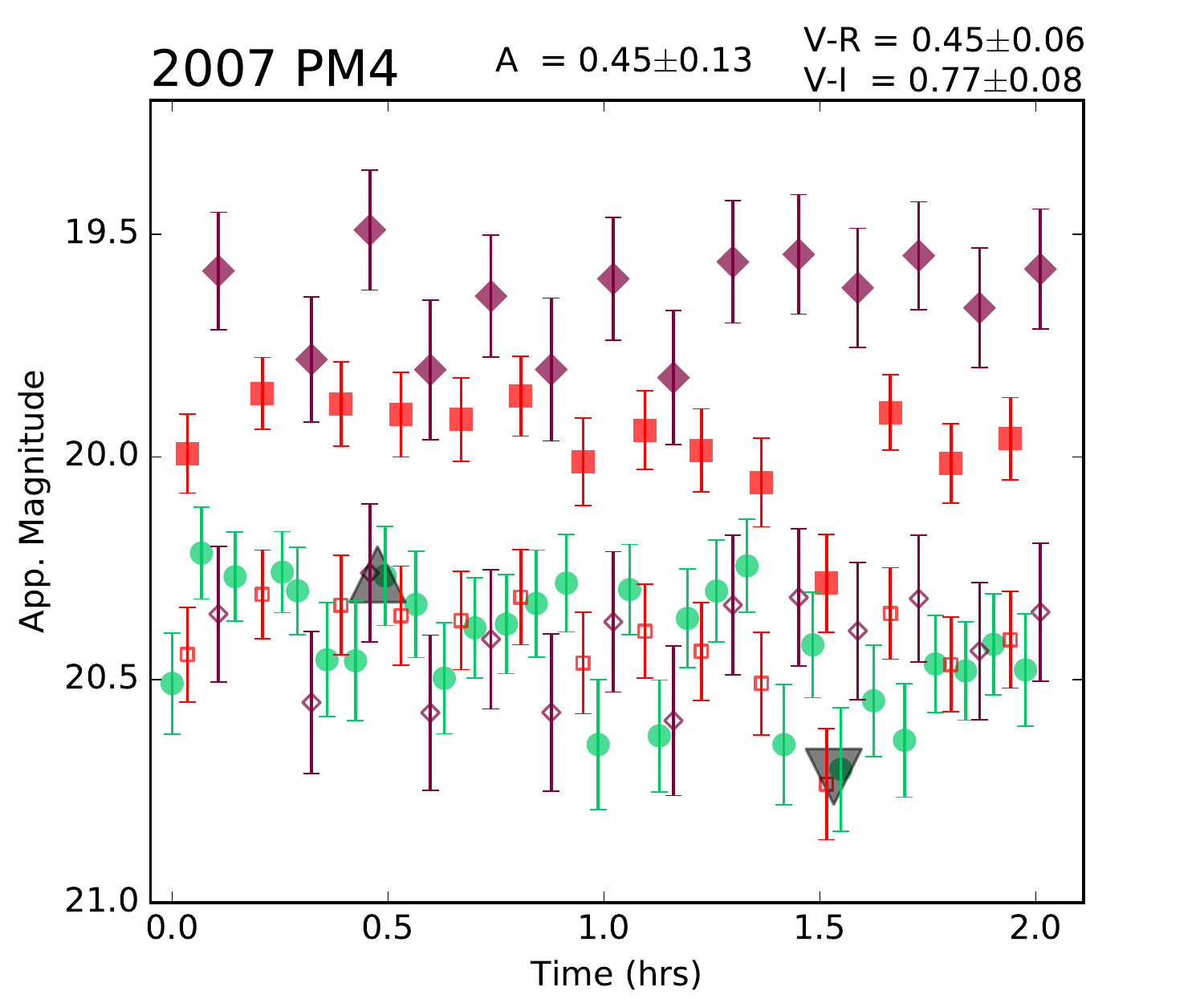}{0.3\textwidth}{(24)}
	}
	\gridline{\fig{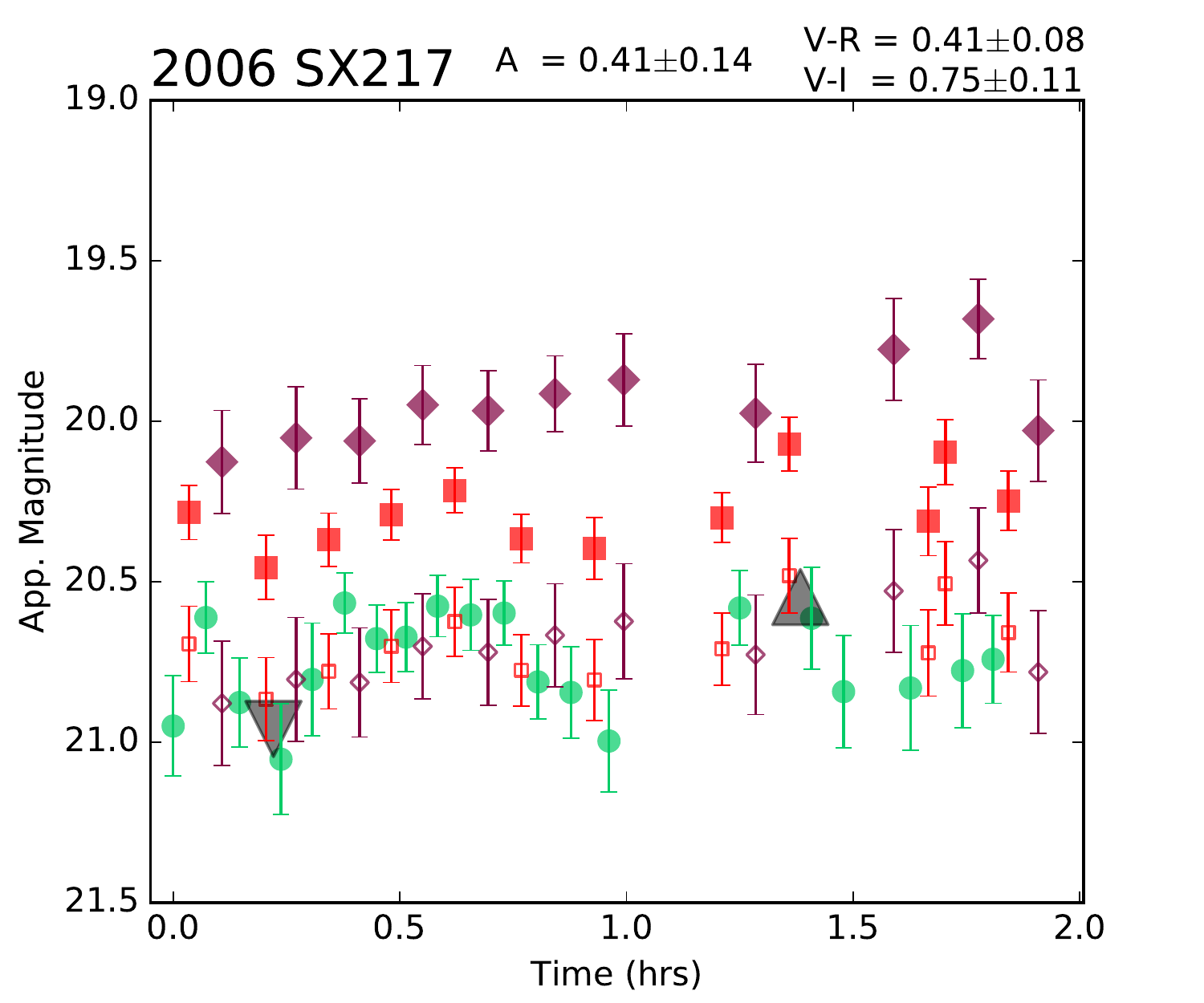}{0.3\textwidth}{(25)}
		\fig{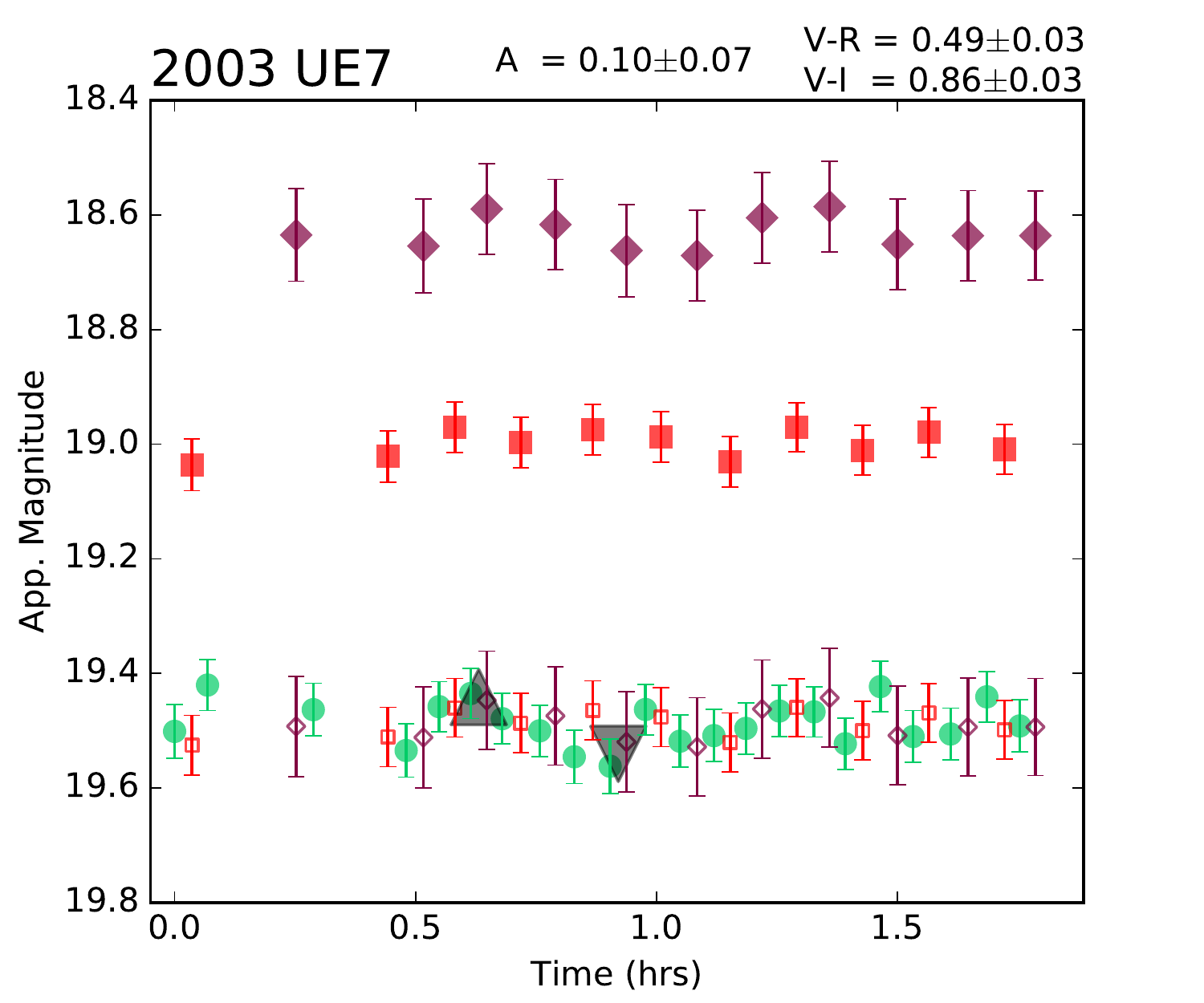}{0.3\textwidth}{(26)}
		\fig{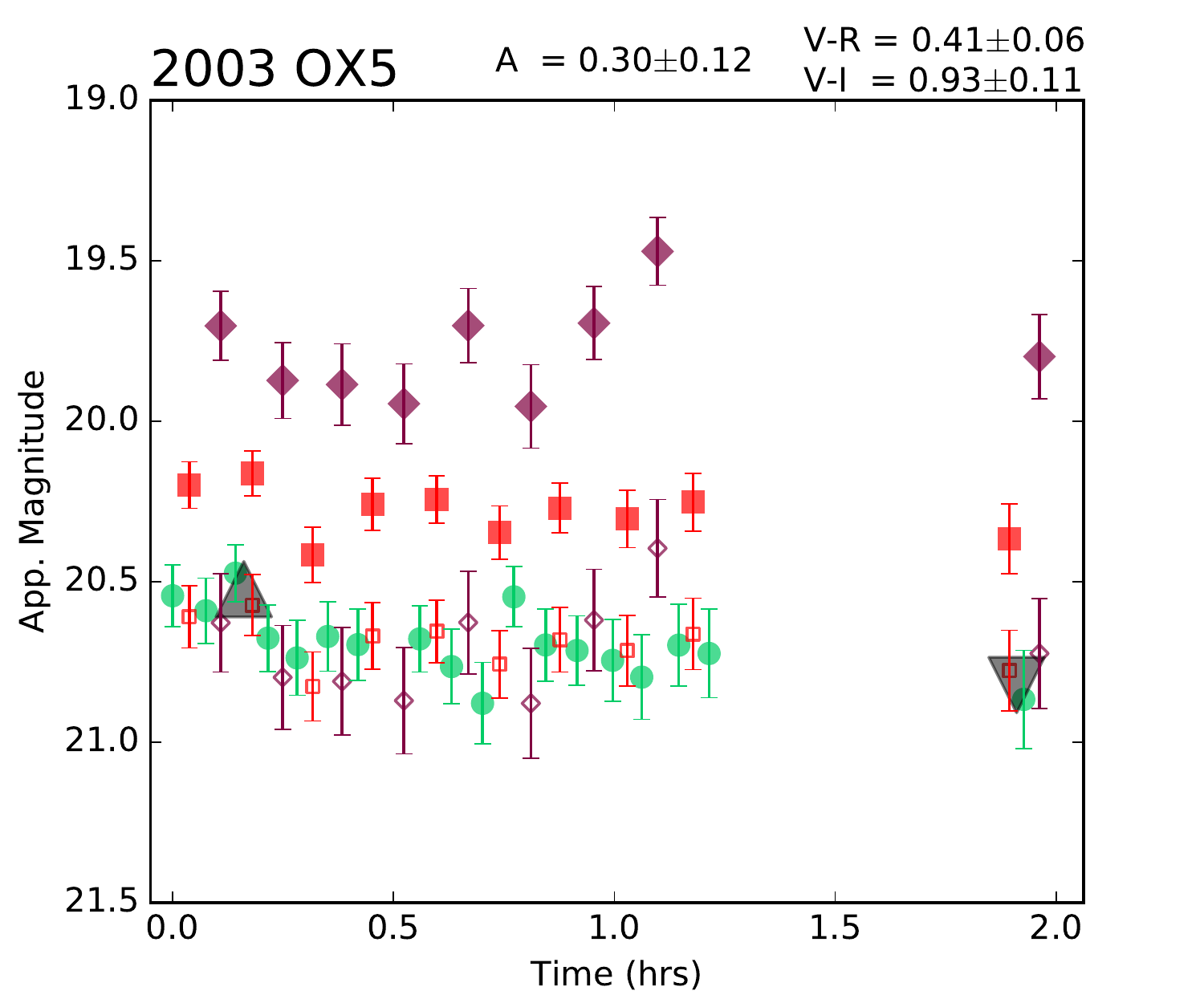}{0.3\textwidth}{(27)}
	}
	\gridline{\fig{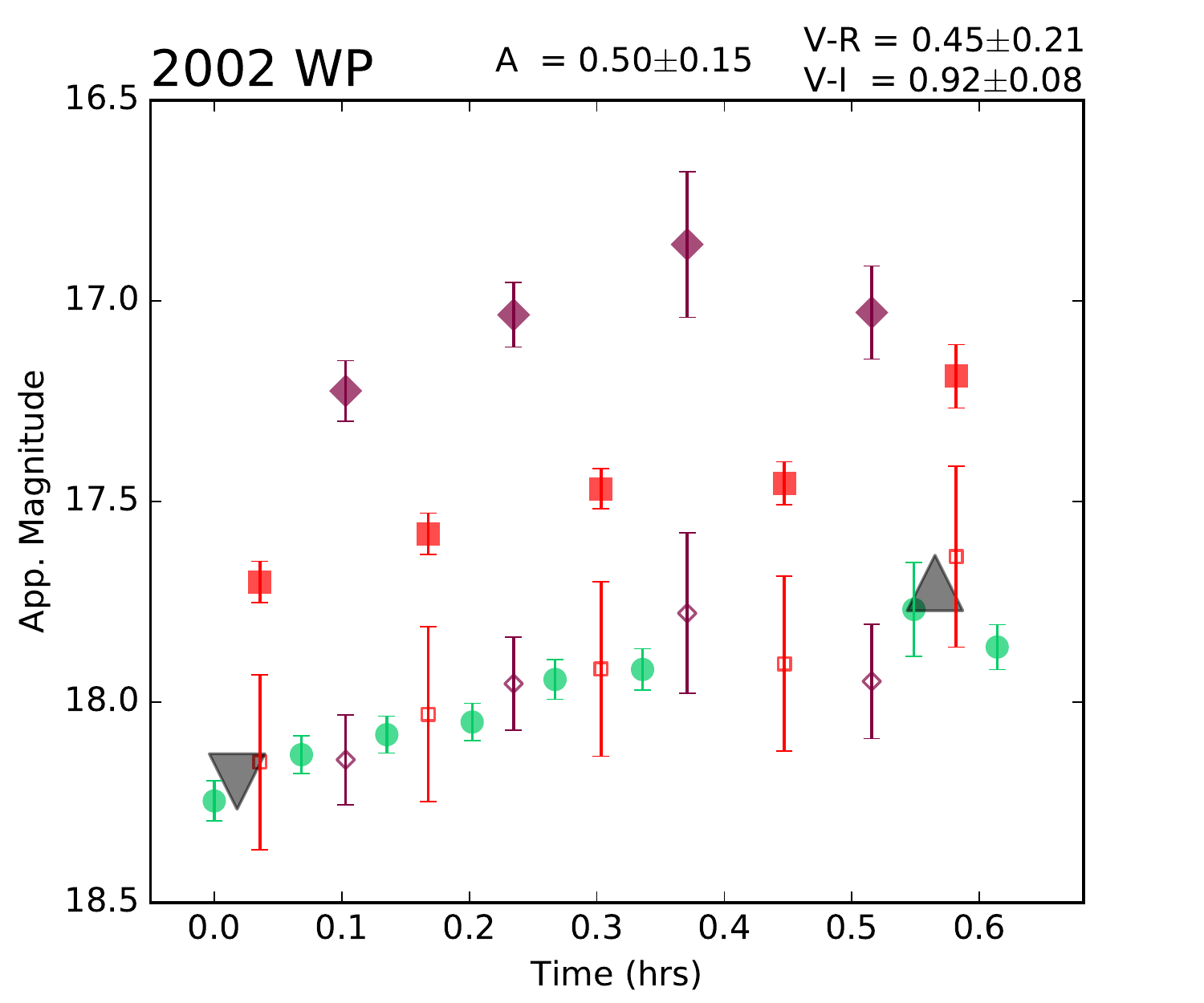}{0.3\textwidth}{(28)}
		\fig{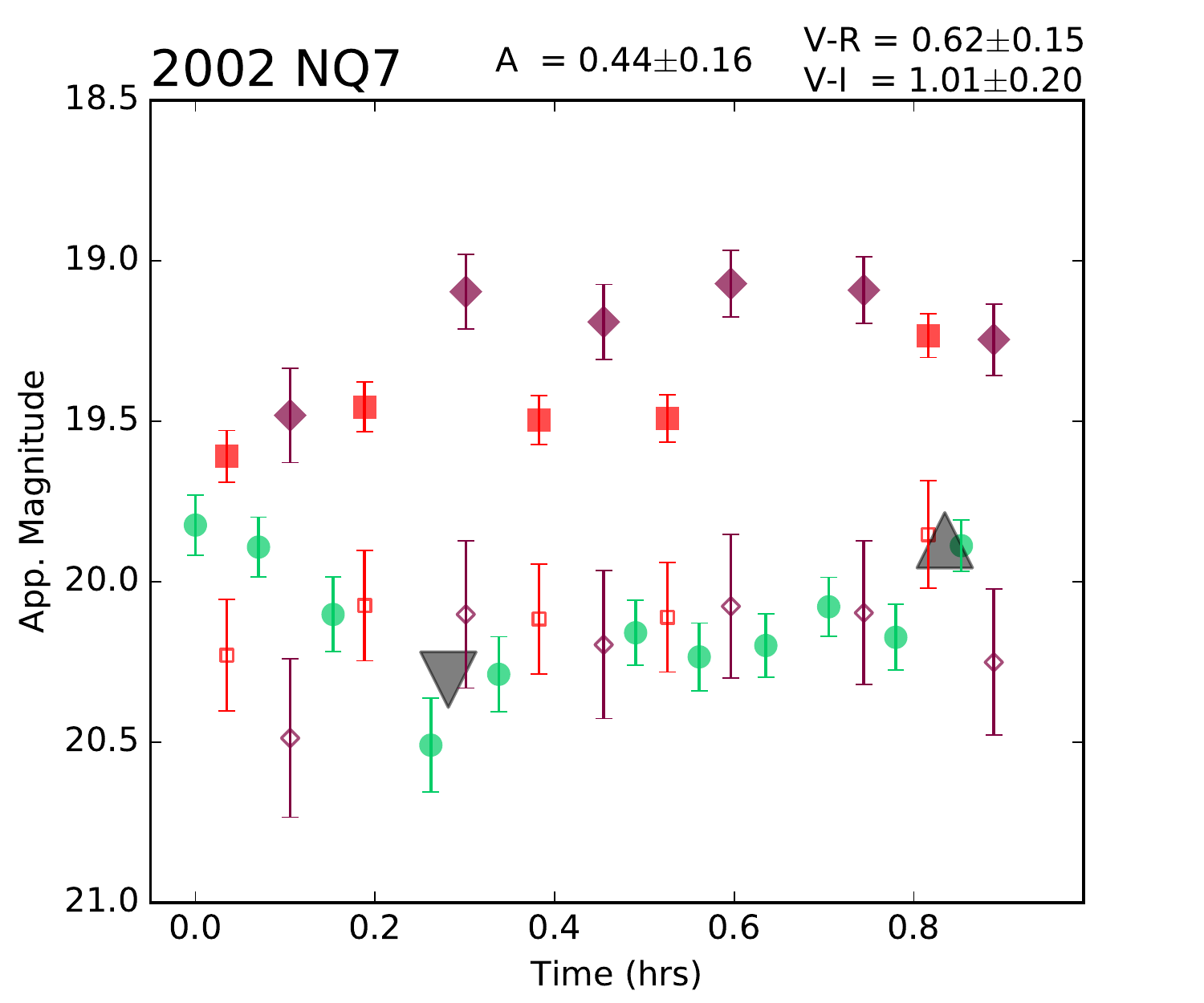}{0.3\textwidth}{(29)}
		\fig{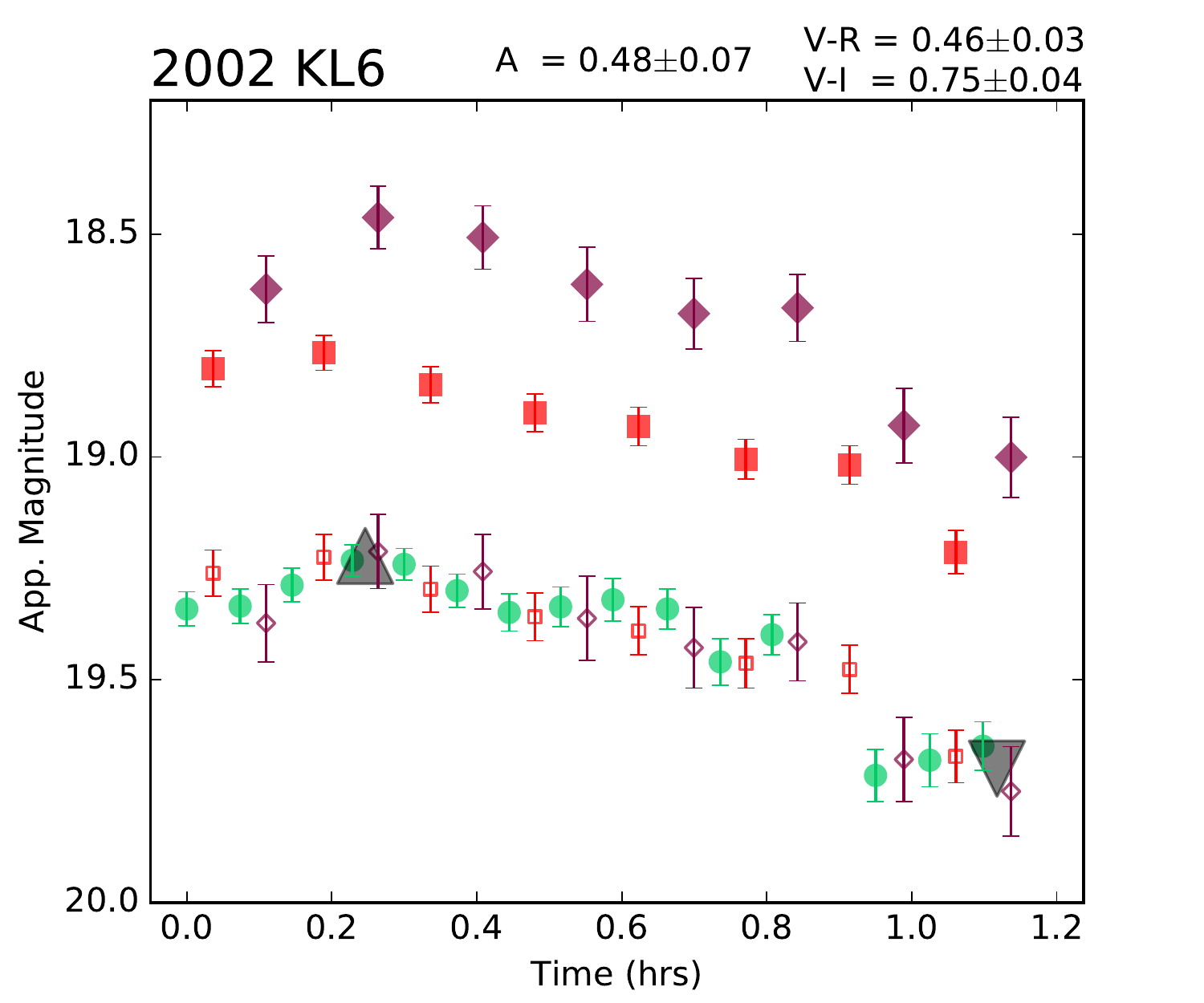}{0.3\textwidth}{(30)}
	}
	\gridline{\fig{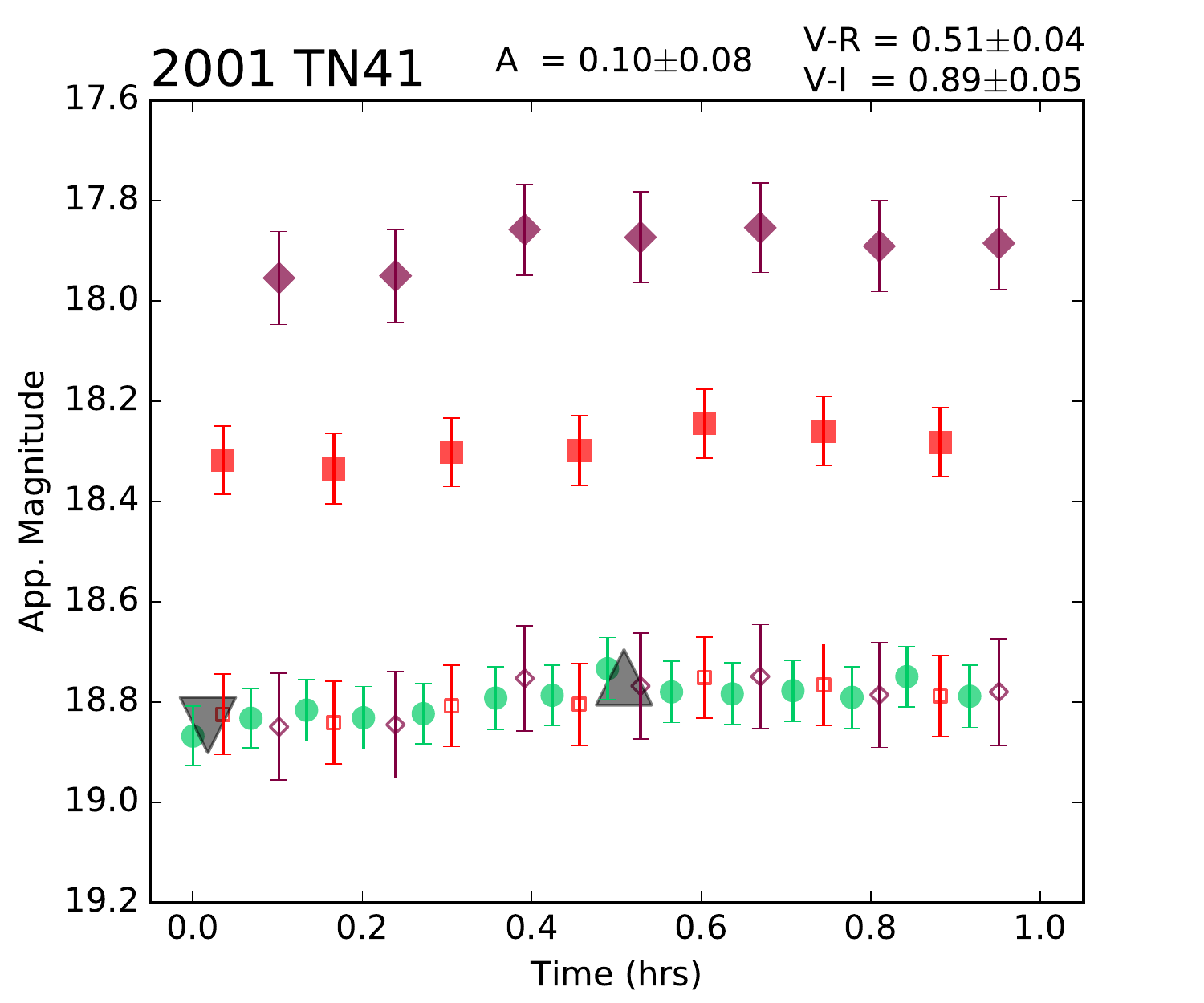}{0.3\textwidth}{(31)}
		\fig{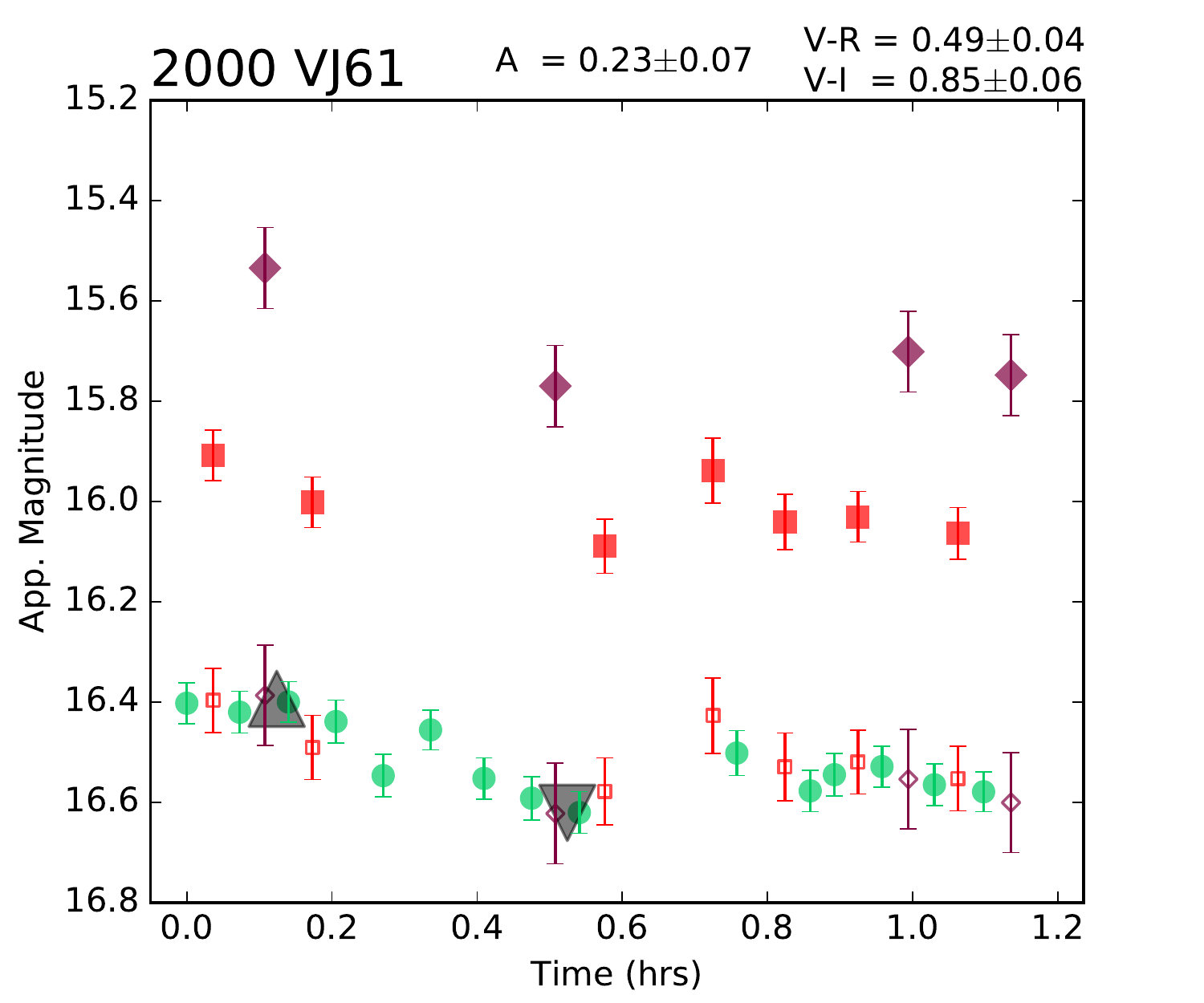}{0.3\textwidth}{(32)}
		\fig{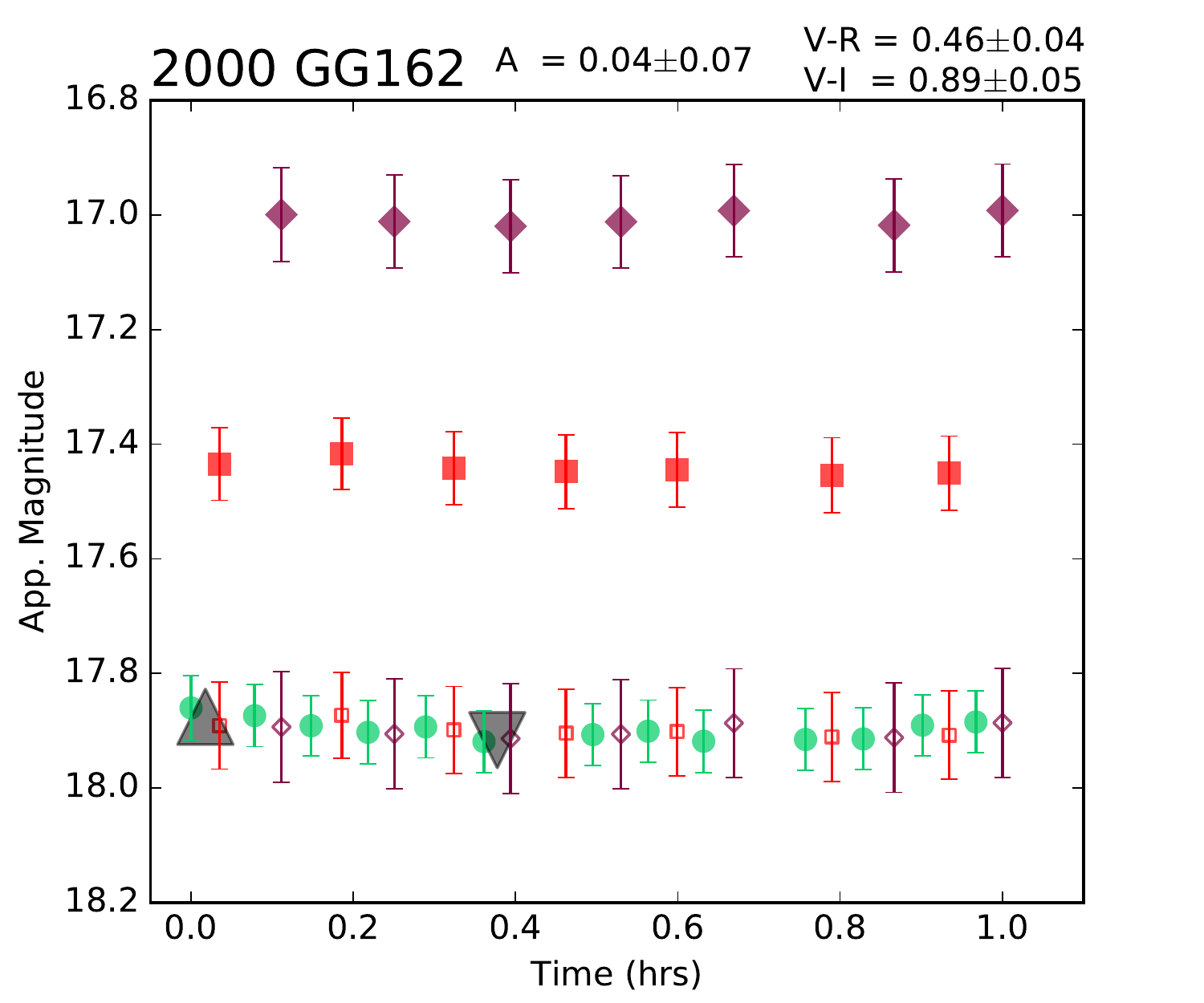}{0.3\textwidth}{(33)}
	}
\end{figure}
\begin{figure}
	\gridline{\fig{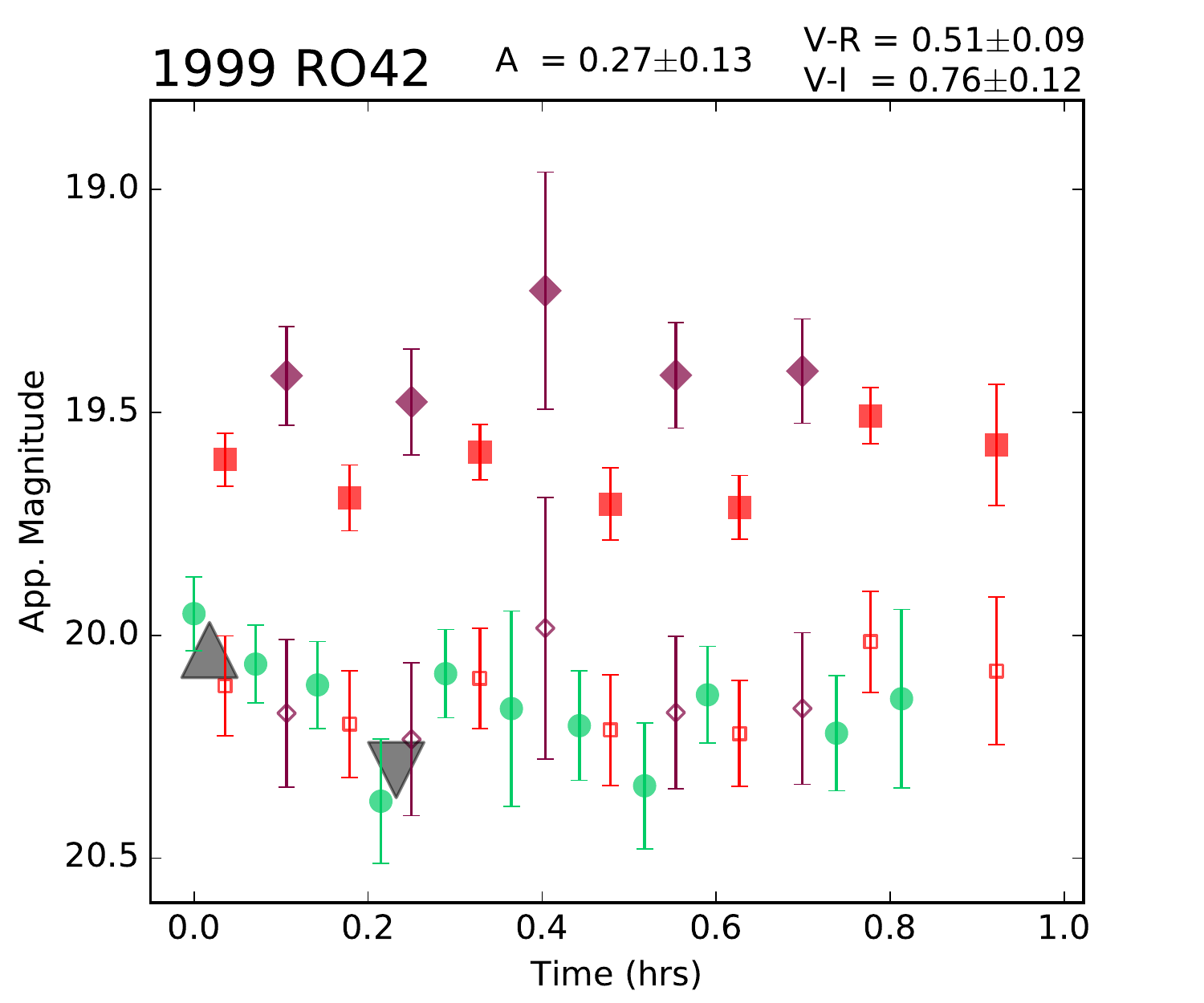}{0.3\textwidth}{(34)}
		\fig{figures/f3_35.pdf}{0.3\textwidth}{(35)}
		\fig{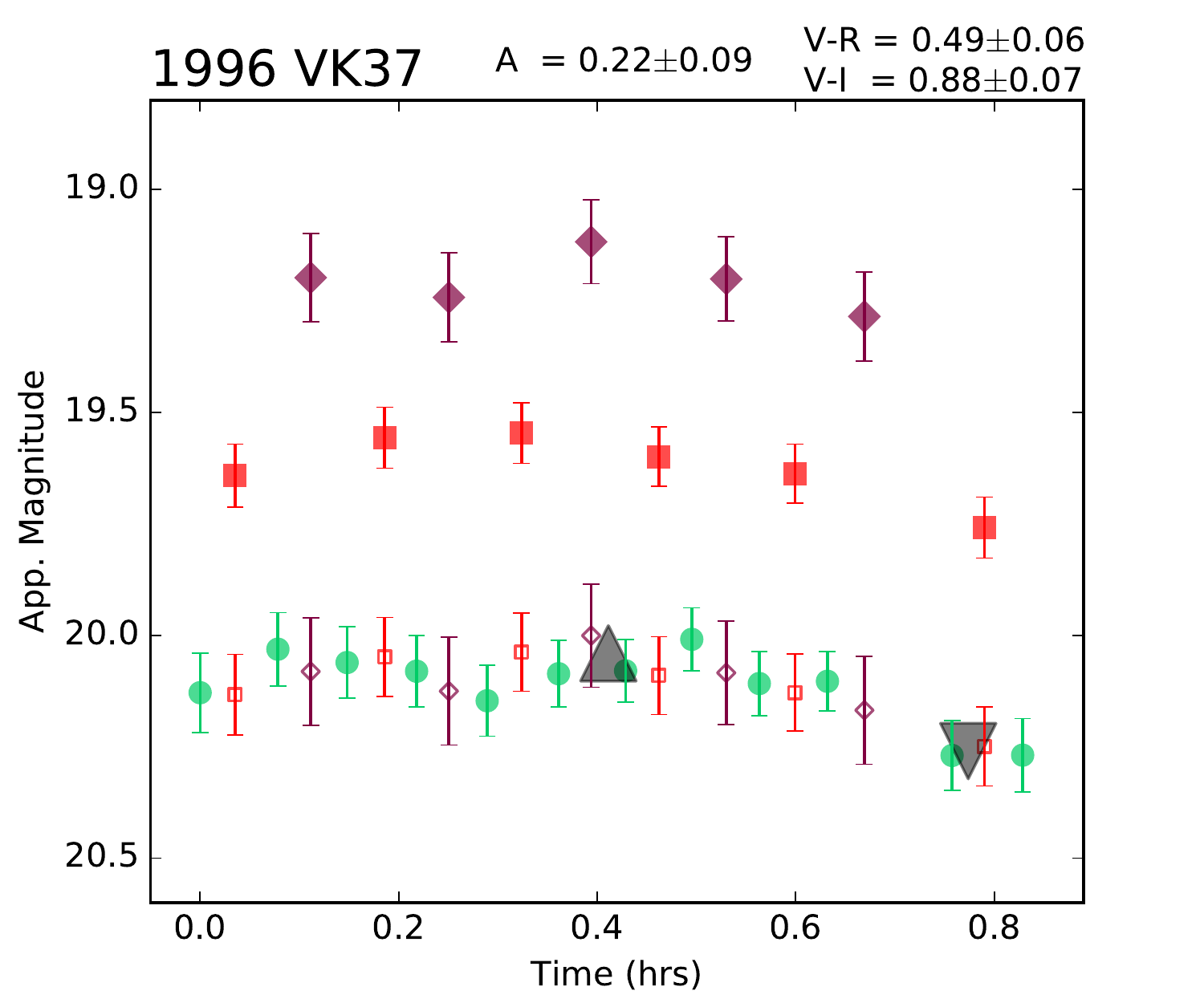}{0.3\textwidth}{(36)}
	}
	\gridline{\fig{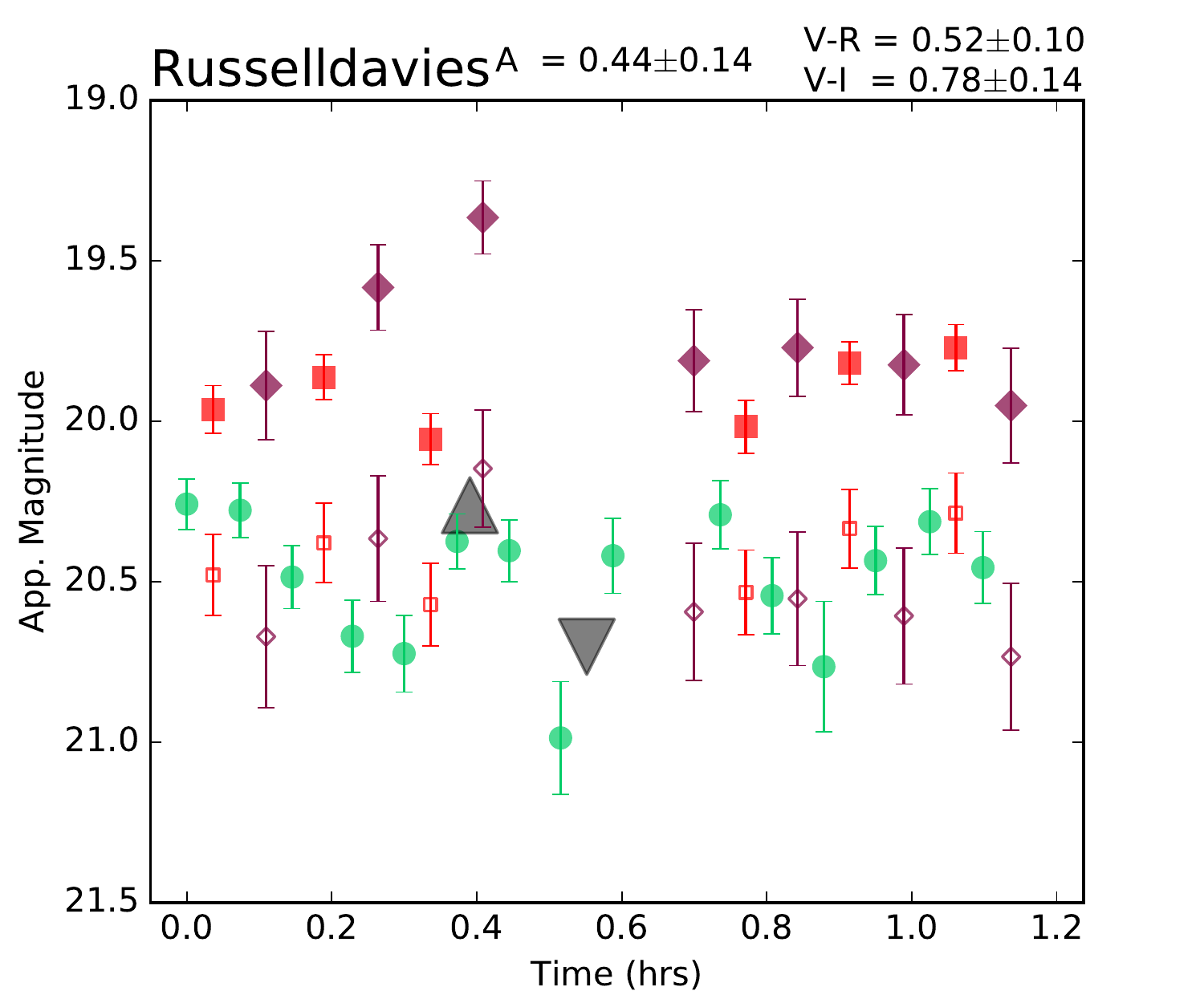}{0.3\textwidth}{(37)}
		\fig{figures/f3_38.pdf}{0.3\textwidth}{(38)}
		\fig{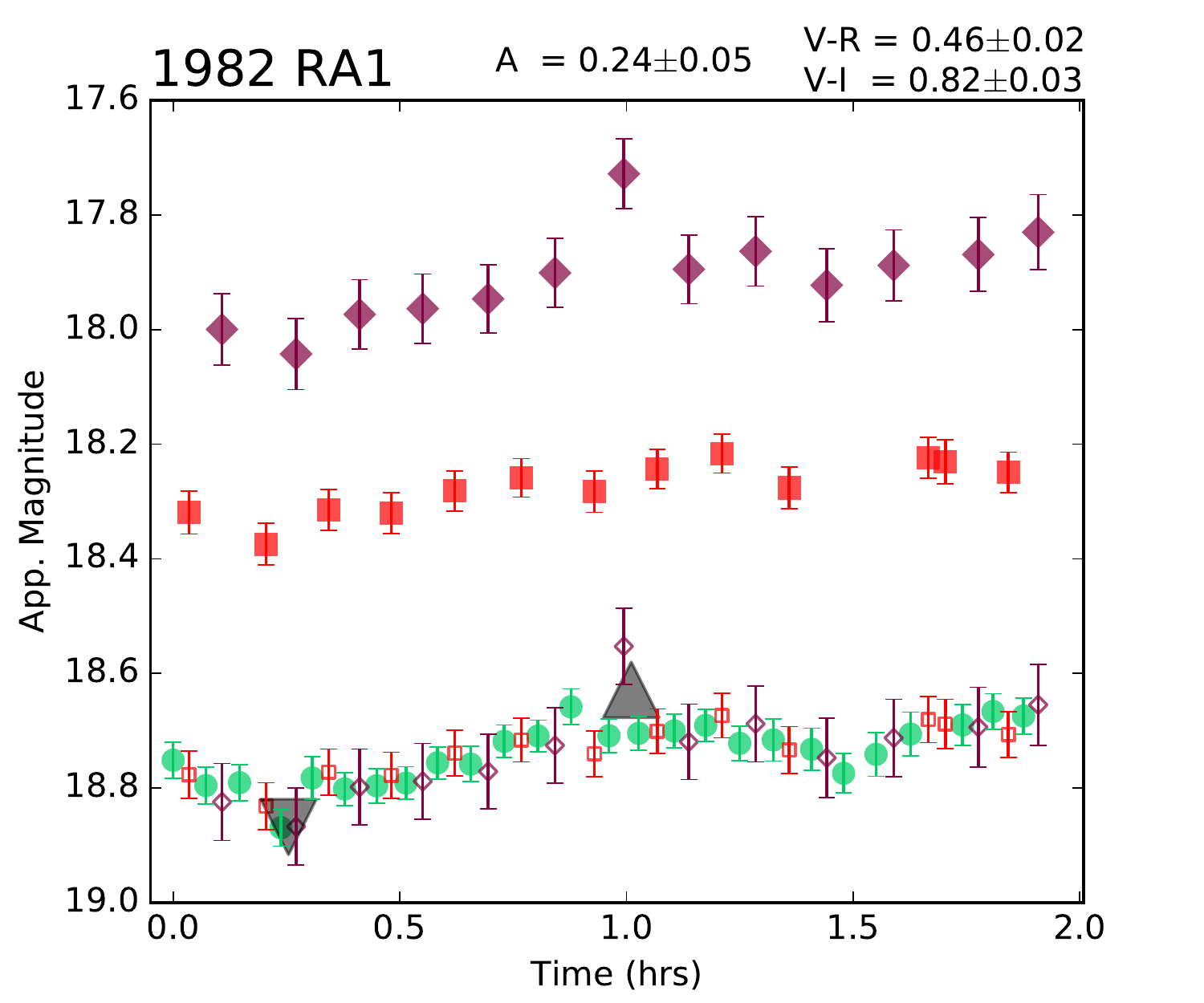}{0.3\textwidth}{(39)}
	}		
	\caption{(1)-(39) The spectrophotometric data for all 39 observed NEAs (same order as listed in Table \ref{table:results}). The data are the calibrated photometric results generated by the PHOTOMETRYPIPELINE  developed by \cite{Mommert2017}. The \textit{V} (green circles), the \textit{R} (red squares) and the \textit{I} (burgundy diamonds) filter data are shown with adjusted \textit{R} and \textit{I} data also indicated resulting in a final light curve (bottom set of data in each plot). The lower limit on the amplitude is calculated using the difference between magnitudes highlighted with the gray triangle symbols. The triangles are situated where the mean of two adjacent data points have the minimum/maximum magnitude. See text and Figure \ref{fig3} for further description.}
	\label{appendix_fig1}
\end{figure}

\begin{figure}
	\gridline{\fig{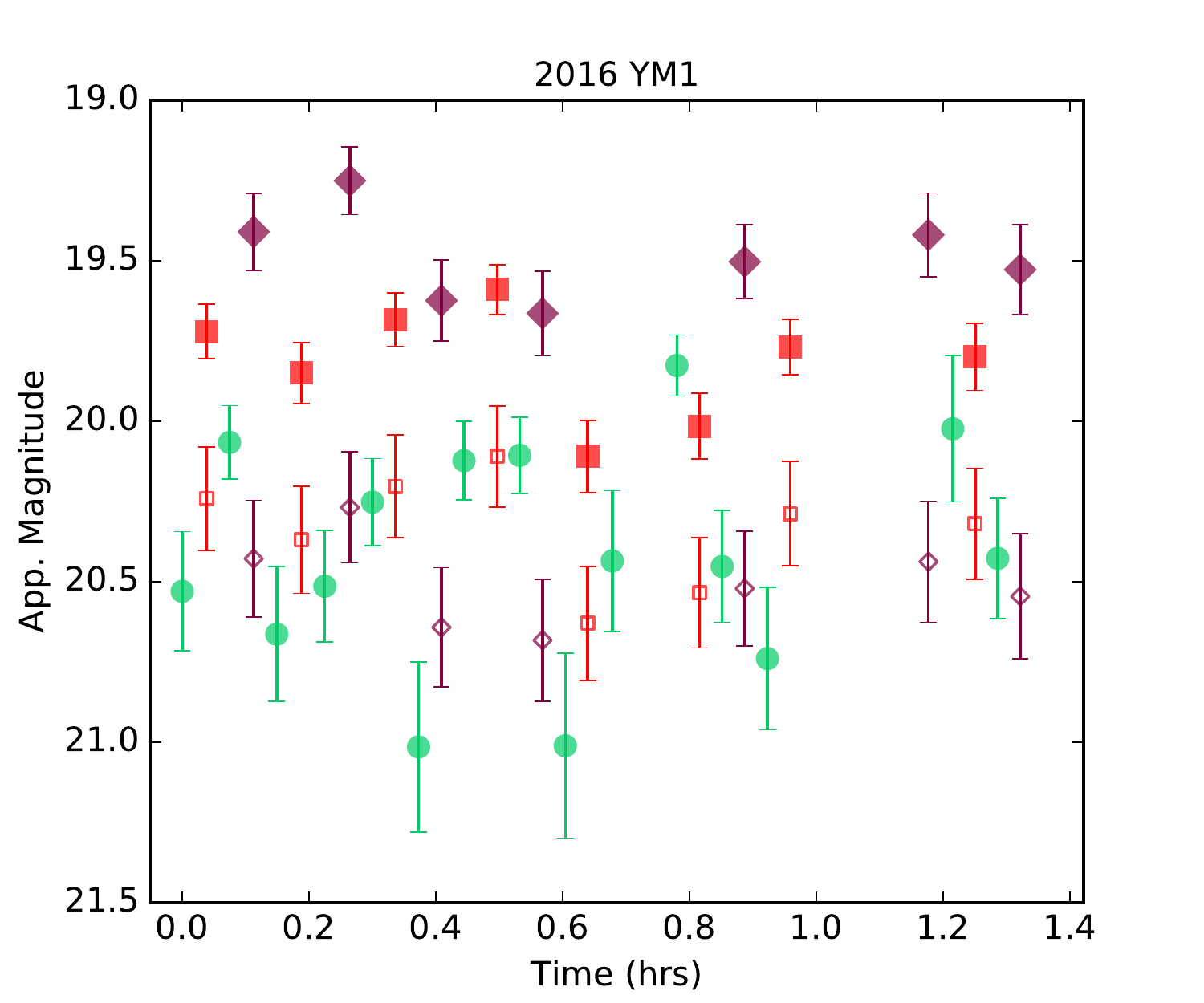}{0.3\textwidth}{}
		\fig{figures/f5_2_part1.pdf}{0.3\textwidth}{}
		\fig{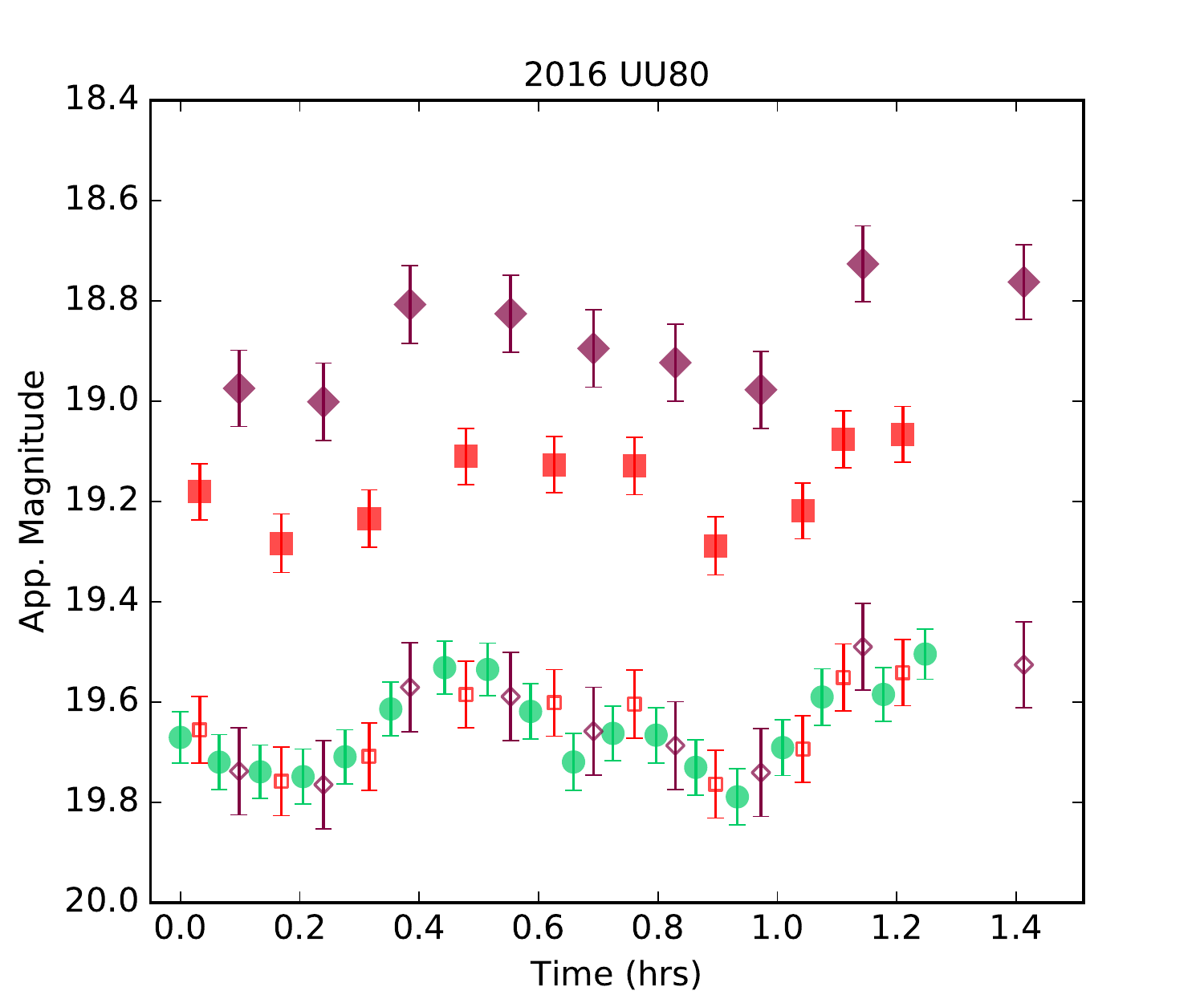}{0.3\textwidth}{}
	}
	\gridline{\fig{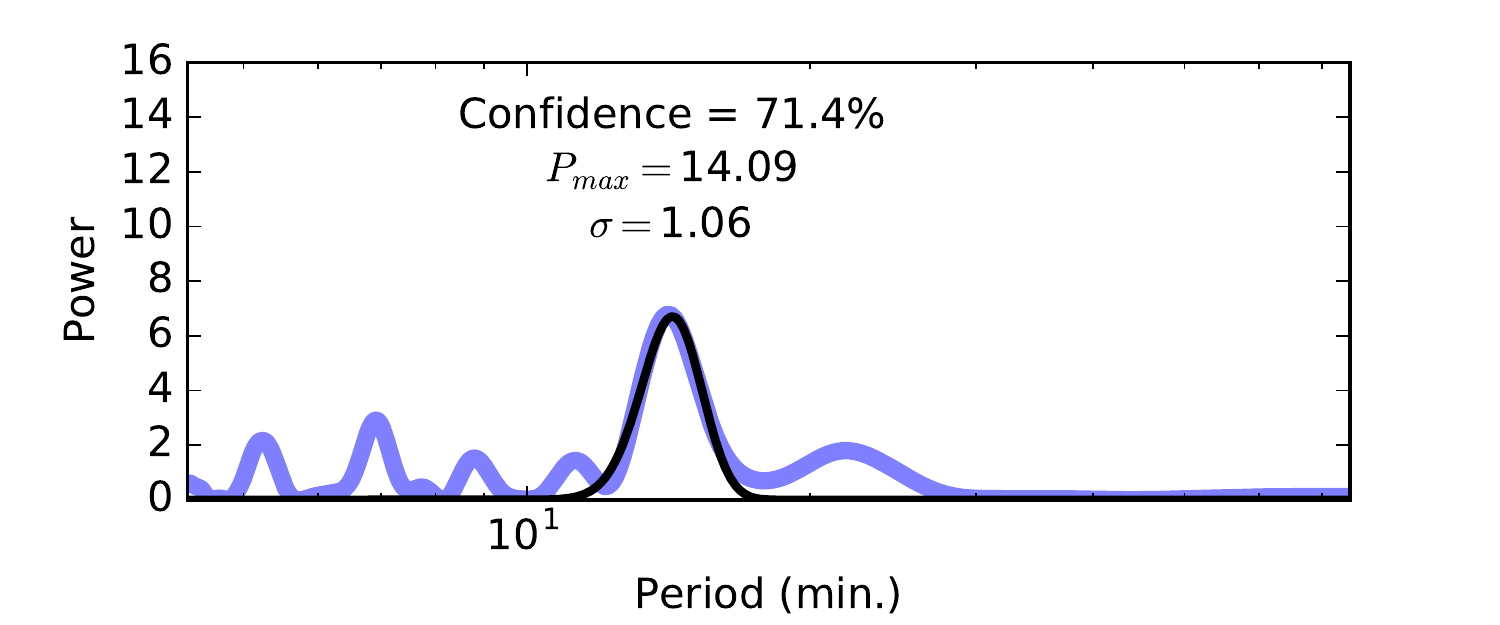}{0.3\textwidth}{}
		\fig{figures/f5_2_part2.pdf}{0.3\textwidth}{}
		\fig{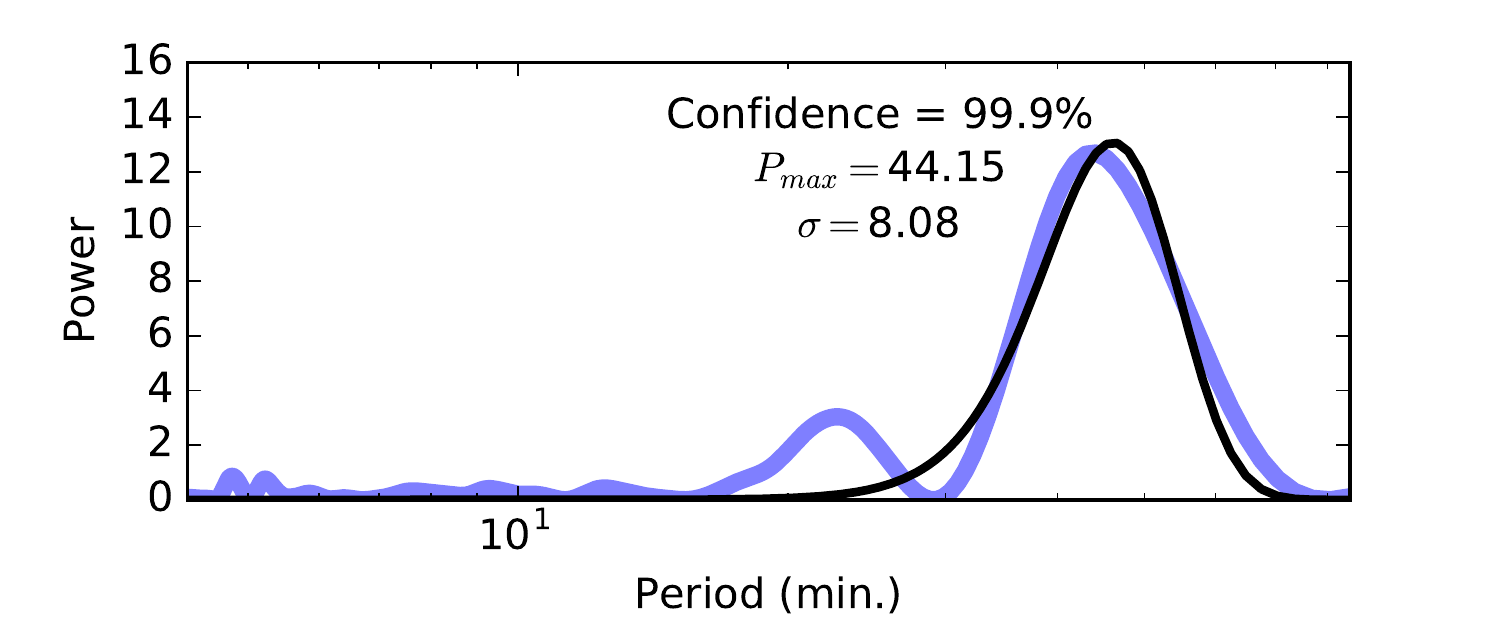}{0.3\textwidth}{}
	}
	\gridline{\fig{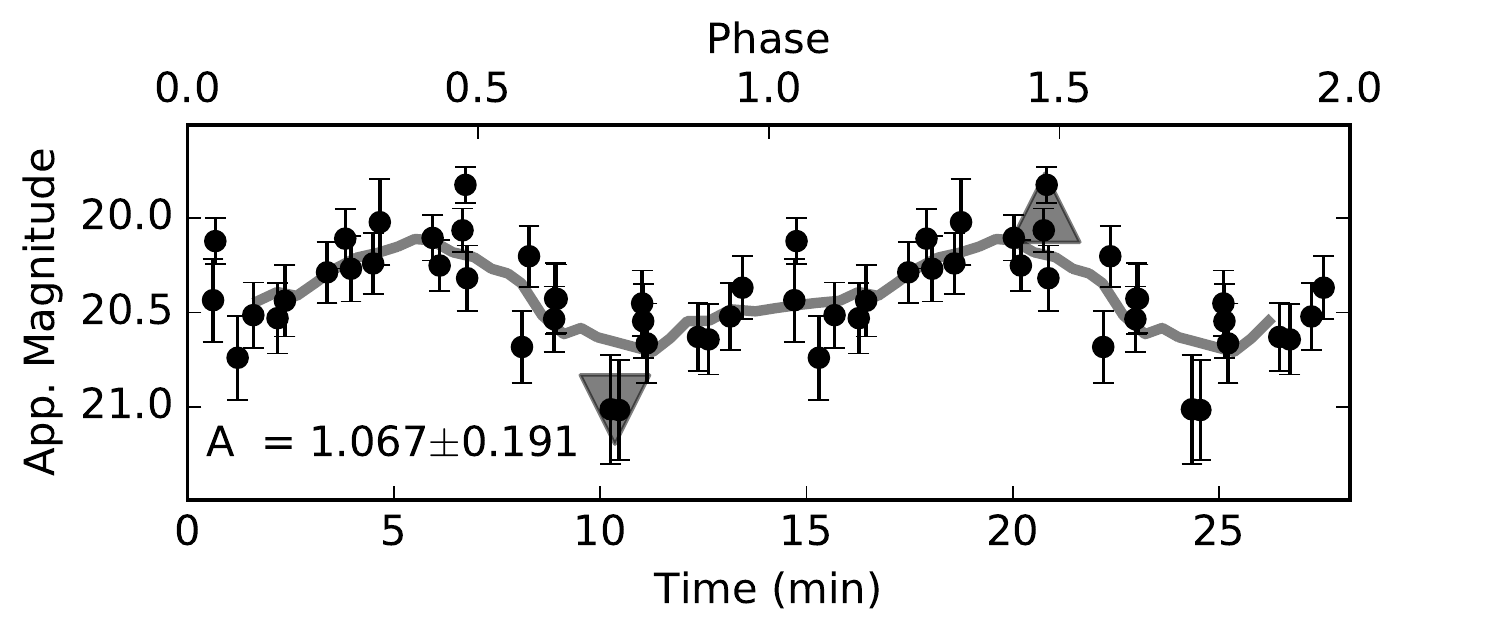}{0.3\textwidth}{(a)}
		\fig{figures/f5_2_part3.pdf}{0.3\textwidth}{(b)}
		\fig{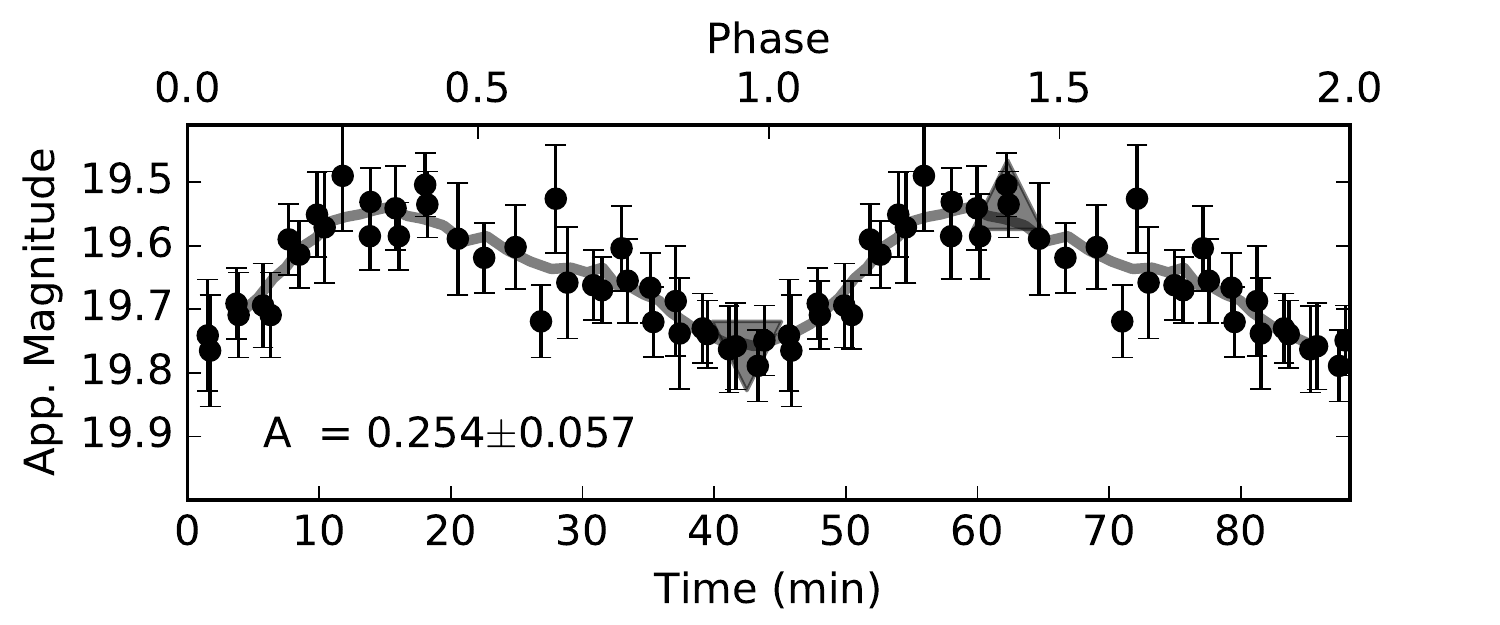}{0.3\textwidth}{(c)}
	}
	\gridline{\fig{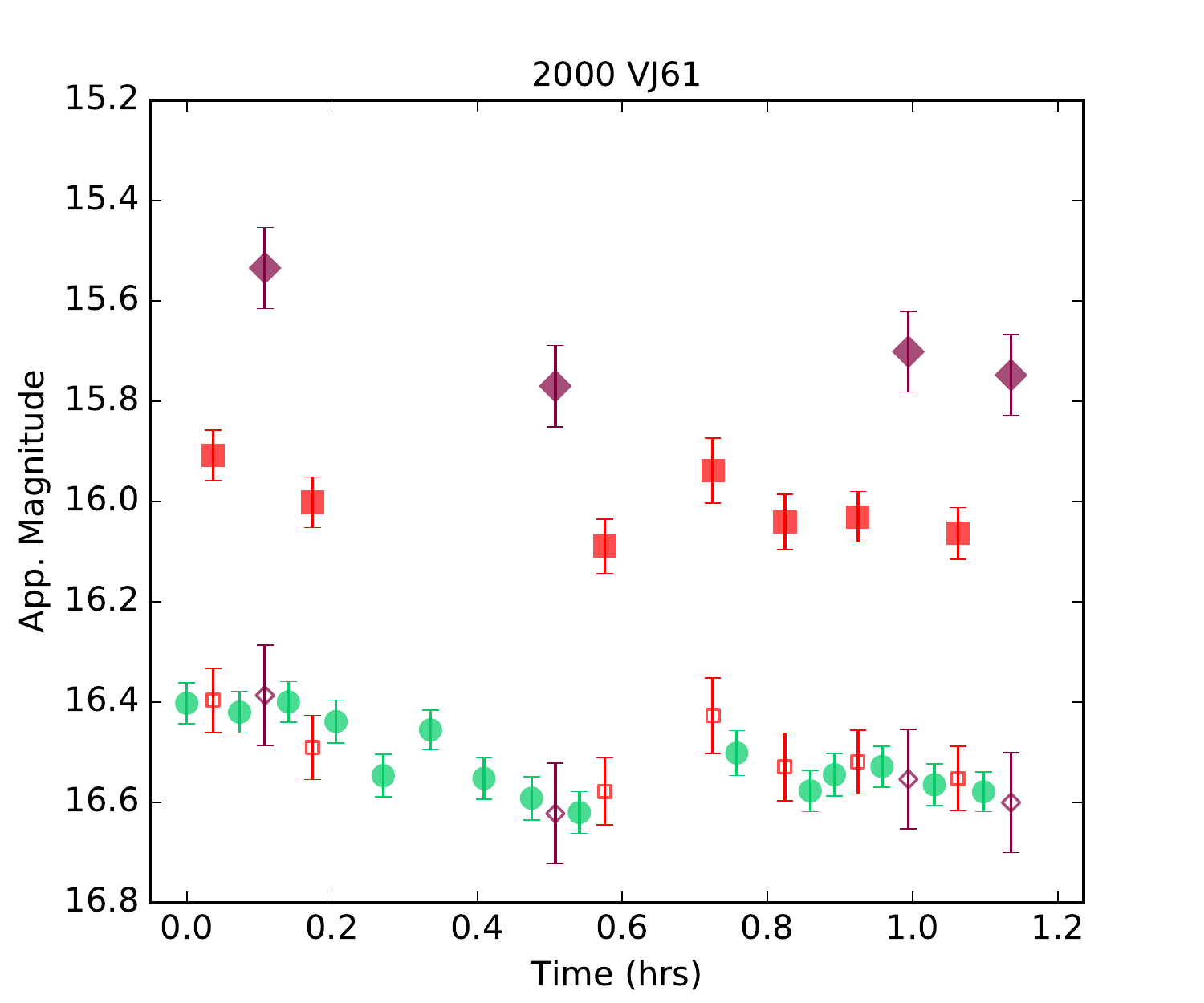}{0.3\textwidth}{}
		\fig{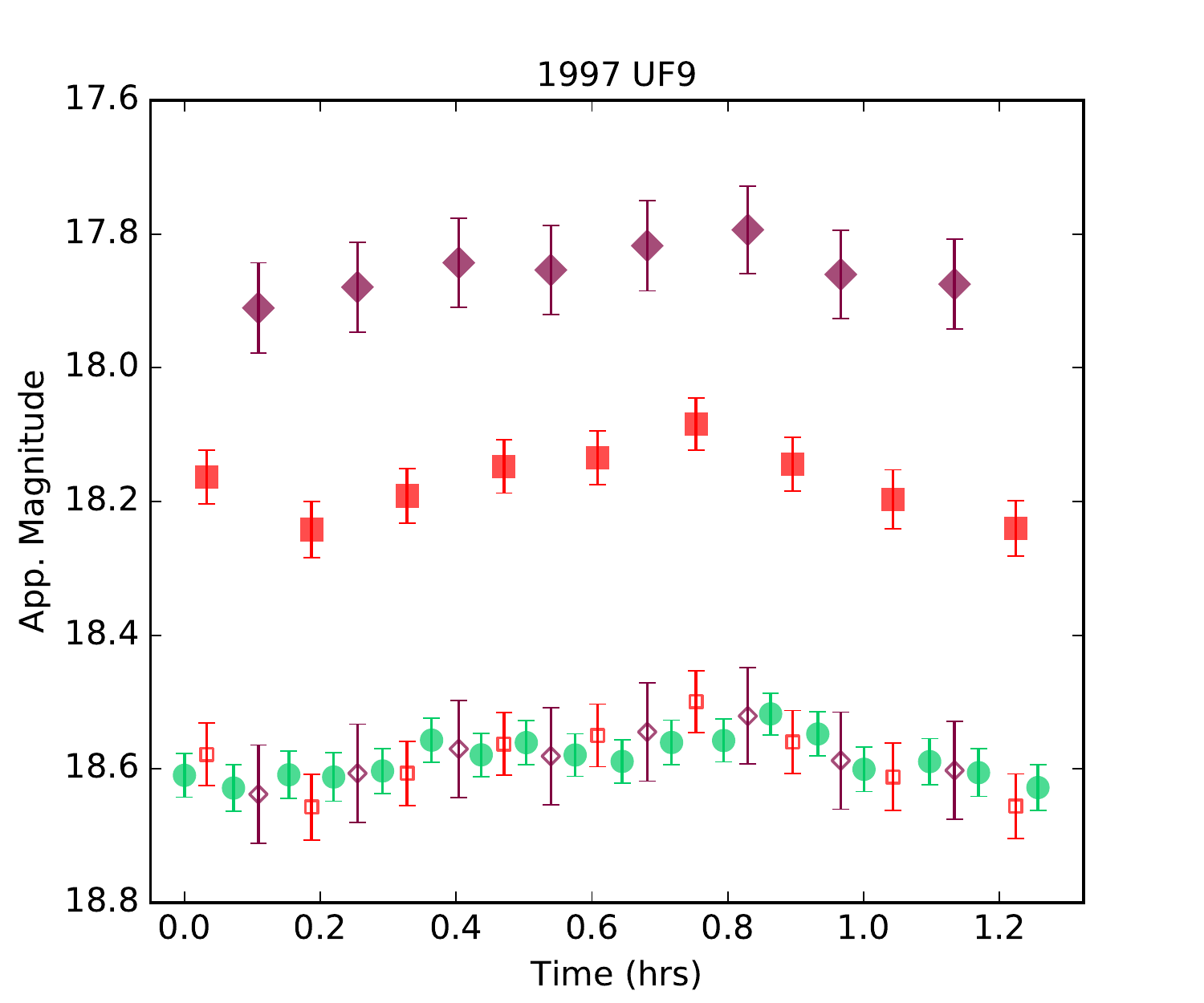}{0.3\textwidth}{}
		\fig{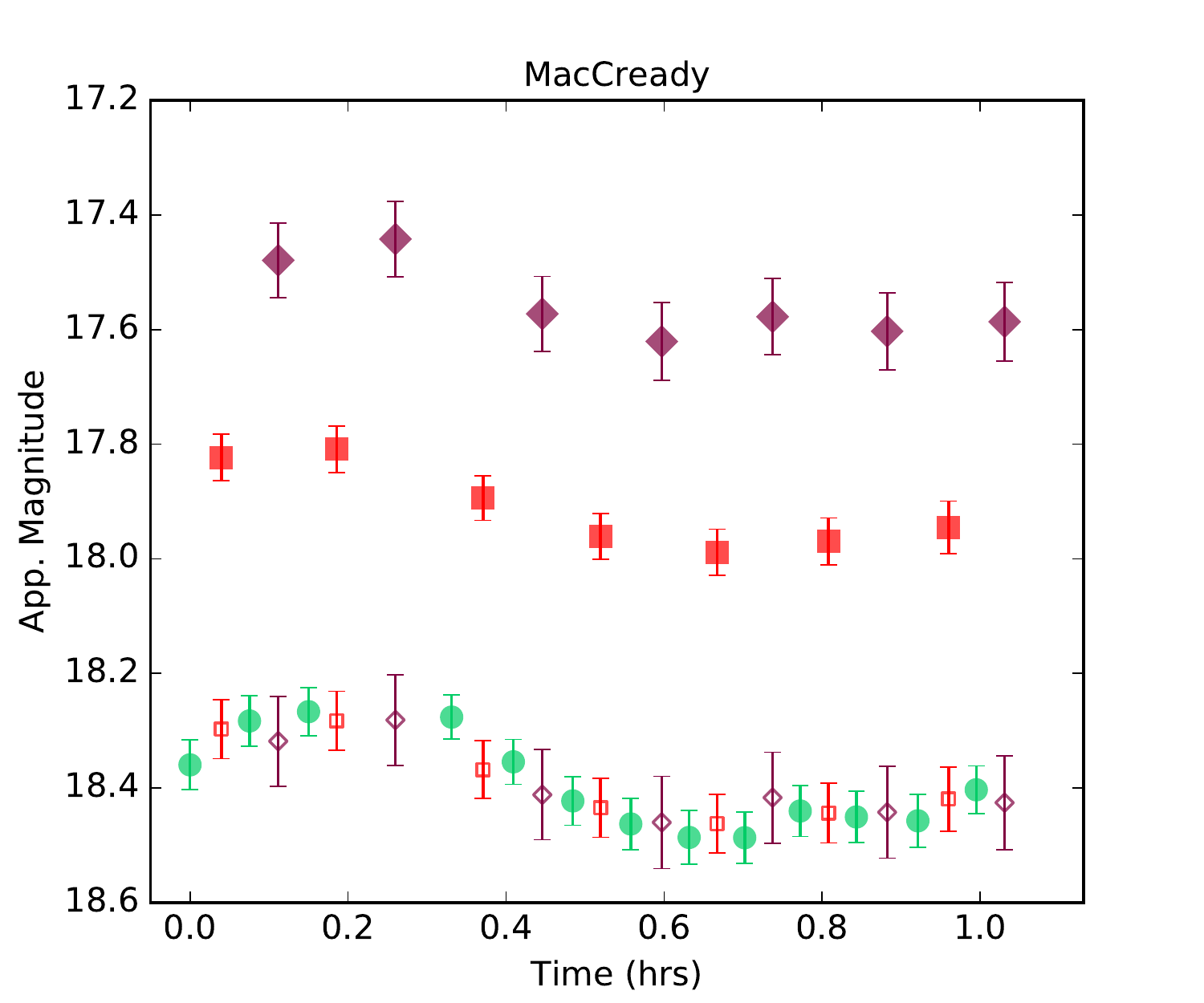}{0.3\textwidth}{}
	}
	\gridline{\fig{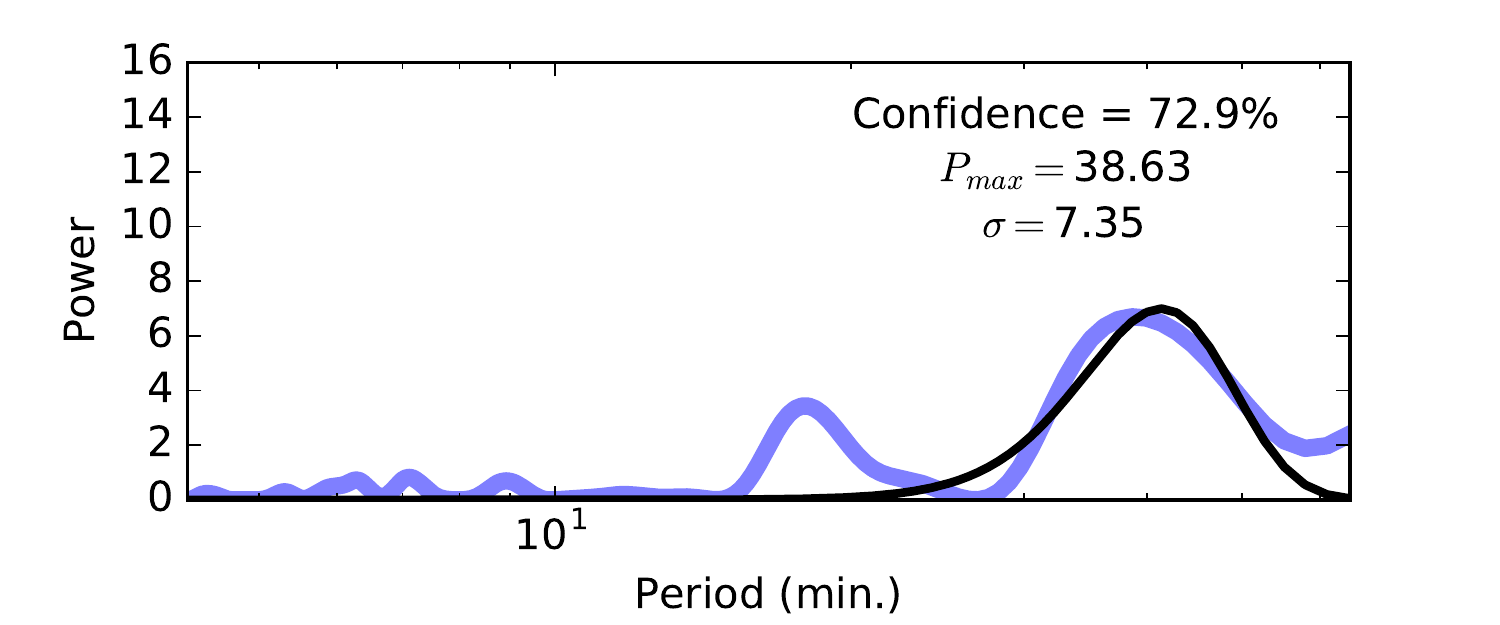}{0.3\textwidth}{}
		\fig{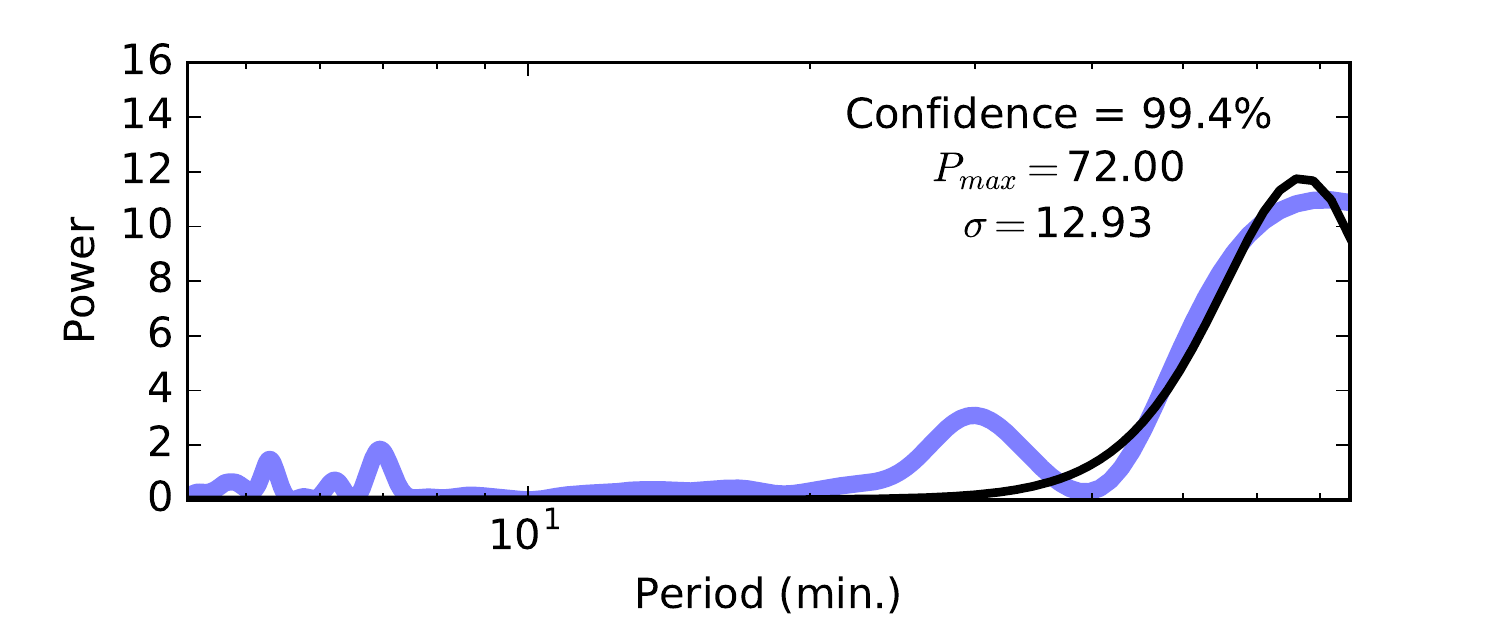}{0.3\textwidth}{}
		\fig{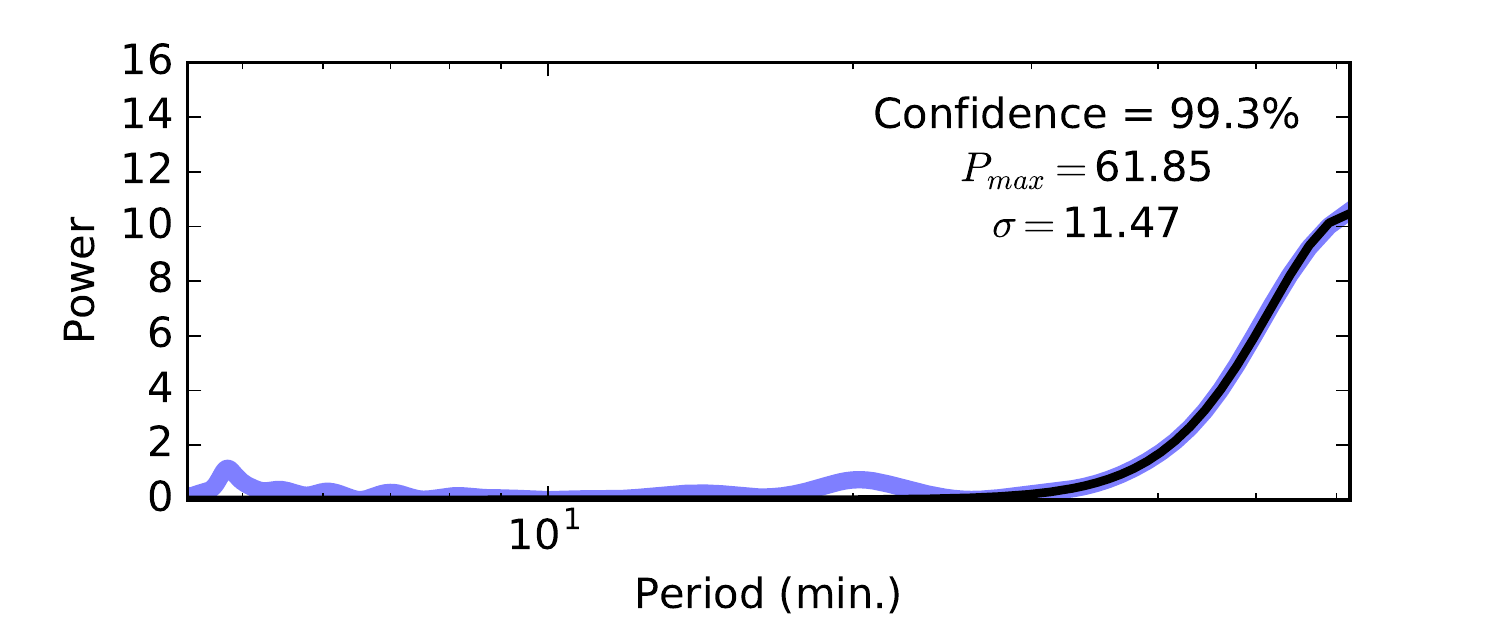}{0.3\textwidth}{}
	}
	\gridline{\fig{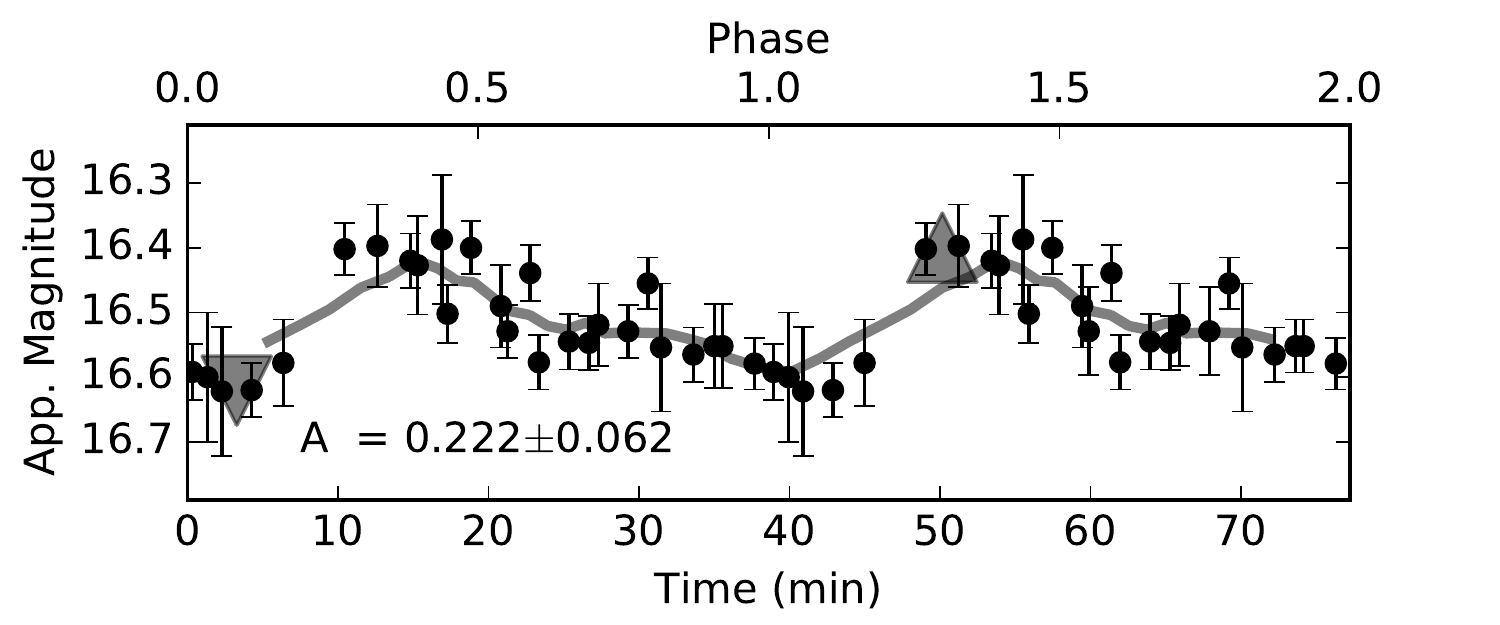}{0.3\textwidth}{(d)}
		\fig{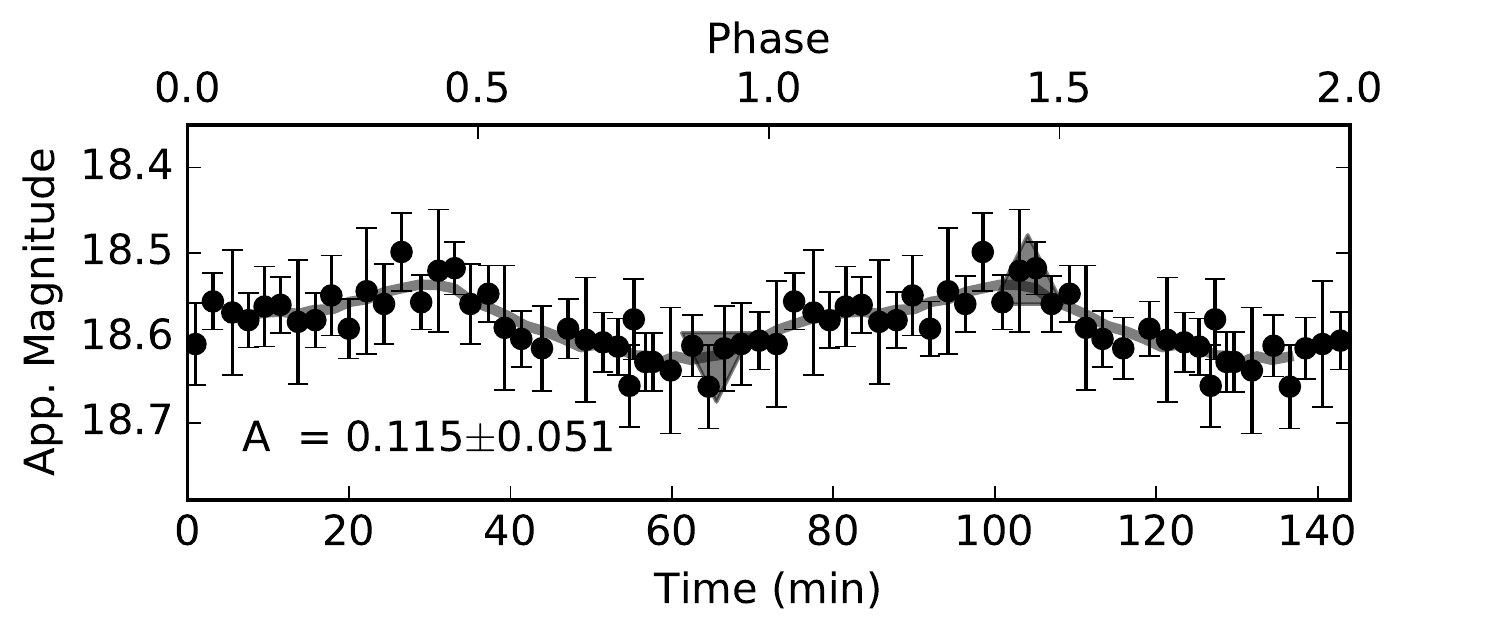}{0.3\textwidth}{(e)}
		\fig{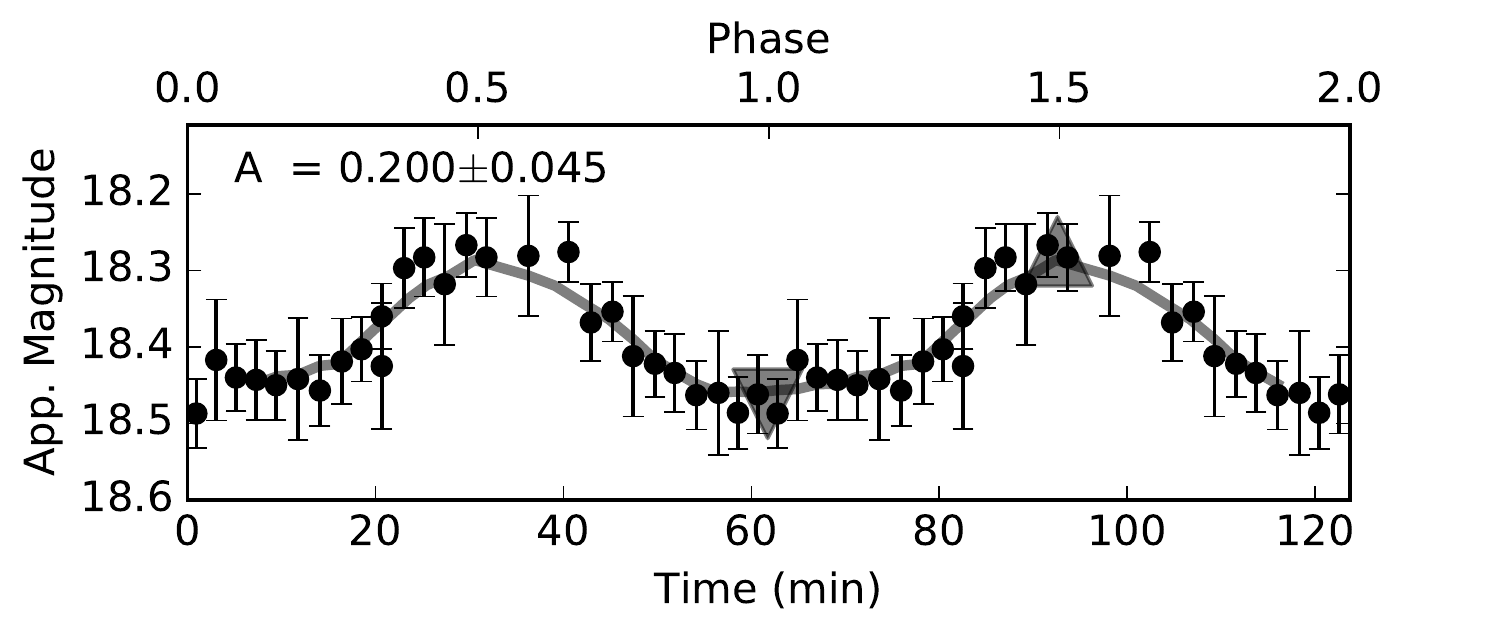}{0.3\textwidth}{(f)}
	}				
	\caption{The calibrated spectrophotometric data, periodogram and folded light curve of (a) 2016 YM1, (b) 2016 WJ1, (c) 2016 UU80, (d) 200 VJ61, (e) 1997 UF9 and (f) 1984 SS (MacCready). The periodogram of each targets shows a distinct peak at the peak position (P$_{max}$), which was also the interval used for folding the original light curve data. Note that the folded data spans two of the best-fit periods. The amplitude (A) is displayed with the triangles indicating the min/max magnitudes used.}
	\label{appendix_fig2}
\end{figure}

\end{document}